\documentclass{lmcs}
\pdfoutput=1

\usepackage{lastpage}
\lmcsdoi{21}{2}{25}
\lmcsheading{}{\pageref{LastPage}}{}{}%
{Feb.~03,~2024}{Jun.~19,~2025}{}

\usepackage{amssymb,amsmath,mathrsfs}
\usepackage{mathtools}
\usepackage{thmtools}
\usepackage{xspace}
\usepackage{dsfont}
\usepackage{stmaryrd}
\usepackage[normalem]{ulem}
\usepackage{tikz}
\usepackage[utf8]{inputenc}
\usepackage{booktabs}
\usetikzlibrary{
    cd,
    fit,
    calc,
    positioning,
    arrows,
    automata,
    shapes
}
\tikzset{
    baseline = (current bounding box.center),
    every state/.append style = {
        rectangle,
		inner sep = 2pt,
		minimum size = 1.5em,
		initial text = {}
	},
	every edge/.append style = {
		->,
		>=stealth,
		bend angle=10,
		thick
	}
}

\usepackage{tikz-cd} 
\usetikzlibrary{arrows,shapes,snakes,automata,backgrounds,petri}
\usepackage{ifthen}
\usepackage[noadjust]{cite}
\usepackage{rotating}
\usepackage{mathpartir}
\usepackage{microtype}
\usepackage{graphicx}
\usepackage{amsbsy,eucal}  
\usepackage{dsfont}
\usepackage{xspace}
\usepackage{xcolor}

\usepackage{hyperref}
\usepackage{cleveref}

\allowdisplaybreaks

\newcommand{\pstar}{\pow_{*}}

\renewcommand{\phi}{\varphi}

\newcommand{\takeout}[1]{\relax}
\renewcommand{\o}{\cdot}
\newcommand{\tr}[1]{\mathrel{\raisebox{-2pt}{\(\xrightarrow{\scriptstyle #1}\)}}}
\newcommand{\Term}{\mathsf{Term}}

\newcommand{\meas}{\operatorname{trm}}
\newcommand{\PTerm}{\mathsf{PTerm}}

\newcommand{\arrowstar}{\mathrel{\to^*}}

\newcommand{\arrowplus}{\mathrel{\to^+}}

\newcommand{\sem}[1]{\left\llbracket{#1}\right\rrbracket}
\newcommand{\bisim}{\mathrel{\raisebox{2pt}{\(\underline{\leftrightarrow}\)}}}
\newcommand{\K}{\mathbf{K}}
\newcommand{\Con}{\operatorname{Con}}

\newcommand{\Set}{\mathsf{Set}}
\newcommand{\reals}{\mathbb{R}}

\newcommand{\trace}{\operatorname{tr}}
\newcommand{\str}{\operatorname{str}}

\newcommand{\pow}{\mathcal{P}}
\newcommand{\set}[1]{\{ #1 \}} 

\newcommand{\axiom}[1]{\mathsf{#1}}
\newcommand{\D}{{\mathcal{D}}}
\newcommand{\Prob}{\mbox{\sf Prob}}
\newcommand{\N}{{\mathbb N}}
\newcommand{\supp}{\operatorname{supp}}
\newcommand{\diam}{\mathsf{dm}}

\renewcommand{\o}{\circ}

\newcommand{\textax}[1]{(\textsf{#1})}

\newcommand{\diff}{\mathsf{d}}

\begin{document}

\title{Fractals from Regular Behaviours}

\author[Todd Schmid]{Todd Schmid\lmcsorcid{0000-0002-9838-2363}}[a]
\author[Victoria Noquez]{Victoria Noquez\lmcsorcid{0000-0001-5517-0929}}[b]
\author[Lawrence S.~Moss]{Lawrence S.~Moss\lmcsorcid{0000-0002-9908-5774}}[c]

\address{Bucknell University, USA}
\email{t.schmid@bucknell.edu}
\address{St.~Mary's College of California, USA}
\email{vln1@stmarys-ca.edu}
\address{Indiana University, USA}
\email{lmoss@iu.edu}

\begin{abstract}
    We forge connections between the theory of fractal sets obtained as attractors of iterated function systems and process calculi. 
    To this end, we reinterpret Milner's expressions for processes as contraction operators on a complete metric space. 
    When the space is, for example, the plane, the denotations of fixed point terms correspond to familiar fractal sets. 
    We give a sound and complete axiomatization of fractal equivalence, the congruence on terms consisting of pairs that construct identical self-similar sets in all interpretations. 
    We further make connections to labelled Markov chains and to invariant measures. 
    In all of this work, we use important results from process calculi. 
    For example, we use Rabinovich's completeness theorem for trace equivalence in our own completeness theorem. 
    In addition to our results, we also raise many questions related to both fractals and process calculi. 
\end{abstract}

\maketitle

%\setcounter{tocdepth}{4}
%\tableofcontents

    \section{Introduction}
\begin{figure}[!b]
        \centering
        \begin{tabular}{c l}
            \(\begin{gathered}
                \begin{tikzpicture}
                    \node[state] (0) at (0,0) {\(x\)};
                    \draw[loop] (0) edge[loop left] node[left] {\(a,b,c\)} (0);
                \end{tikzpicture} \\\\
                \sigma_a\begin{bmatrix}
                    r \\ s
                \end{bmatrix} = \begin{bmatrix}
                    \frac12r + \frac14 \\ \frac12s + \frac{\sqrt3}4
                \end{bmatrix}
                \qquad 
                \sigma_b\begin{bmatrix}
                    r \\ s
                \end{bmatrix} = \begin{bmatrix}
                    \frac12r \\ \frac12s
                \end{bmatrix}
                \\
                \sigma_c\begin{bmatrix}
                    r \\ s
                \end{bmatrix} = \begin{bmatrix}
                    \frac12r + \frac12 \\ \frac12s
                \end{bmatrix}
            \end{gathered}\)
            &
            \raisebox{-2cm}{\includegraphics[height=4cm]{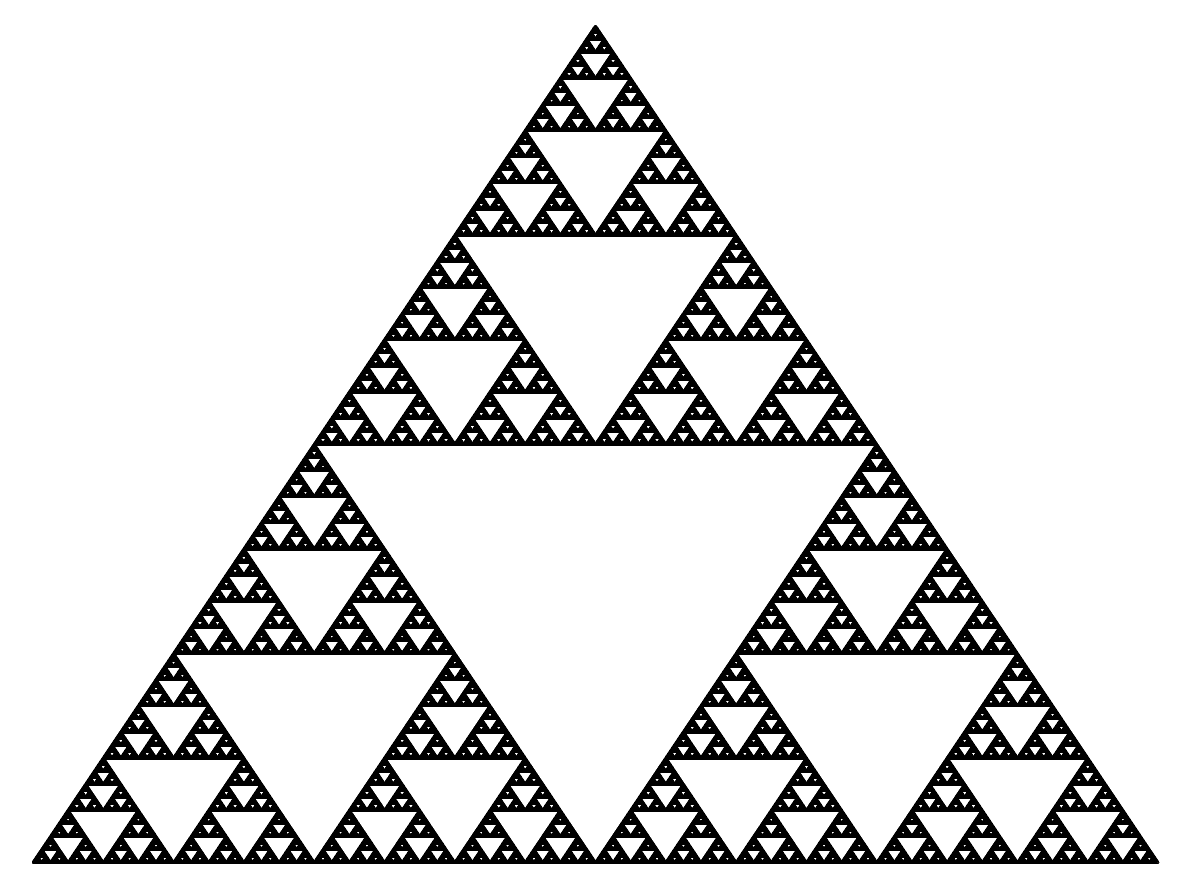}}
        \end{tabular}
        \caption{\label{fig:Sierpinski gasket} The Sierpi\'nski gasket is the unique non-empty compact subset \(\mathbf S\) of \(\reals^2\) such that \(\mathbf S = \sigma_a(\mathbf S) \cup \sigma_b(\mathbf S) \cup \sigma_c(\mathbf S)\). Each of its points corresponds to a stream emitted by the state \(x\).}
    \end{figure}

    Hutchinson noticed in~\cite{Hutchinson1981Fractals} that many familiar examples of fractals can be captured as the set-wise fixed-point of a
    finite family of contraction (i.e., distance shrinking) operators on a metric space.
    He called these spaces \emph{self-similar}, since the intuition behind the contraction operators is that they are witnesses for the appearance of the fractal in a proper (smaller) subset of itself.
    For example, the famous Sierpi\'nski gasket is the unique non-empty compact subset of the plane left fixed by the union of the three operators \(\sigma_a,\sigma_b,\sigma_c : \reals^2 \to \reals^2\) in \autoref{fig:Sierpinski gasket}.

    The self-similarity of Hutchinson's fractals hints at an algorithm for constructing them: Each point in a self-similar set is the limit of a sequence of points obtained by applying the contraction operators one after the other to an initial point.
    In the Sierpi\'nski gasket, the point \((1/4,\sqrt{3}/4)\) is the limit of the sequence
    \begin{equation}\label{eq:first stream eg}
        p,\ 
        \sigma_b(p),\ 
        \sigma_b \sigma_a(p),\ 
        \sigma_b \sigma_a \sigma_a(p),\ 
        \sigma_b \sigma_a \sigma_a \sigma_a(p),\ \dots
    \end{equation}
    where the initial point \(p\) is an arbitrary element of \(\reals^2\) (note that \(\sigma_b\) is applied last).
    Hutchinson showed in~\cite{Hutchinson1981Fractals} that the self-similar set corresponding to a given family of contraction operators is precisely the collection of points obtained in the manner just described.
    The limit of the sequence in~\eqref{eq:first stream eg} does not depend on the initial point \(p\) because \(\sigma_a,\sigma_b,\sigma_c\) are contractions.
    Much like digit expansions to real numbers, every stream of \(a\)'s, \(b\)'s, and \(c\)'s corresponds to a unique point in the Sierpi\'nski gasket, as we have seen in (\ref{eq:first stream eg}).
    The point \((1/4, \sqrt3/4)\), for example, corresponds to the stream \((b,a,a,a,\dots)\) ending in an infinite sequence of \(a\)'s.
    Conversely, every point in the Sierpi\'nski gasket comes from (in general more than one) corresponding stream.

\begin{figure}[!ht]
        \centering
        \begin{tabular}{c l}
            \(\begin{gathered}
                \begin{tikzpicture}
                    \node[state] (0) at (0,0) {\(x\)};
                    \node[state] (1) at (2,0) {\(y\)};
                    \draw (0) edge[out=30, in=150] node[above] {\(b,c\)} (1);
                    \draw[loop] (0) edge[loop left] node[left] {\(a\)} (0);
                    \draw (1) edge[out=210, in=330] node[above] {\(b,c\)} (0);
                \end{tikzpicture} \\
                \begin{aligned}
                	\sigma_a\begin{bmatrix}
                    r \\ s
                \end{bmatrix} &= \begin{bmatrix}
                    \frac12r + \frac14 \\ \frac12s + \frac{\sqrt3}4
                \end{bmatrix}
                \\
                \sigma_b\begin{bmatrix}
                    r \\ s
                \end{bmatrix} &= \begin{bmatrix}
                   \frac r2 \\ \frac s2
                \end{bmatrix}
                \\
                \sigma_c\begin{bmatrix}
                    r \\ s
                \end{bmatrix} &= \begin{bmatrix}
                    \frac r2 + \frac12 \\ \frac s2
                \end{bmatrix}
                \end{aligned}
            \end{gathered}\)
            &
            \begin{tabular}{c c}
	            \raisebox{-2cm}{\includegraphics[height=3.3cm]{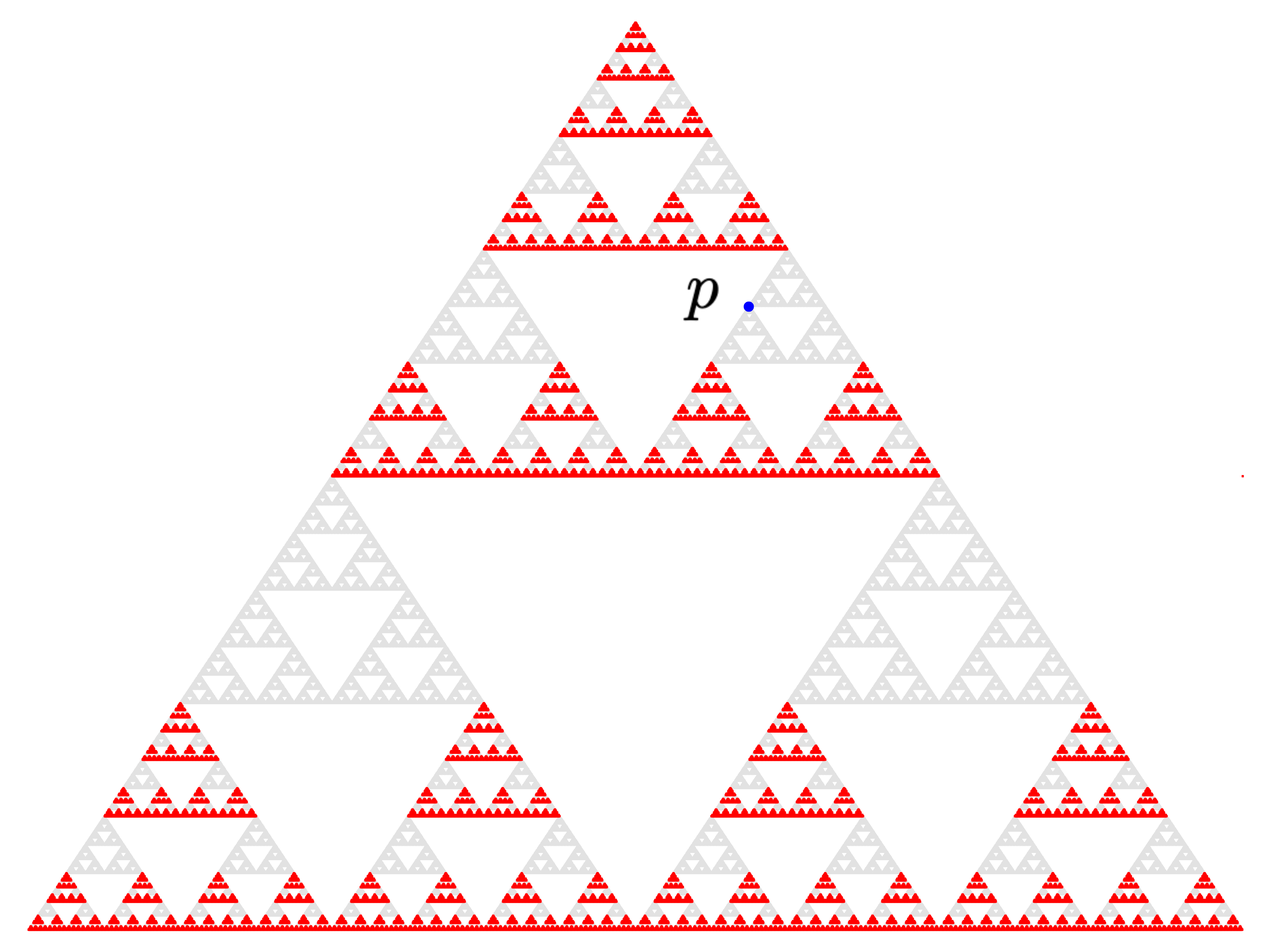}} 
	            &
	            \raisebox{-2cm}{\includegraphics[height=3.3cm]{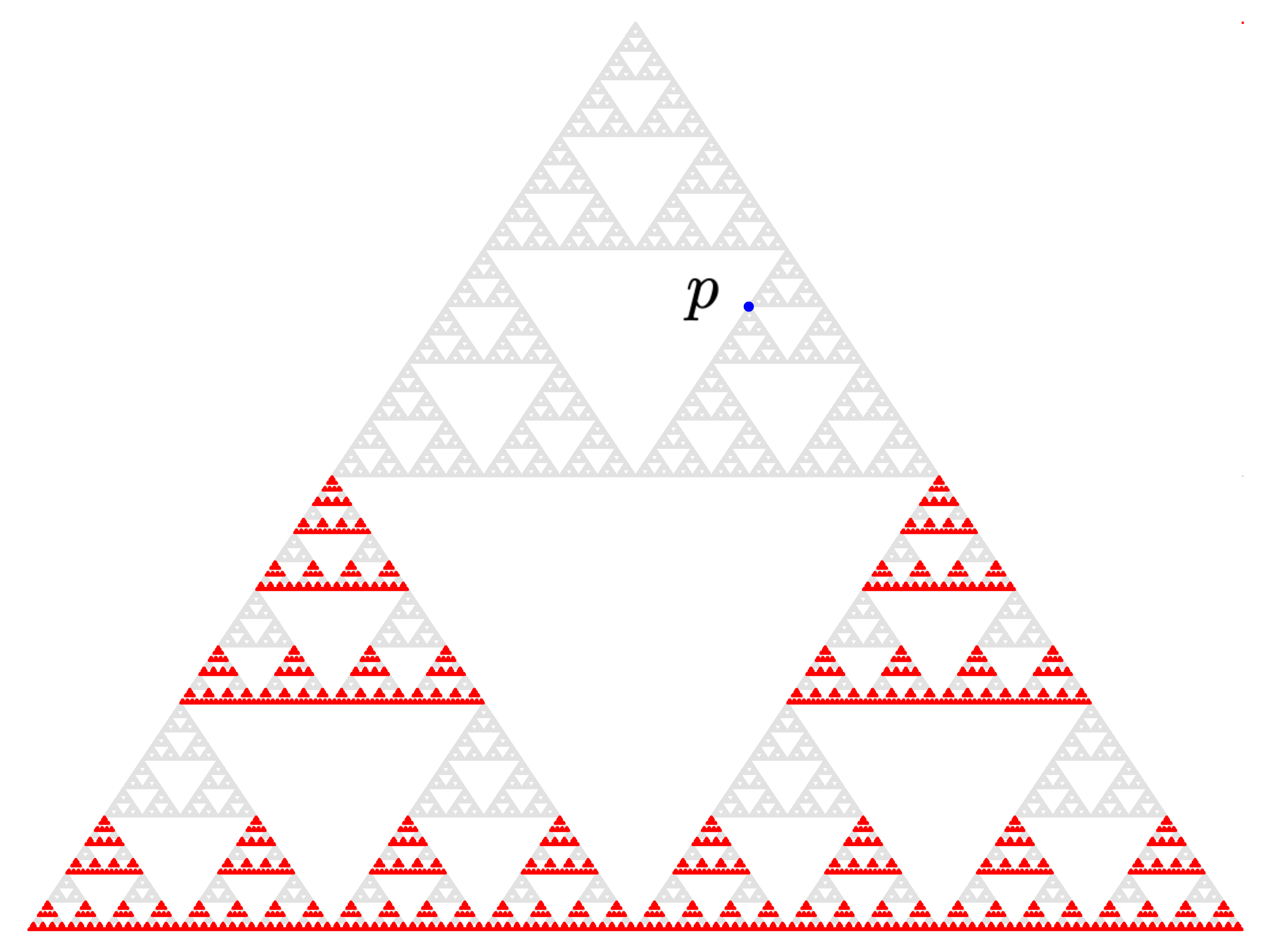}} \\
	            Generated by \(x\)
	            & Generated by \(y\)
            \end{tabular}
        \end{tabular}
        \caption{\label{fig:Twisted Sierpinski gasket} 
            Twisted Sierpi\'nksi gaskets, depicted in {\color{red}red}, and the point \(p = (\frac{19}{32}, \frac{11\sqrt{3}}{32})\).
            The fractal set generated by \(x\) is the unique compact set fixed by taking the union of the image of \(\sigma_a\) with the images of the four compositions \(\sigma_b\sigma_b\), \(\sigma_b\sigma_c\), \(\sigma_c\sigma_b\), and \(\sigma_c\sigma_c\).
            Intuitively, this is the set obtained from the Sierpinski gasket by recursively removing the upper triangle in the bottom two thirds.
            The set generated by \(y\) is the union of \(\sigma_b\) and \(\sigma_c\) applied to the set generated by \(x\).
        }
    \end{figure}

    From a computer science perspective, the languages of streams considered by Hutchinson are the \emph{traces} observed by one-state labelled transition systems, like the one in \autoref{fig:Sierpinski gasket}.
    We investigate whether one could achieve a similar effect with languages of streams obtained from labelled transition systems having more than one state.
    Observe, for example, \autoref{fig:Twisted Sierpinski gasket}.
    These twisted versions of the Sierpi\'nski gasket are constructed from the streams emitted by the two-state labelled transition system in \autoref{fig:Twisted Sierpinski gasket}, starting from the states \(x\) and \(y\) respectively.

    Each point in a twisted Sierpi\'nski gasket corresponds to a stream of \(a\)'s, \(b\)'s, and \(c\)'s, but not every stream corresponds to a point in the set: for example, the limit corresponding to the stream \((a, c, a, b, a, a, a, \dots)\) 
    is \(p = (19/32, 11\sqrt{3}/32)\).
    This stream is not emitted by either of the states \(x\) or \(y\), and consequently the point \(p\) does not appear in either of the twisted Sierpi\'nski gaskets generated by \(x\) or \(y\). 

    In analogy with the theory of regular languages, we call the fractals generated by finite labelled transition systems \emph{regular subfractals}, and give a logic for deciding if two labelled transition systems represent the same recipe under all interpretations of the labels, allowing both the underlying space and the chosen contractions to vary.
    By identifying points in the fractal set generated by a labelled transition system with traces observed by the labelled transition system, it is reasonable to suspect that two labelled transition systems represent equivalent fractal recipes---i.e., they represent the same fractal under every interpretation---if and only if they are trace equivalent.
    This is the content of \autoref{thm:completeness}, which allows us to connect the theory of fractal sets to mainstream topics in computer science.

    Labelled transition systems are a staple of theoretical computer science, especially in the area of process algebra~\cite{Baeten2005History}, where a vast array of different notions of equivalence and axiomatization problems have been studied.
    We specifically use a syntax introduced by Milner in~\cite{Milner1984Complete} to express labelled transition systems as terms in an expression language with recursion.
    This leads us to a fragment of Milner's calculus consisting of just the terms that constitute recipes for fractal constructions.
    Using a logic of Rabinovich~\cite{Rabinovich1993Traces} for deciding trace equivalence in Milner's calculus, we obtain a complete axiomatization of fractal recipe equivalence.
    As a consequence of this completeness theorem, we are also able to show that Rabinovich's logic is complete for the specific interpretation of process terms as fractal subsets of the Cantor set.

    In his study of self-similar sets, Hutchinson also makes use of probability measures supported on self-similar sets, called \emph{invariant measures}.
    Each invariant measure is specified by a probability distribution on the set of contractions generating its support. 
    In the last technical section of the paper, we adapt the construction of invariant measures to a probabilistic version of labelled transition systems called \emph{labelled Markov chains}.
    These allow us to give a measure-theoretic semantics to terms in a probabilistic version of Milner's specification language, the calculus introduced by Stark and Smolka~\cite{Stark2000Probabilistic}.
    Specifically, we interpret each probabilistic process term as a probability measure supported by a regular subfractal.
    Our measure-theoretic semantics of probabilistic process terms can be seen as a generalization of the trace measure semantics of Kerstan and K\"onig~\cite{Kerstan2013Trace}.
    We offer a sound axiomatization of equivalence under this semantics and, calling on a result proved in a separate paper~\cite{cirstea_et_al2025} using other techniques, also 
    prove completeness.
    Finally, we show that our proposed axiomatization of fractal measure equivalence is also complete for the specific interpretation of probabilistic process terms as fractal measures on the Cantor set.

    In sum, the contributions of this paper are as follows.
    \begin{itemize}
        \item In \cref{sec:fractals from LTS}, we give a fractal recipe semantics to process terms using a generalization of iterated function systems.
        \item In \cref{sec:fractal and trace}, we show that two process terms agree on all fractal interpretations if and only if they are trace equivalent.
        This implies that fractal recipe equivalence is decidable for process terms, and it allows us to derive a complete axiomatization of fractal recipe equivalence from Rabinovich's axiomatization~\cite{Rabinovich1993Traces} of trace equivalence of process terms.
        \item We adapt the fractal semantics of process terms to the probabilistic setting in \cref{sec:calculus of measures}.
        We furthermore show that probabilistic fractal recipe equivalence coincides with Kerstan and K\"onig's notion of trace measure equivalence~\cite{Kerstan2013Trace}.
        \item Finally, we propose an axiomatization of probabilistic fractal recipe equivalence in Sections~\ref{sec:syntax for prob} and~\ref{sec:axioms for prob}.
        We prove that our axioms are sound  and complete with respect to our semantics.
    \end{itemize}

    % \textcolor{black}{    
    % The rest of this Introduction is background material.      
    % The paper begins in earnest with a brief overview of trace semantics in process algebra and Rabinovich's Theorem (\autoref{thm:Rabinovich}) in \cref{sec:LTS and trace}.
    % }

    % SECTION
    \section{Labelled Transition Systems and Trace Semantics}%
    \label{sec:LTS and trace}
    Labelled transition systems are a widely used model of nondeterminism.
    We are  given a fixed finite set \(A\) of \emph{action labels}.
    A \emph{labelled transition system} (LTS) is a pair \((X, \alpha)\) consisting of a set \(X\) of \emph{states} (also called the \emph{state space}) and a \emph{transition function} \(\alpha \colon X \to \mathcal P(A \times X)\).
    We generally write \(x \tr{a}_\alpha y\) if \((a, y) \in \alpha(x)\), or simply \(x \tr{a} y\) if \(\alpha\) is clear from context, and say that \emph{\(x\) emits \(a\) and transitions to \(y\)}.

    Given a state \(x\) of an LTS \((X, \alpha)\), we write \(\langle x \rangle _\alpha\) for the LTS obtained by restricting the relations \(\tr{a}\) to the set of states \emph{reachable} from \(x\), meaning there exists a \emph{path} of the form \(x \tr{a_1} x_1 \tr{ } \cdots \tr{ } x_{n-1} \tr{a_n} x_n\). 
    We refer to \(\langle x \rangle_\alpha\) as either the LTS \emph{generated by \(x\)}, or the \emph{process starting at \(x\)}.
   	An LTS \((X, \alpha)\) is \emph{locally finite} if \(\langle x \rangle_\alpha\)  is finite for every state $x$.
 	The reader should note that there exist locally finite LTSs with infinite state spaces, and that one such LTS will be important in this paper;
	see~\autoref{def:term}.
    % Traces observed by a process
    \paragraph*{Traces}
    In the context of the current work, nondeterminism occurs when a process branches into multiple threads that execute in parallel.
    Under this interpretation, to an outside observer (without direct access to the implementation details of an LTS), two processes that emit the same set of sequences of action labels are indistinguishable.

    Formally, let \(A^*\) be the set of finite words formed from the alphabet \(A\).
    Given a state \(x\) of an LTS \((X, \alpha)\), the set \(\trace_\alpha(x)\) of \emph{traces  emitted by \(x\)} is the set of words \(a_1\cdots a_n \in A^*\) such that  there is a path of the form \(x \tr{a_1} x_1 \tr{a_2} \cdots \tr{a_{n-1}} x_{n-1} \tr{a_n} x_n\) through \((X, \alpha)\).
    Given LTSs \((X, \alpha)\) and \((Y, \beta)\), states \(x \in X\) and \(y \in Y\) are called \emph{trace equivalent} if \(\trace_\alpha(x) = \trace_\beta(y)\).
    If the transition structure is clear from context, we drop it from the notation and write simply \(\trace(x)\) for \(\trace_\alpha(x)\).
    It should be noted that every trace language $\trace_\alpha(x)$ is \emph{prefix-closed}, which for a language \(L\) means that \(w \in L\) whenever \(wa \in L\) for some \(a \in A\).

    Trace equivalence is a well-documented notion of equivalence for processes, dating back at least to the work of Hoare and Milner in the late 1970s~\cite{Hoare1978Traces,Milner78,BPS2001Handbook,Baeten2005History}.
    We shall see it in our work on fractals as well.
   
    \begin{defi}
        A \emph{stream} is an infinite sequence \((a_i)_{i \in \N} = (a_1, a_2, \dots)\) of letters from \(A\).
        The set of streams is denoted \(A^\omega\).
        A state \(x\) in an LTS \((X, \alpha)\) \emph{emits} a stream \((a_i)_{i \in \N}\) if there is an infinite path of the form \(x \tr{a_1} x_1 \tr{a_2} \cdots  \tr{a_n} x_n \tr{a_{n+1}} \cdots\) through \((X, \alpha)\). 
        We write $\str_\alpha(x)$ for the set of streams emitted by \(x\).
        Two states \(x,y\) are said to be \emph{stream equivalent} if \(\str_\alpha(x) = \str_\alpha(y)\). 
    \end{defi}

    \begin{rem}
        We use both \((a_i)_{i \in \N}\) and \((a_1, a_2, \dots)\) as notations for streams. 
        These two are intended to denote the same stream of letters. 
        Note that indexing starts at \(1\) rather than \(0\).
    \end{rem}

    In our construction of fractals from LTSs, points are represented only by (infinite) streams.
    We therefore focus primarily on LTSs with the property that for all states $x$, \(\trace(x)\) is precisely the set of finite prefixes of streams emitted by \(x\).
    We refer to an LTS \((X, \alpha)\) satisfying this condition as \emph{productive}.
    Productivity is equivalent to the absence of \emph{deadlock} states, which are states with no outgoing transitions.
    
    \begin{lem}
        \label{lem:trace stream equivalence}
    	Let \((X, \alpha)\) be a finitely branching LTS.
    	Then the following are equivalent: 
        \begin{enumerate}
        	\item \((X, \alpha)\) is productive, i.e., for any \(x \in X\), 
	       	\begin{align}
	    		\trace_\alpha(x) 
	    		&= \{a_1\dots a_n \in A^* \mid \exists (b_i)_{i \in \N} \in \str_\alpha(x) ~\text{such that}~ b_1 = a_1, \dots, b_n = a_n\} \label{eq:first of 1}\\
	    		\str_\alpha(x) 
	    		&= \{(a_i)_{i \in \N} \in A^\omega \mid \forall n \in \mathbb N,~ a_1\cdots a_n \in \trace_\alpha(x)\} \label{eq:second of 1}
	    	\end{align}
        	\item For any \(x \in X\), \(\alpha(x) \neq \emptyset\).
        \end{enumerate} 
        Consequently, if \((X, \alpha)\) is productive and finitely branching, and \(x,y \in X\), then \(\str_\alpha(x) = \str_\alpha(y)\) if and only if \(\trace_\alpha(x) = \trace_\alpha(y)\).
        In other words, for productive and finitely branching LTSs, trace equivalence and stream equivalence coincide.
    \end{lem}
    
    \begin{proof}
    	To see that 1 implies 2, observe that trivially, \(\epsilon \in \trace_\alpha(x)\) for any \(x \in X\) (\(\epsilon\) is the empty word).
    	This implies there is a stream \((a_i)_{i \in \N} \in \str_\alpha(x)\) such that \(\epsilon\) is an initial segment of \((a_i)_{i \in \N}\), or in other words, \(\str_\alpha(x) \neq \emptyset\).
    	Hence, \(\alpha(x) \neq \emptyset\), as \(x\) emits a stream.
    	
    	To see that 2 implies 1, assume \(\alpha(x) \neq \emptyset\) for all \(x \in X\).
    	Then for any \(x \in X\), a simple inductive argument shows that there are paths of arbitrary length starting from \(x\). 
    	It follows from K\"onig's lemma~\cite{kleene2002mathematical} (every infinite but finitely branching tree has an infinite branch) that there is therefore an infinite path starting at \(x\).
    	This allows us to argue as follows:
    	To see~\eqref{eq:first of 1}, it suffices to show that \(\trace_\alpha(x)\) is contained in the right hand side, as the reverse containment is clear.
    	So, let \(x \in X\) and suppose that \(a_1 \dots a_n \in \trace_\alpha(x)\). 
		Then there is a path \(x \tr{a_1} \cdots \tr{a_n} x_n\), and by the observation above, there is an infinite path \(x_n \tr{a_{n+1}} x_{n+1} \tr{a_{n+1}} \cdots\).
		Hence, \((a_1, \dots, a_n, a_{n+1}, a_{n+2}, \dots) \in \str_\alpha(x)\), as desired. 
		This establishes~\eqref{eq:first of 1}.

	    To see~\eqref{eq:second of 1}, it suffices to see that \(\str_\alpha(x)\) contains the right hand side, as the forward containment is clear. 
        So, let \((a_i)_{i \in \N} \in A^\omega\) such that for any \(n \in \mathbb N\), \(a_1 \cdots a_n \in \trace_\alpha(x)\). 
	    We are going to use K\"onig's lemma again: Let \(x \in X\) and let \(\mathfrak P\) be the set of all paths of the form \(x \tr{a_1} \cdots \tr{a_n} x_n\). 
        We define a tree structure on \(\mathfrak P\) as follows: Given \(P_1, P_2 \in \mathfrak P\), write \(P_1 \to P_2\) if \(P_1 = (x \tr{a_1} \cdots \tr{a_n} x_n)\) and \(P_2 = (x \tr{a_1} \cdots \tr{a_n} x_n \tr{a_{n+1}} x_{n+1})\) for some \(x_{n+1}\). 
	    Then \((\mathfrak P, \to)\) is an infinite but finitely branching tree, since every initial segment of \((a_i)_{i \in \N}\) is in \(\trace_\alpha(x)\) and \((X, \alpha)\) is finitely branching. 
	    By K\"onig's lemma, there is an infinite path \(P_1 \to P_2 \to \cdots\) in \((\mathfrak P, \to)\).
	    By definition, this means there is an infinite path \(x \tr{a_1} x_1 \tr{a_2} \cdots\) in \((X, \alpha)\).
	    Hence, \((a_i)_{i \in \N} \in \str_\alpha(x)\).
	    This establishes~\eqref{eq:second of 1}.

        Let us now establish the stated consequence.
        In one direction, let \(\trace_\alpha(x) = \trace_\alpha(y)\). 
        Then 
        \begin{align*}
            \str_\alpha(x)
            &= \{(a_i)_{i \in \N} \in A^\omega \mid \forall n \in \mathbb N,~ a_1\cdots a_n \in \trace_\alpha(x)\} \\
            &= \{(a_i)_{i \in \N} \in A^\omega \mid \forall n \in \mathbb N,~ a_1\cdots a_n \in \trace_\alpha(y)\} \\
            &= \str_\alpha(y)
        \end{align*}
        In the other, if \(\str_\alpha(x) = \str_\alpha(y)\), then 
        \begin{align*}
            \trace_\alpha(x)
            &= \{a_1\dots a_n \in A^* \mid \exists (b_i)_{i \in \N} \in \str_\alpha(x) ~\text{such that}~ b_1 = a_1, \dots, b_n = a_n\} \\
            &= \{a_1\dots a_n \in A^* \mid \exists (b_i)_{i \in \N} \in \str_\alpha(y) ~\text{such that}~ b_1 = a_1, \dots, b_n = a_n\} \\
            &= \trace_\alpha(y) \qedhere
        \end{align*}
    \end{proof}
    
    \begin{rem}
    	The finite branching condition is necessary in \autoref{lem:trace stream equivalence}.
    	For example, consider the LTSs in \autoref{fig:hedgehog}.
        Between \(x_0\) and \(y_0\), only \(y_0\) emits the stream \((a, b, a, b, a, b, \dots)\) of alternating letters \(a\) and \(b\), despite both states 
        emitting the trace language \(\{(ab)^n \textcolor{black}{a^m}\mid  \textcolor{black}{m}, n\in \mathbb N\}\).
        This is possible because the LTS on the left is not finitely branching: Since \(x_0 \tr{a} x_{i,1}\) for every \(i \in \N\), \(x_0\) has infinitely many outgoing paths.
    \end{rem}

	\begin{figure}[ht]
		\begin{center}
			\begin{tikzpicture}
				\node[state] at (0,1) (0) {\(x_0\)};
				\node[state] at (2,0) (1) {\(x_{1,1}\)};
				\node[state] at (2,1) (2) {\(x_{2,1}\)};
				\node[state] at (4,1) (21) {\(x_{2,2}\)};			
				
                \node at (2, 1.85) (dots) {\(\vdots\)};
                \node at (1.5, 3.65) (void) {};

				\node[state] at (2,2.5) (3) {\(x_{n,1}\)};	
				\node[state] at (4,2.5) (31) {\(x_{n,2}\)};		
				\node at (6,2.5) (newdots) {\(\cdots\)};		
				\node[state] at (8,2.5) (32) {\(x_{n,n}\)};
				\node at (2, 3.35) (dots2) {\(\vdots\)};
				
				\draw (0) edge[->] node[below] {\(a\)} (1);
				\draw (0) edge[->] node[above] {\(a\)} (2);
				\draw (0) edge[->] node[above] {\(a\)} (3);
				\draw (2) edge[->] node[above] {\(b\)} (21);
				\draw (3) edge[->] node[above] {\(b\)} (31);
				\draw (31) edge[->] node[above] {\(a\)} (newdots);
				\draw (newdots) edge[->] node[above] {} (32);
				\draw[loop] (32) edge[loop right] node[right] {\(a\)} (32);
				\draw[loop] (21) edge[loop right] node[right] {\(a\)} (21);
				\draw[loop] (1) edge[loop right] node[right] {\(a\)} (1);	
                \draw (0) edge[gray!20] (void);
			\end{tikzpicture}
			\hspace*{-10em}
            \raisebox{-3em}{
			\begin{tikzpicture}
				\node[state] at (0,0) (0) {\(y_0\)};
				\node[state] at (2,0) (1) {\(y_1\)};
				\node[state] at (4,0) (2) {\(y_2\)};
				\draw (0) edge[->, bend right] node[below] {\(a\)} (1);
				\draw (1) edge[->, bend right] node[above] {\(b\)} (0);
				\draw (1) edge[->] node[above] {\(a\)} (2);
				\draw[loop] (2) edge[loop right] node[right] {\(a\)} (2);
			\end{tikzpicture}}
		\end{center}
		\caption{\label{fig:hedgehog}(A hedgehog model) 
        In the LTS on the left in the above picture, the paths starting from \(x_0\) are all of the form \(x_0 \tr{a} x_{n,1} \tr{b} x_{n,2} \tr{a} x_{n, 3} \tr{b} \cdots \tr{\phantom{a}} x_{n, n} \tr{a} \cdots\) (beginning with alternating \(a\)s and \(b\)s). 
        Thus, the state \(x_0\) emits every finite initial segment of \((a,b,a,b,a,b,\dots)\), but does not emit this stream.
		Therefore, \(\trace(x_0) = \trace(y_0)\) and yet \(\str(x_0) \neq \str(y_0)\).
		\label{fig-hedghehog}}
	\end{figure}

    % Specification language for processes (states of an LTS)
    \paragraph*{Specification}
    We use the following language for specifying processes:
    Starting with a fixed countably infinite set \(\{v_1, v_2, \dots\}\) of \emph{variables}, the set of \emph{\(\mu\)-expressions} is generated by the grammar
    \[
        v \mid ae \mid e_1 + e_2 \mid \mu v ~ e
    \] 
    where \(v\) is  \(v_i\) for some \(i \in \mathbb N\), \(a \in A\), and \(e,e_1,e_2\) are \(\mu\)-expressions.

    Intuitively, the process \(ae\) emits \(a\) and then turns into \(e\), and \(e_1 + e_2\) is the process that nondeterministically branches into \(e_1\) and \(e_2\).
    The process \(\mu v~e\) is like \(e\), but with instances of \(v\) that appear free in \(e\) acting like \texttt{goto} expressions that return the process to \(\mu v~e\).

    \begin{defi}\label{def:term}
        A \emph{process term} is a \(\mu\)-expression \(e\) in which every occurrence of a variable \(v\) appears both 
        within the scope of a \emph{recursive call} \(\mu v~(-)\) (\(e\) is \emph{closed}) and within the scope of an \emph{action prefix operator} \(a(-)\) (\(e\) is \emph{guarded}).
        For short, we call these \emph{terms}.
        The set of terms is written \(\Term\).
        It is the set underlying the LTS \((\Term, \gamma)\) defined in \autoref{fig:transitionrelation}.    
    \end{defi}
    
    \begin{figure}[ht]
        \begin{mathpar}
            \infer{}{ae \tr{a} e}
            \and 
            \infer{e_1 \tr{a} f}{e_1 + e_2 \tr{a} f}
            \and
            \infer{e_2 \tr{a} f}{e_1 + e_2 \tr{a} f}
            \and 
            \infer{e[\mu v~e/v] \tr{a} f}{\mu v~e \tr{a} f}
        \end{mathpar}
        \caption{
            \label{fig:transitionrelation}
            The relation \({\tr{a}} \subseteq \Term \times \Term\) defining \((\Term, \gamma)\).
        }
    \end{figure}

    Call a variable \(v\) appearing in an expression \(e\) \emph{bound} if it appears withing the scope of a recursive call \(\mu v~(-)\) and \emph{free} if it does not.
    In \autoref{fig:transitionrelation}, we use the notation \(e[g/v]\) to denote the expression obtained by replacing each free occurrence of \(v\) in \(e\) with the expression $g$.
    Note that the expression \(e[g/v]\) is only well-defined if no free variable in \(g\) appears bound in \(e\).
    Given \(e \in \Term\), the \emph{process specified by \(e\)} is the LTS \(\langle e \rangle _\gamma\).

    \begin{exa}
   	    Let \(e = av\), and let \(f = \mu v~e = \mu v~a v\).  
	    Then \(e[f/v] = af \tr{a} f\).
		By the last rule in \autoref{fig:transitionrelation}, \(\mu v~e \tr{a} f = \mu v~e\). 
		By looking at the other rules, we see that none apply.
		Hence, if  \(\mu v~e \tr{b} g\), then \(b = a\) and \(g = \mu v~e\).
		The upshot is that \(\langle f \rangle_\gamma\), the process specified by \(f\), is the one-point process which can perform $a$ but no other action: 
		its state set is \(\set{f} \), and \(\gamma(f) = \set{(a,f)}\).
		\[\begin{tikzpicture}
			\node[state] at (0,0) (0) {\(\mu v~av\)};
			\draw[loop] (0) edge[loop left] node[left] {\(a\)} (0);
		\end{tikzpicture}
		\qquad 
		\gamma(\mu v~av) = \gamma(a(\mu v~av)) = \{(a, \mu v~av)\}\]
		
	\end{exa}

    \begin{rem}
        The set of process terms, as we have named them, is the fragment of Milner's fixed-point calculus from~\cite{Milner1984Complete} consisting of only the \(\mu\)-expressions that specify  productive LTSs.
    \end{rem}

    Labelled transition systems specified by process terms are finite and productive, and conversely, every finite productive process is trace-equivalent to some process term.

    \begin{lem}
        \label{lem:local finiteness}
        The LTS \((\Term, \gamma)\) is productive and locally finite, i.e., for any \(e \in \Term\), the set of terms reachable from \(e\) in \((\Term, \gamma)\) is finite.
        Conversely, if \(x\) is a state in a productive and locally finite LTS \((X, \alpha)\), then there is a process term \(e\) such that \(\trace(e) = \trace_\alpha(x)\).
    \end{lem}

    \begin{proof}
        The proof that \((\Term, \gamma)\) is productive is a routine induction on \(e \in \Term\) that shows that \(\gamma(e) \neq \emptyset\).
        To see that \((\Term, \gamma)\) is locally finite, we start by reproducing Milner's~\cite[Proposition 5.1]{Milner1984Complete} below, to which we have made only small changes to reflect the minor differences between our syntax and Milner's.
        
        Consider the LTS consisting of arbitrary \(\mu\)-expressions (as opposed to just process terms).
        Let us write \((E, \bar\gamma)\) for the set of \(\mu\)-expressions \(E\) equipped with the transition structure \(\bar\gamma\) defined as follows: \(\bar\gamma(v_i) = \emptyset\) for all \(i \in \mathbb N\), and all other transition relations are those derivable from the rules in \autoref{fig:transitionrelation}.       
        Write \(\to^+\) and \(\to^*\) for the transitive and reflexive-transitive closures of the transition relation \(\to\) respectively.
        For any \(e \in E\), define the set of all \(\mu\)-expressions reachable from \(e\) by \(\langle e \rangle = \{f \in E \mid e \arrowstar f\}\).  
        We proceed with a proof by induction on \(\mu\)-expressions that \(\langle e\rangle\) is finite for every \(e \in E\), i.e., \((E, \bar\gamma)\) is locally finite.%
        \footnote{And in fact, the size of \(\langle e \rangle\) is linear with respect to the length of the \(\mu\)-expression \(e\).} 
        
        In the base case, we have \(e = v_i\) for some \(i \in \mathbb N\). 
        Here, \(|\langle v_i\rangle| = |\{v_i\}| = 1\).
        For the inductive step, let \(e,e_1,e_2\) be guarded \(\mu\)-expressions and assume that \(|\langle e_i \rangle|\) for \(i \in \{1, 2\}\) and \(|\langle e\rangle|\) are all finite. 
        \begin{itemize}
            \item By definition of \(\bar\gamma(ae)\), \(|\langle ae\rangle| = |\{ae\} \cup \langle e\rangle| \le 1 + |\langle e \rangle|\).
            The induction hypothesis therefore tells us that \(\langle ae\rangle\) is finite.
            
            \item The set \(\langle e_1 + e_2\rangle\) precisely consists of \(e_1 + e_2\) and all of the \(\mu\)-expressions \(f\) such that either \(e_1 \arrowplus f\) or \(e_2 \arrowplus f\). 
            It follows that \[
                |\langle e_1 + e_2\rangle| = |\{e_1 + e_2\} \cup \langle e_1 \rangle \cup \langle e_2 \rangle| \le 1 + |\langle e_1 \rangle| + |\langle e_2\rangle|
            \]
            By the induction hypothesis, \(\langle e_1 + e_2\rangle\) is finite. 

            \item For the final induction step, observe that \(\mu v~e \arrowplus f\) if and only if \(e[\mu v~e/v] \arrowplus f\).
            It follows that \(\langle \mu v~e\rangle \setminus \{\mu v~e, e[\mu v~e/v]\} = \langle e[\mu v~e/v]\rangle \setminus \{\mu v~e, e[\mu v~e/v]\}\).
            Clearly then, \(\langle \mu v~e\rangle\) is finite if and only if \(\langle e[\mu v~e/v]\rangle\) is finite.
            Now, every expression \(f\) for which \(e[\mu v~e/v] \arrowstar f\) is necessarily either of the form \(g[\mu v~e/v]\) with \(e \arrowstar g\) or is \(\mu v~e\) itself. 
            This tells us that \[
                |\langle e[\mu v~e/v]\rangle| 
                \le |\{\mu v~e\} \cup \{g[\mu v~e/v] \mid e \arrowstar g\}|
                = 1 + |\{g \mid e \arrowstar g\}|
                = 1 + |\langle e\rangle|
            \]
            The induction hypothesis now tells us that \(\langle e[\mu v~e/v]\rangle\) is finite, and therefore \(\langle \mu v~e\rangle\) is also finite.
        \end{itemize}
        \noindent %FIXED indent
        This concludes the proof that \((E, \bar\gamma)\) is locally finite.
        To see that \((\Term, \gamma)\) is locally finite, observe from the definition of \(\bar \gamma\) that for each process term \(e \in \Term\) we have \(\bar\gamma(e) = \gamma(e) \subseteq A \times \Term\).
        It follows that \(\langle e\rangle \subseteq \Term\) for each \(e \in \Term\). 
        Since \((E, \bar\gamma)\) is locally finite, \(\langle e\rangle\) is finite.
        This shows that \((\Term, \gamma)\) is locally finite.
        
        The converse statement, that if \(x\) is a state in a finite LTS \((X, \alpha)\), then \(\trace_\alpha(x) = \trace(e)\) for some \(e \in \PTerm\), is a direct consequence of~\cite[Theorem 5.7 and Corollary 5.8]{Milner1984Complete}.%
        \footnote{In fact, the mentioned proof shows something stronger: that every state in a productive and locally finite LTS is \emph{bisimilar} to a process term. It is well-known that bisimilarity implies trace equivalence~\cite{BPS2001Handbook}.}
    \end{proof}
    
    % Axioms for trace equivalence 
    \paragraph*{Axiomatization of trace equivalence}
    Given an interpretation of process terms as states in an LTS, and given the notion of trace equivalence, one might ask if there is an algebraic or proof-theoretic account of trace equivalence of process terms.
    Rabinovich showed in~\cite{Rabinovich1993Traces} that a complete inference system for trace equivalence can be obtained by adapting earlier work of Milner~\cite{Milner1984Complete}.
    The axioms of the complete inference system include equations like \(e_1 + e_2 = e_2 + e_1\) and \(a(e_1 + e_2) = ae_1 + ae_2\), which are intuitively true for trace equivalence.

    \begin{figure}[t]
        \begin{gather*}
            \begin{array}{c r l}
                (\axiom{I}) &               e + e &\hspace*{-0.8em}\equiv e \\
                (\axiom{C}) &           e_2 + e_1 &\hspace*{-0.8em}\equiv e_1 + e_2 \\
                (\axiom{A}) &   e_1 + (e_2 + e_3) &\hspace*{-0.8em}\equiv (e_1 + e_2) + e_3 \\
                (\axiom{D}) &        a(e_1 + e_2) &\hspace*{-0.8em}\equiv ae_1 + ae_2 \\
                (\axiom{F}) &             \mu v~e &\hspace*{-0.8em}\equiv e[\mu v~e/v] 
            \end{array}
            \qquad
            \begin{gathered}
	            \begin{array}{c c}
                    (\axiom{\alpha}) & \mu w~e \equiv \mu v~e[v/w] \\
	                (\axiom{Cong}) & \infer{e_1 \equiv f_1 ~ \cdots ~ e_n \equiv f_n}{g[\vec e/\vec v] \equiv g[\vec f/\vec v]}  \\ 
	                (\axiom{RSP}) & \infer{g \equiv e[g/v]}{g \equiv \mu v~e}
	            \end{array}
            \end{gathered}
        \end{gather*}
        \caption{
            \label{fig:axioms}
            The names for the axioms on the left stand for \emph{idempotence, commutativity, associativity, distributivity}, and \emph{fixed point} respectively.
            On the right, we have \emph{\(\alpha\)-equivalence, congruence}, and the \emph{recursive specification principle}.
            The axioms and rules of the provable equivalence relation are those listed here, in addition to those of equational logic (not shown). 
            Here, \(e,e_i,f,f_i,g \in \Term\) for all \(i\).
            In \((\axiom{Cong})\), \(g\) has precisely the free variables \(v_1, \dots, v_n\), and no variable that appears free in \(f_i\) is bound in \(g\) for any \(i\). 
            In \((\axiom{\alpha})\), \(v\) does not appear free in \(e\). 
        }
    \end{figure}

    \begin{defi}
        Given \(e_1, e_2 \in \Term\), we say that \(e_1\) and \(e_2\) are \emph{provably equivalent} if \(e_1\equiv e_2\) can be derived from the axioms in \autoref{fig:axioms}, and call \(\equiv\) \emph{provable equivalence}.
    \end{defi}
 %FIXED changed to thmC to avoid double parantheses
	\begin{thmC}[Rabinovich~\cite{Rabinovich1993Traces}]\label{thm:Rabinovich}
        Let \(e_1,e_2 \in \Term\).
        Then \(e_1 \equiv e_2\) iff \(\trace(e_1) = \trace(e_2)\).
    \end{thmC}

    \begin{exa}
    \label{example-proof-system}
        Consider the processes specified by \(e_1 = \mu w~\mu v~(a_1 a_2 v + a_1a_3 w)\) and \(e_2 = \mu v~(a_1(a_2v + a_3v))\) below. 
        \[
        \begin{gathered}
            \begin{tikzpicture}
                \node[state] (e1) at (0,0) {\(e_1\)};
                \node[state] (a3e1) at (0,-1.2) {\(a_3e_1\)};
                \node[state] (a2f1) at (2,0) {\(a_2f\)};
                \node[state] (f1) at (2,-1.2) {\(f\)};
                \draw (e1) edge[bend left] node[right] {\(a_1\)} (a3e1);
                \draw (e1) edge node[above] {\(a_1\)} (a2f1);
                \draw (a3e1) edge[bend left] node[left] {\(a_3\)} (e1);
                \draw (a2f1) edge[bend left] node[right] {\(a_2\)} (f1);
                \draw (f1) edge[bend left] node[left] {\(a_1\)} (a2f1);
                \draw (f1) edge node[below] {\(a_1\)} (a3e1);
            \end{tikzpicture}
            \qquad \qquad
            \begin{tikzpicture}
                \node[state] (e2) at (0,0) {\(e_2\)};
                \node[state] (a3e2) at (0,-1.2) {\(a_2e_2 + a_3e_2\)};
                \draw (e2) edge[bend left] node[right] {\(a_1\)} (a3e2);
                \draw (a3e2) edge[bend left] node[left] {\(a_2,a_3\)} (e2);
            \end{tikzpicture}
        \end{gathered}
        \]
        The term $f$ is $\mu v~(a_1 a_2 v + a_1 a_3 e_1)$.    
        The traces emitted by both \(e_1\) and \(e_2\) are those that alternate between \(a_1\) and either \(a_2\) or \(a_3\). %, which we can see from the diagrams of their LTSs in \\autoref{fig:example deduction}.  
       
        By \autoref{thm:Rabinovich}, it should be possible to prove that \(e_1 \equiv e_2\).
        Indeed, this can be done as follows:  
    	Applying $\text{(\(\axiom{F}\))}$ to $e_1 = \mu w~\mu v ~(a_1a_2v + a_1a_3w)$, we start with 
    	\[ 
    		e_1 
    		= \mu w~\mu v~(a_1 a_2 v + a_1a_3 w) 
    		\stackrel{(\axiom{F})}\equiv  \mu v ~(a_1 a_2 v + a_1 a_3 e_1) = f
    	\]
    	Applying $(\axiom{F})$ again, and then \((\axiom{Cong})\), we get 
        \[
            f \stackrel{(\axiom{F})}\equiv  a_1 a_2 f + a_1 a_3 e_1 \stackrel{(\axiom{Cong})}\equiv  a_1 a_2 e_1 + a_1 a_3 e_1
        \]	
    	Applying $\text{(\(\axiom{D}\))}$ and using our substitution notation gives us 
    	\[
    		a_1a_2e_1 + a_1a_3 e_1
    		\stackrel{(\axiom{D})}\equiv a_1(a_2e_1 + a_3e_1) 
    		= (a_1(a_2v + a_3v))[e_1/v]
    	\]
        Thus $e_1 \equiv  (a_1(a_2v + a_3v))[e_1/v]$.   We apply the rule (RSP) to see that 	
    	\[
    		e_1 
    		\equiv \mu v~(a_1(a_2v + a_3v)) = e_2
    	\]
        as desired.
	\end{exa}
	
%	\begin{figure}[!ht]
%        
%        \caption{\label{fig:example deduction} 
%	        Diagrams of $\langle e_1\rangle$ and $\langle e_2\rangle$. 
%	        Above, $f_1 = \mu v~(a_1a_2 v + a_1a_3e_1)$.
%        }
%    \end{figure}

    Rabinovich's theorem tells us that, up to provable equivalence, our specification language consisting of process terms is really a specification language for languages of traces.
    In what follows, we are going to give an alternative semantics to process terms by using LTSs to generate fractal subsets of complete metric spaces.
    The main result of our paper is that the equivalence relations coming from
     these two semantics coincide: Two process terms are trace equivalent if and only if they generate the same fractals in all spaces.
    This is the content of \cref{sec:fractals from LTS,sec:fractal and trace} below.

    % SECTION
    \section{Fractals from Labelled Transition Systems}
    \label{sec:fractals from LTS}

    In the Sierpi\'nski gasket \(\mathbf S\) from \autoref{fig:Sierpinski gasket}, every point of \(\mathbf S\) corresponds to a stream of letters from the alphabet \(\{a,b,c\}\), and every stream corresponds to a unique point.
    To obtain the point corresponding to a particular stream \((a_1, a_2, a_3, \dots)\) with each \(a_i \in \{a,b,c\}\), start with any \(p \in \mathbb R^2\) and compute the limit \(\lim_{n\to\infty} \sigma_{a_1}\cdots\sigma_{a_n}(p)\). 
    The point in the fractal corresponding to \((a_1, a_2, a_3, \dots)\) does not depend on \(p\) because \(\sigma_a,\sigma_b, \sigma_c\) in \autoref{fig:Sierpinski gasket} are \emph{contraction operators}.

    \begin{defi}
        Given a metric space \(M\), a \emph{contraction operator} on \(M\) is a function \(h : M \to M\) such that for some \(c \in [0, 1)\), \(d(h(x), h(y)) \le c~d(x, y)\) for any \(x, y \in M\).
        The number \(c\) is called a \emph{contraction coefficient} of \(h\). 
        The set of contraction operators on \(M\) is written \(\Con(M)\).\footnote{ In general, with a metric space $(M,d)$, we usually omit the metric and just write $M$.
        Also note that our notation $\Con(M)$ should not be read as suggesting that $\Con$ is a functor. It is not. 
        Finally, we shall always assume that our metric spaces are non-empty.}
    \end{defi}

    For example, with the Sierpi\'nski gasket (Figure~\ref{fig:Sierpinski gasket}) associated to the contractions \(\sigma_a\), \(\sigma_b\), and \(\sigma_c\), \(r = 1/2\) is a contraction coefficient for all three maps.
    Now, given \(p, q \in \mathbb R^2\), 
    \[
        d(\sigma_{a_1} \cdots \sigma_{a_{n}}(p), \sigma_{a_1} \cdots \sigma_{a_{n}}(q)) \le \frac1{2^n}~d(p, q)
    \] 
    for all \(n\), so it follows that
    \(
        \lim_{n\to\infty} \sigma_{a_1}\cdots\sigma_{a_n}(p)
        = \lim_{n\to\infty} \sigma_{a_1}\cdots\sigma_{a_n}(q)
    \).
For any finite set of contraction operators %\(\{\sigma_{a_1},\dots, \sigma_{a_n}\}\) 
indexed by \(A\) and acting on a complete metric space \(M\), every stream  from \(A\) corresponds to a unique point in \(M\).

    % Define contraction operator interpretations of actions
    \begin{defi}
        \label{def:contra operator interpretation}
        A \emph{contraction operator interpretation} is a function \(\sigma \colon A \to \Con(M)\).
        We usually write \(\sigma_a = \sigma(a)\).
        Given \(\sigma\colon A \to \Con(M)\) and a stream \((a_i)_{i \in \N}\) from \(A\), define 
        \begin{equation}\label{eq:lim of stream def}
            \sigma_\omega \colon A^\omega \to M
            \qquad
            \sigma_\omega(a_1, a_2, \dots) = \lim_{n\to\infty} \sigma_{a_1}\cdots\sigma_{a_n}(p)
        \end{equation}
        where \(p \in M\) is arbitrary.
        The \emph{self-similar set} corresponding to a contraction operator interpretation \(\sigma\) is the set 
        \begin{equation}\label{def-S-sigma}
        \mathbf S_\sigma = \{\sigma_\omega(a_1, a_2, \dots) \mid (a_1, a_2, \dots)\text{ is a stream from }A\}.
        \end{equation}
    \end{defi}

    \begin{rem} 
        Note that in (\ref{eq:lim of stream def}), the contraction operators corresponding to the initial trace \((a_1,\dots,a_n)\) are applied in \emph{reverse} order.
        That is, \(\sigma_{a_n}\) is applied before \(\sigma_{a_{n-1}}\), \(\sigma_{a_{n-2}}\) is applied before \(\sigma_{a_{n-1}}\), and so on, and \(\sigma_{a_1}\) is applied last.     
    \end{rem}

    It should also be noted that contractions are \emph{continuous}, which implies that for any stream \((a_1, a_2, \dots) \in A^\omega\), \(\sigma \colon A \to \Con(M)\), and \(p \in M\),
    \[
    	\sigma_\omega(a_1, a_2, \dots) 
    	= \lim_{n\to\infty} \sigma_{a_1}\sigma_{a_2}\cdots \sigma_{a_n}(p) 
    	= \sigma_{a_1}(\lim_{n\to\infty} \sigma_{a_2}\cdots \sigma_{a_n}(p))
    	= \sigma_{a_1}\circ \sigma_\omega(a_2, \dots)
    \] 
    If we write \(a(-)\) for the prefixing map 
    \begin{equation}\label{prefixing} 
    a(a_i)_{i \in \N} = (a, a_1, \dots),
    \end{equation}
     this implies the following statement: Let \(M\) be a complete metric space and \(\sigma \colon A \to \Con(M)\).
    Then for any \(a \in A\), 
    \begin{equation}
        \label{lem:a dot commutes}
        \begin{gathered}
            \begin{tikzcd}
                A^\omega \ar[r, "a"] \ar[d, "\sigma_\omega"'] & A^\omega \ar[d, "\sigma_\omega"]\\
                M \ar[r, "\sigma_a"] & M
            \end{tikzcd}
            \qquad
            \sigma_\omega \circ a (-) = \sigma_a \circ \sigma_\omega
        \end{gathered}
    \end{equation}

    \paragraph*{Regular Subfractals}
    Generalizing the fractals of Mandelbrot~\cite{Mandelbrot1977Fractals}, Hutchinson introduced self-similar sets in~\cite{Hutchinson1981Fractals} and gave a comprehensive account of their theory.
    In op.\ cit., Hutchinson defines a self-similar set to be the invariant set of an \emph{iterated function system} (or \emph{IFS}).
    In our terminology, an iterated function system is equivalent to a contraction operator interpretation of a finite set  $A$ of actions, and the invariant set is the total set of points obtained from
    all streams from $A$.
    The fractals constructed from an LTS paired with a contraction operator interpretation generalize Hutchinson's self-similar sets to non-empty compact sets of points obtained from certain subsets of the streams, namely the subsets emitted by single states in the LTS.

    % Give the stream->point correspondence. Maybe comment on how we think that the multi-state fractal generation we've defined is more expressive?

    \begin{defi}\label{defn:subfractal from process}
        Given a productive LTS \((X, \alpha)\) and a contraction operator interpretation \(\sigma : A \to \Con(M)\), we define %\(\sem{-}_{\alpha, \sigma}\colon X \to \K(M)\) by
        \begin{equation}\label{eq:subfractal from process}
            \sem{x}_{\alpha, \sigma} = \sigma_\omega(\str_\alpha(x)) = \left\{\sigma_\omega((a_i)_{i \in \N}) \mid \text{\((a_i)_{i \in \N}\) is emitted by \(x\)}\right\}
        \end{equation}
        and call this the set \emph{generated by the state \(x\)}.
    \end{defi}
    
    It is worth noting that \(\sem{x}_{\alpha,\sigma}\) is always a subset of \(\mathbf S_\sigma\), since \(\str_\alpha(x) \subseteq A^\omega\) and therefore \(\sem{x}_{\alpha,\sigma} = \sigma_\omega(\str_\alpha(x)) \subseteq \sigma_{\omega}(A^\omega) = \mathbf S_\sigma\).
      
    \begin{defi}
        \label{def:regular subfractal}
        Given a process term \(e\in \Term\) and a contraction operator interpretation \(\sigma : A \to \Con(M)\), the \emph{regular subfractal semantics of \(e\)} corresponding to \(\sigma\) is \(\sem{e}_\sigma = \sem{e}_{\gamma,\sigma}\), where \(\gamma\) is the structure map on $\Term$ given in~\autoref{def:term}.
    \end{defi}

    \begin{exa}
        The regular subfractal generated by \(x\) in Figure~\ref{fig:Twisted Sierpinski gasket} is the regular subfractal semantics of \(\mu v~(av + b(bv + cv) + c(bv + cv))\) corresponding to the interpretation \(\sigma\) given in that figure.
        The regular subfractal semantics of \(e\) is a proper subset of the Sierpi\'nski gasket, as it does not contain the point \(p = (19/32, 11\sqrt3/32)\) corresponding to \((a,c,a,b,a,a,\dots)\).  
    \end{exa}
    
\begin{figure}[ht]
        \begin{tabular}{c l}
            \(\begin{gathered}
                \begin{tikzpicture}
                    \node[state] (0) at(0,0) {\(e\)};
                    \node[state] (1) at (2,0) {\(f\)};   
                    \draw[loop] (0) edge[loop left] node[left] {\(a\)} (0);
                    \draw[loop] (1) edge[loop right] node[right] {\(b\)} (1);
                    \draw (0) edge[bend left] node[above] {\(b,c\)} (1);
                    \draw (1) edge[bend left] node[below] {\(b,c\)} (0);
                \end{tikzpicture}
                \\
                \begin{aligned}
                e &= \mu v~(av + bg + cg) \\
                g &= \mu u ~(bu + bv + cv) \\
                f &= g[e/v] = \mu u~(bu + be + ce) 
                \end{aligned}
            \end{gathered}\)
            & 
            \raisebox{-2cm}{\includegraphics[scale=0.2]{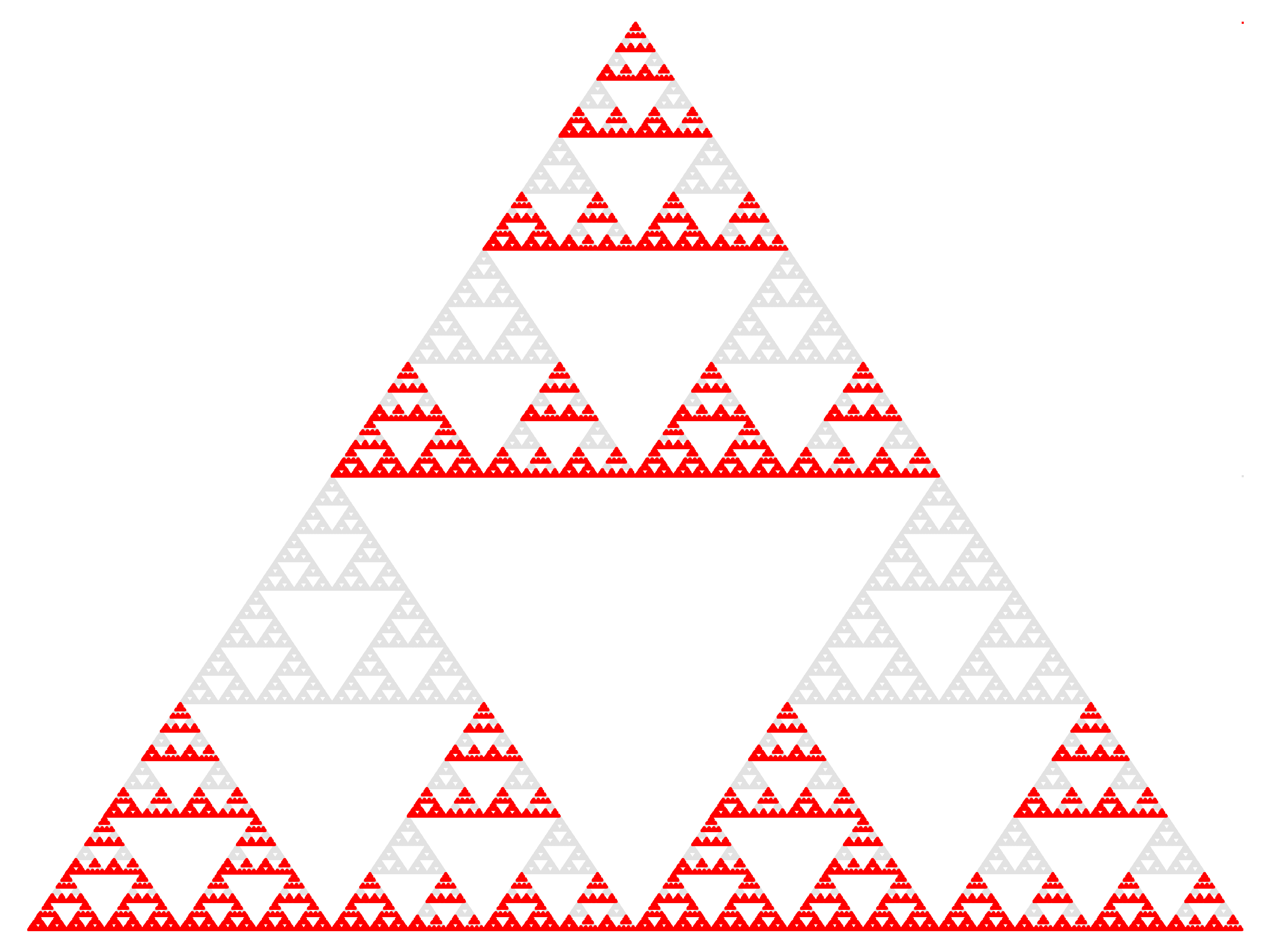}}
        \end{tabular}
        \caption{
            \label{fig:asymmetric sierpinski}
            An asymmetrical regular subfractal of the Sierpi\'nski gasket generated by the process term \(e = \mu v~(av + bg + cg)\), where \(g = \mu u ~(bu + bv + cv)\), corresponding to the same contraction operator interpretation as in \autoref{fig:Twisted Sierpinski gasket}.
        }
    \end{figure}

    \begin{exa}
        For another example involving the Sierpi\'nski gasket, see \autoref{fig:asymmetric sierpinski}. 
        Let \(g = \mu u ~(bu + bv + cv)\), and note that \(v\) is free in \(g\).
        The asymmetrical regular subfractal appearing in \autoref{fig:asymmetric sierpinski} is the regular subfractal semantics of the term \(e = \mu v~(av + bg + cg)\) corresponding to the same contraction operator interpretation as in Figures~\ref{fig:Sierpinski gasket} and~\ref{fig:Twisted Sierpinski gasket}.
    \end{exa}

    \paragraph*{Systems and Solutions}
    \label{sec:regular subfractals are solutions}
    Self-similar sets are often characterized as the unique non-empty compact sets that solve systems of equations of the form 
    \[
        K = \sigma_1(K) \cup \cdots \cup \sigma_n(K)
    \]
    with each \(\sigma_i\) a contraction operator on a complete metric space. 
    For example, the Sierpi\'nski gasket is  the unique non-empty compact solution to \(K = \sigma_a(K) \cup \sigma_b(K) \cup \sigma_c(K)\) in \autoref{fig:Sierpinski gasket}. 
    In this section, we are going to provide a similar characterization for regular subfractals that will play an important role in the completeness proof in \cref{sec:fractal and trace}. 
    The main connection of the characterization above to trace semantics arises from thinking about an \(n\)-state LTS \((X, \alpha)\) as a system of formal equations for the trace sets associated to the states:
    \[
        x_i = a_{i1} x_{i1} + a_{i2} x_{i2} + \cdots + a_{im_i}x_{im_i}
    \]
    indexed by \(X = \{x_1, \dots, x_n\}\), where \(x_i \tr{a_{ij}}_\alpha x_{ij}\) for \(i \le n\) and \(j \le m_i\).\medskip

    For the next few definitions and results, fix a non-empty metric space \(M\).
    
    \begin{defi} 
        Write \(\K(M)\) for the set of non-empty compact subsets of \(M\).   
        We equip \(\K(M)\) with the structure of a metric space using the \emph{Hausdorff metric} on \(\K(M)\)~\cite{Hausdorff}, defined by
        \begin{equation}\label{eqH}
            d_H(K_1, K_2) = \max\left\{\sup_{u \in K_1} \inf_{v \in K_2} d(u,v), \sup_{v \in K_2}\inf_{u \in K_1} d(u,v)\right\}
        \end{equation}
        If $M$ is complete, so is \(\K(M)\).
        Incidentally, we need to restrict to \emph{non-empty} sets in (\ref{eqH}), since otherwise we would worry about infima over the empty set.
        This is the primary motivation for the guardedness condition which we imposed on our terms.
    \end{defi}   
    
    We also recall the \emph{Banach fixed-point theorem}, which allows for the computation of fixed-points by iteration.
%FIXED
    \begin{thmC}[Banach~\cite{Banach1922Fixedpoint}]
        \label{thm:Banach fixed-point}
        Let \(f \colon M \to M\) be a contraction map. 
        % Then the following two statements are true.
        \begin{enumerate}
            \item The map \(f\) has at most one fixed-point. That is, if \(p = f(p)\) and \(q = f(q)\), then \(p = q\).
            \item If \(M\) is complete, then for any \(p \in M\), \(\lim_{n \to\infty} f^n(p)\) is the unique fixed-point of \(f\).
        \end{enumerate}
    \end{thmC}

    \begin{defi}\label{def:solution}
        Given a contraction operator interpretation \(\sigma : A \to \Con(M)\),
        and an LTS \( (X, \alpha)\), we call a function \(\varphi : X \to \K(M)\) a \emph{(\(\sigma\)-)solution} to \((X, \alpha)\) if for any \(x \in X\), 
        \[
            \varphi(x) = \bigcup_{x \tr{a} y} \sigma_a(\varphi(y))
        \]
    \end{defi}

    \begin{exa}
        Let \(\mathbf S\) be the Sierpi\'nski gasket as a subset of \(\reals^2\).
        Let \((X, \alpha)\) be the LTS in \autoref{fig:Sierpinski gasket}.  
        Then we have a single state, \(x\), with \(x \tr{a,b,c} x\).
        The function \(\phi\colon X \to \K(\reals^2)\) given by \(\phi(x) = \mathbf{S}\) is a solution to \((X, \alpha)\), because \(\mathbf S = \sigma_a(\mathbf S) \cup \sigma_b(\mathbf S) \cup \sigma_c(\mathbf S)\).
        Self-similar sets, in the sense of Hutchinson~\cite{Hutchinson1981Fractals}, are regular subfractals generated by one-state LTSs.
    \end{exa}
    
    We are very interested in the existence and uniqueness of solutions to finite productive LTSs in $\K(M)$ for a given space \(M\), since these are our connection to subfractal subsets of \(M\).
    We shall obtain a general result in this direction in \autoref{thm:fractal semantics coincide} below.   In fact, we shall show that \emph{locally finite} productive LTSs have unique solutions.
    Our work is an application of the Banach Fixed Point Theorem, and it also calls on general metric and topological properties of the set \(A^{\omega}\) of streams of actions.

    \paragraph*{Topology of Streams}
    Recall that, given a contraction operator interpretation \(\sigma \colon A \to \Con(M)\), the subset of \(M\) generated by a state \(x\) in a LTS \((X, \alpha)\) is given by taking the image of the set of streams \(\str_\alpha(x)\) emitted by \(x\) under the map \(\sigma_\omega \colon A^\omega \to M\) defined in~\eqref{eq:lim of stream def}.
    As we are about to see, if \(M\) is complete, then subsets of \(M\) generated by states of \((X, \alpha)\) are always compact in the topology of \(M\) induced by the metric.
    Topological results like these follow from familiar topological properties of \(A^\omega\) and the sets \(\str_\alpha(x)\), which we discuss next.

    \begin{defi}
        \label{def:cantor sets}
        Let \(A\) be a finite set of actions. 
        For each \(a \in A\), let \(0 < c_a < 1\), and write \(c \colon A \to (0, 1)\) for the map \(a \mapsto c_a\).
        Let \(c_{\diam} > 0\).
        Define \((A^\omega, d_c)\) to be the space of streams from \(A\) equipped with the metric \(d_c\), defined
        \begin{equation}
            \label{eq:Cantor metric wrt sigma}
            d_c((a_1, a_2, \dots), (b_1, b_2, \dots))
            = \begin{cases}
                0 & \text{if } a_n = b_n \text{ for all }n \\
                c_\diam\prod_{i=1}^n c_{a_i} &n = \min\{n \mid a_{n+1} \neq b_{n+1}\} \text{, otherwise}
            \end{cases}  
        \end{equation}
        We call \((A^\omega, d_c)\) the \emph{diameter-\(c_\diam\) Cantor set equipped with the \(c\)-metric}. 
    \end{defi}
    
    For any choice in \(c \colon A \to (0,1)\) and \(c_{\diam} > 0\), \((A^\omega, d_c)\) is a compact totally disconnected metric space with no isolated points~\cite{counterexamples}.
    The diameter-\(1\) Cantor set equipped with the \(1/2\)-metric (\(c_a = 1/2\) for all \(a\in A\)) is referred to as simply \emph{the Cantor set}.
    The topology on \(A^\omega\) induced by any \(c\)-metric has a basis of open sets consisting of the \emph{cylinder sets}, subsets of \(A^\omega\) of the form
    \begin{equation}
        \label{eq:cylinder set}
        B_{a_1\cdots a_n} = \big\{(b_i)_{i \in \N} \mid b_1 = a_1, \dots, b_n = a_n\big\}
    \end{equation}
    Note that above, the case of \(n = 0\) has \(B_\epsilon = A^\omega\), where \(\epsilon\) is the empty word.

    \begin{thm}
        \label{thm:fractal semantics coincide}
        Let \((X, \alpha)\) be a productive LTS, \(M\) be a complete metric space, and \(\sigma\colon A \to \Con(M)\) a contraction operator interpretation.
        Then
        \begin{enumerate} 
            \item \label{item-nonempty-closed} For all \(x \in X\), \(\str_\alpha(x)\) is non-empty and closed in \(A^\omega\).
            \item \label{item-compactness} For all \(x \in X\), \(\sem{x}_{\alpha,\sigma}\) is non-empty and compact in \(M\).
            \item \label{item-solution} \(\sem{-}_{\alpha,\sigma} \colon X \to \K(M)\) is a  solution to \((X, \alpha)\).
            \item\label{item-needy} If \((X, \alpha)\) is locally finite, then \(\sem{-}_{\alpha,\sigma} \colon X \to \K(M)\) is the unique solution to \((X, \alpha)\).
        \end{enumerate}
    \end{thm}

    \begin{proof}
        To see item~\ref{item-nonempty-closed}, first observe that \(\str_\alpha(x) \neq \emptyset\) because \((X, \alpha)\) is productive.
        Let \(c \colon A \to (0,1)\) be any function and consider the diameter-\(1\) Cantor set \((A^\omega, d_c)\).
        Now consider a Cauchy sequence \(\{(a_1^{(i)}, a_2^{(i)}, \dots)\}_{i \in \mathbb N}\) in \(\str_\alpha(x)\), and let \((a_i)_{i \in \N}\) be its limit in \((A^\omega, d_c)\).
        Then \(x\) emits every finite initial segment of \((a_i)_{i \in \N}\), because for any \(N \in \mathbb N\), there is an \(m > N\) such that \((a_1, \dots, a_m, a_{m+1}^{(m)}, \dots) \in\str_\alpha(x)\).
        Since \((X, \alpha)\) is finitely branching and productive, \autoref{lem:trace stream equivalence} tells us that \((a_1, a_2, \dots) \in \str_\alpha(x)\). 
        Therefore, \(\str_\alpha(x)\) is a non-empty and also closed set. 

        \newcommand{\cmax}{c_{\mathsf{max}}}
        \newcommand{\lmax}{\ell_{\mathsf{max}}}
        \newcommand{\lstar}{\ell^\star}

        Item~\ref{item-compactness} is standard in the literature on fractals: it follows from the fact that (in our terminology) $\sigma_\omega$ is a surjective map whose image is \(\mathbf S_\sigma \subseteq M\) as defined in (\ref{def-S-sigma}), and this last set is the self-similar set defined by a contraction operator interpretation (in terminology from the fractals literature, an \emph{iterated function system}).   
        Such sets are known to be compact. 
        Nevertheless, for the convenience of the reader we provide a proof from first principles.

        For each $a\in A$, let \(c_a \in (0, 1)\) be a contraction coefficient for \(\sigma_a\) and let $m_a\in M$ be the fixed point of $\sigma_a$.
        Let
        \[
            \cmax = \max_{a \in A} c_a
            \qquad 
            \lmax = \max_{a, b\in A} d(m_a, m_b)
        \]     
        For each stream \((b_i)_{i \in \mathbb N} \in A^{\omega}\), $p\in M$, and $k\in \mathbb{N}$, let 
        \[ 
            h(p,(b_i)_{i \in \N}, k) = \sigma_{b_1}\cdots \sigma_{b_{k}}(p)
        \]
        Then by definition, $\sigma_{\omega}((b_i)_{i \in \N}) = \lim_{k \to \infty} h(p, (b_i)_{i \in \N}, k)$ for all $p$.
        Now fix a point \(p \in M\) and let
        \[
            \ell_p = \min\{d(p, m_b) \mid b \in A\}
        \]  
        Fix \(a \in A\) such that \(d(p, m_a) = \ell_p\).
        We show by induction on $k\geq 1$ that for $(b_i)_{i \in \N} \in A^\omega$, 
        \begin{equation}\label{estminate}
            d(h(p, (b_i)_{i \in \N},k), m_{b_1}) \leq \cmax^k \ell_p + \sum_{i=0}^{k-1} \cmax^i \lmax 
        \end{equation} 
        Here is the argument for $k = 1$:
        We have $h(p, (b_i)_{i \in \N}, 1) = \sigma_{b_1}(p)$ and
        \[ 
            d(\sigma_{b_1}(p), m_{b_1}) 
            = d(\sigma_{b_1}(p), \sigma_{b_1}(m_{b_1})) 
            \leq \cmax(d(p, m_{a}) + d(m_a, m_{b_1})) 
            \leq \cmax(\ell_p + \lmax)
        \]
        Assuming that (\ref{estminate}) holds for $k$, we fix  $(b_i)_{i \in \N}$ and apply our induction hypothesis to the tail $(b_2, b_3, \ldots)$.  
        We have $d(h(p, (b_2, b_3, \dots),k), m_{b_2}) \leq \cmax^k \ell_p + \sum_{i=0}^{k-1} \cmax^i \lmax$. 
        Then 
        \begin{align*}
            & d(h(p, (b_i)_{i \in \N},k+1), m_{b_1})\\
            &= d(\sigma_{b_1}(h(p, (b_2, b_3, \dots),k), \sigma_{b_1}(m_{b_1}))  
                & \tag{$m_{b_1}$ is the fixed point of $\sigma_{b_1}$} \\
            &\leq  c_{b_1}  d(h(p, (b_2, b_3, \dots),k), m_{b_1})  
                & \tag{defn.~of $c_{b_1}$} \\
            &\leq  \cmax( d(h(p, (b_2, b_3, \dots),k), m_{b_2}) + d(m_{b_2}, m_{b_1} )) 
                 \tag{triangle inequality $+$ defn.~of $\cmax$}  \\
            &\leq  \cmax ((\cmax^k \ell_p + \sum_{i=0}^{k-1} \cmax^i \lmax) +\lmax ) 
                 \tag{induction hypothesis $+$ defn.~of $\lmax$} \\
            &\leq \cmax^{k+1} \ell_p  +  \sum_{i=0}^{k-1} \cmax^{i+1} \lmax + \cmax^0 \lmax 
            \tag{since $\cmax\lmax < \lmax = \cmax^0 \lmax$}\\
            &\leq \cmax^{k+1} \ell_p  +  \sum_{i=1}^{k} \cmax^{i} \lmax + \cmax^0 \lmax \\
            &= \cmax^{k+1} \ell_p  +  \sum_{i=0}^{(k+1) - 1} \cmax^i \lmax
        \end{align*}
        This completes the induction.   
        Let $k$ be large enough so that $\cmax^k \ell_p < 1$.   
        Then we see that for large enough $k$, $d(h(p, (b_i)_{i \in \N},k),m_{b_1}) \leq \lstar$, where
        \[ 
            \lstar = 1 +  \sum_{i=0}^{\infty} \cmax^i \lmax= 1 + \frac{\lmax}{1-\cmax}
        \]
        This is independent of $p$.   
        We conclude that $d(\sigma_{\omega}((b_i)_{i \in \N}), m_{b_1}) \leq \lstar$.
        Now let us obtain the diameter of the Cantor set we have in mind.
        Let $c_\diam = 2\lstar + \lmax$.
        For two streams in $A^{\omega}$, say $(b_1,b_2, \ldots)$ and $(b'_1, b'_2, \ldots)$, we have 
        \begin{align*}
            & d(\sigma_{\omega}(b_1, b_2, \ldots), \sigma_{\omega}(b'_1, b'_2, \ldots))  \\ 
            &\leq d(\sigma_{\omega}(b_1, b_2, \dots),m_{b_1})  + d(m_{b_1}, m_{b'_1}) + d(m_{b'_1},\sigma_{\omega}(b'_1, b_2', \ldots))\\
            &\leq \lstar +  \lmax + \lstar \\
            &= c_{\diam}
        \end{align*}
        With this estimate in hand, we can now let \(c \colon A \to (0,1)\) be the constant function \(c(a) = \cmax\) and consider the diameter-\(c_\diam\) Cantor set \((A^\omega, d_c)\) equipped with the \(c\)-metric.
        
        Now we show that \(\sigma_\omega \colon A^\omega \to M\) is nonexpanding with respect to \(d_c\). 
        Consider \((a_i)_{i \in \N}\) and \((b_i)_{i \in \N}\) in \(A^\omega\). 
        We would like to show that \(d_c(\sigma_\omega((a_i)_{i \in \N}), \sigma_\omega((b_i)_{i \in \N})) \le d_c((a_i)_{i \in \N}, (b_i)_{i \in \N})\).
        If \(a_i = b_i\) for all \(i\), then \((a_i)_{i \in \N} = (b_i)_{i \in \N}\) and we have 
        \[ 
            d_c((a_i)_{i \in \N}, (b_i)_{i \in \N}) = 0 = d_M(\sigma_\omega((a_i)_{i \in \N}), \sigma_\omega((b_i)_{i \in \N}))
        \]
        Otherwise, let \(n = \min\{i \in \N \mid a_i \neq b_i\} - 1\). 
        Then 
        \(
            d_c((a_i)_{i \in \N}, (b_i)_{i \in \N}) = c_{\diam}\prod_{i=1}^{n} c_{a_i}
        \)
        since \(b_i = a_i\) for all \(i \le n\). 
        On the other side, we have 
        \begin{align*}
            &d_M(\sigma_\omega((a_i)_{i \in \N}), \sigma_\omega((b_i)_{i \in \N})) \\
            &= d_M\big(
                \sigma_{a_1} \cdots \sigma_{a_n} \sigma_\omega((a_{i+n})_{i \in \N}),
                \sigma_{a_1} \cdots \sigma_{a_n} \sigma_\omega((b_{i+n})_{i \in \N})
            \big) \\
            &\le \Big(\prod_{i=1}^n c_{a_i}\Big)~d_M\big(
                \sigma_\omega((a_{i+n})_{i \in \N}),
                \sigma_\omega((b_{i+n})_{i \in \N})
            \big) \\
            &\le \Big(\prod_{i=1}^n c_{a_i}\Big)c_{\diam} \\
            &= d_c((a_i)_{i \in \N}, (b_i)_{i \in \N})
        \end{align*}
        Thus, \(\sigma_\omega\) is nonexpanding.
        Since \(\sigma_\omega \colon A^\omega \to M\) is nonexpanding, it is continuous.
        Then, by item~\ref{item-nonempty-closed}, $\sem{x}_{\alpha,\sigma} = \sigma_\omega(\str_\alpha(x))$ is the image of the non-empty compact set $\str_{\alpha}(x)$ under the continuous function $\sigma_{\omega}$.
        It follows that $\sem{x}_{\alpha,\sigma}$ is non-empty and compact.
        
        We now show item~\ref{item-solution}, that \(\sem{x}_{\alpha,\sigma}\colon X\to \K(M)\) is a solution to \((X,\alpha)\).
        The idea is to construct a function  
        \[
            F\colon  \K(M)^X\to \K(M)^X
        \]
        then to show that fixed points of \(F\) are exactly solutions to \((X,\alpha)\), and finally to show that 
        \(\sem{x}_{\alpha,\sigma}\colon X\to \K(M)\) is a fixed point of \(F\).  
        We take 
        \begin{equation}
            \label{eq:definition of F}
            F(f)(x) = \bigcup_{x \tr{a} y} \sigma_a(f(y)) 
        \end{equation}
        Let us check that each set \(F(f)(x)\) really is a compact subset of \(M\).    
        There are two reasons for this.
        First, each set $\sigma_a(K)$ is compact, being the image of a compact set under a continuous function.
        Second, the finite union of compact sets is compact.
        
        It is clear that a solution to \((X,\alpha)\) is exactly a fixed point of \(F\). 
        Let us check next that the map \(\sem{x}_{\alpha,\sigma}\) is a fixed point.
        That is, we must show that for all \(x\in X\), 
        \begin{equation}
            \label{goal_unique_stream_soln}
            \sem{x}_{\alpha,\sigma} = \bigcup_{x \tr{a} y} \sigma_a(\sem{y}_{\alpha,\sigma})
        \end{equation}
        We are going to drop the subscripts and write \(\sem{x}\).
        Let \(\sigma_{\omega}((a_i)_{i \in \N}) \in \sem{x}\).  
        This stream \((a_i)_{i \in \N}\) is emitted by \(x\).  
        Fix a corresponding infinite path \(x \tr{a_1} x_1 \tr{a_2} \cdots  \tr{a_n} x_n \tr{a_{n+1}} \cdots\).
        This path begins with \(x \tr{a_1} x_1\).
        Its tail shows that the stream \((a_{i+1})_{i \in \N}\) is emitted by \(x_1\) and hence belongs to \(\str_{\alpha}(x_1)\).
        Thus, \(\sigma_{\omega}((a_{i+1})_{i \in \N})\in \sem{x_1}\).  
        It follows from the commutativity of the diagram in~\eqref{lem:a dot commutes} that
        \[
            \sigma_\omega((a_i)_{i \in \N})
            = (\sigma_\omega \circ \sigma_{a_1}) ((a_{i+1})_{i \in \N}) 
            = \sigma_{a_1}(\sigma_\omega ((a_{i+1})_{i \in \N}))
            \in \{\sigma_a(\sem{y}_{\alpha,\sigma}) \mid x \tr{a} y\}
        \]
        As desired, \(\sigma_\omega((a_i)_{i \in \N})\) belongs to the right-hand side of~\eqref{goal_unique_stream_soln}.
        This shows that \(\sem{x}_{\alpha,\sigma} \subseteq \bigcup_{x \tr{a} y} \sigma_a(\sem{y}_{\alpha,\sigma})\).
        The reverse inclusion is similar.

        We now turn to item~\ref{item-needy}. 
        Let us first assume that $X$ is finite.
        The more general case of locally finite $X$ is considered at the very end of this proof.  
        It is here that we use the fact that the set \(\K(M)^X\) of functions \(f\colon X\to \K(M)\) is a complete metric space with respect to the product metric \(d(f,g) = \max_{x \in X} d_H(f(x),g(x))\), where \(d_H\) is the Hausdorff metric.
        By the previous part, we need to verify that the function \(F\) from~\eqref{eq:definition of F} is contractive.
        Then, by the Banach Fixed Point Theorem, \(F\) has just one fixed point.
        Let $f,g \colon X\to \K(M)$, and let $\ell = d(f,g) = \max_x d(f(x), g(x))$.
        By the definition of the Hausdorff metric we have the following for all $x\in X$: for all $z\in f(x)$ there is some $w\in  g(x)$ with $d(x,w) \leq \ell$, and vice-versa.  
        Let $\cmax < 1$ be the maximum of the contraction coefficients of all $\sigma_a$ for $a\in A$.
 
        We claim that $d(F(f), F(g)) \leq \ell\cmax $.   
        Let $x\in X$, and let $z\in F(f)(x)$.  
        Then there is some $(a,y)\in \alpha(x)$ such that $z\in \sigma_a(f(y))$.
        For $z$, there is some $w\in f(y)$ such that $z = \sigma_a(w)$.
        Let $v\in g(y)$ be such that $d(w,v) < \ell$.   
        Then $\sigma_a(v) \in \sigma_a(g(y))\subseteq F(g)(x)$.
        Write $z' = \sigma_a(v)$.
        Then we have
        \[
            d(z, z') =  d(\sigma_a(w), \sigma_a(v)) < \cmax d(w, v) < \ell\cmax 
        \]
        So, for every $z\in F(f)(x)$, we have some $z'\in F(g)(x)$ with $d(z,z') < \ell\cmax $.  
        The converse assertion also holds.
        Thus, $d(F(f)(x), F(g)(x)) < \ell\cmax $.  
        Since $x$ is arbitrary, we obtain $d(F(f), F(g)) < \ell\cmax $.
        
        At this point, we know that \(\sem{-}_{\alpha,\sigma}\) is the only solution to \((X,\alpha)\) when \(X\) is finite.
        We also want to show that \(F\) has a unique solution when $X$ is merely locally finite, but perhaps infinite.
        Fix $x\in X$, and also take any solution $\phi: X\to \K(M)$.   
        Let us write the finite sub-LTS of $X$ determined by $x$ as $(\langle x \rangle, \alpha_0)$.
        The restriction $\phi|_{\langle x \rangle}$ is a solution to $(\langle x \rangle, \alpha_0)$.  
        By what we already know, $(\phi|_{\langle x \rangle})(x) = \sem{x}_{\alpha_0,\sigma}$.
        Now
        \[
            \phi(x) = (\phi|_{\langle x \rangle})(x) = \sem{x}_{\alpha_0,\sigma}
        \]
        Since the right-hand side above does not involve $\phi$, we see that $\phi$ is unique.
    \end{proof}
        
    \begin{rem}
        \label{rem:sigma-continuous}
        One step of the proof above was to show that for any given contraction operator interpretation \(\sigma \colon A \to \Con(M)\) there is a map \(c \colon A \to (0,1)\) such that \(\sigma_\omega \colon A^\omega \to M\) is nonexpanding with respect to the \(c\)-metric \(d_c\) on \(A^\omega\). 
        This implies that \(\sigma_\omega\) is continuous with respect to \(d_c\).
        But any two \(c\)-metrics induce the same topology on \(A^\omega\), so this actually tells us that \(\sigma_\omega\) is continuous for any contraction operator interpretation \(\sigma\).
        Of course, there is another way to arrive at the same fact, using an interchange of limits: 
        given a sequence of streams \(\{(a_i^{(j)})_{i \in \N}\}_{j \in \N}\) that converges to \((a_i)_{i \in \N}\), we know that for any \(i \in \N\), there is an \(l > 0\) such that for any \(j > l\), \(a_i^{(j)} = a_i\). 
        For arbitrary \(p \in M\), this implies that for any \(i \in \N\), \(\lim_{j \to \infty} \sigma_{a_1^{(j)}} \cdots \sigma_{a_i^{(j)}}(p) = \sigma_{a_1} \cdots \sigma_{a_i}(p)\).
        We already know that \(\sigma_\omega((a_i^{(j)})_{i \in \N}) = \lim_{i \to \infty} \sigma_{a_1^{(j)}} \cdots \sigma_{a_i^{(j)}}(p)\) converges for each \(j \in \N\).
        Thus, the double sequence \((\sigma_{a_1^{(j)}} \cdots \sigma_{a_i^{(j)}}(p))_{i,j \in \N}\) converges along every row and column.
        By the Moore-Osgood theorem, this allows us to exchange limits: 
        \begin{align*}
            \lim_{j \to \infty} \sigma_\omega((a_i^{(j)})_{i \in \N})
            &= \lim_{j \to \infty} \lim_{i \to \infty} \sigma_{a_1^{(j)}} \cdots \sigma_{a_i^{(j)}}(p) \\
            &= \lim_{i \to \infty} \lim_{j \to \infty} \sigma_{a_1^{(j)}} \cdots \sigma_{a_i^{(j)}}(p) \\
            &= \lim_{i \to \infty} \sigma_{a_1} \cdots \sigma_{a_i}(p) \\
            &= \sigma_\omega((a_i)_{i \in \N}) \\
            &= \sigma_\omega\big(\lim_{j \to \infty} \{(a_i^{(j)})_{i \in \N}\}_{j \in \N}\big) 
        \end{align*}
        This provides a second argument showing that \(\sigma_\omega\) is continuous. 
    \end{rem}

    \begin{exa}
        To see why some extra assumption is needed in item~\ref{item-needy},  consider the LTS \(x_0 \tr{a} x_1 \tr{a} x_2 \tr{a} \cdots\). 
        The smallest sub-LTS containing \(x_0\) is the whole LTS, so this LTS is not locally finite.
        Let $M = \mathbb{R}$, and let $\sigma_a(x) = \frac{1}{2}x$.
        One solution is \(\sem{x_n}_{\alpha,\sigma} = \set{0}\) for all $n$.   
        Another solution is $\phi(x_n) = \set{2^n}$.  
        Indeed, for all $r\in \mathbb{R}$, we have a solution  \(\phi(x_n) = \set{r2^n}\). 
        What is going on in this example is that \(\K(M)^X\) is not a metric space under our definitions: the singleton map is an isometric embedding of any space $M$ into $\K(M)$.  Thus $d(\set{2^n},\set{0}) = 2^n$.   
        So $\sup_n d(\set{2^n},\set{0}) =\infty$.
        But all distances are finite in this paper, as this is one of the requirements of the Banach fixed point theorem.
    \end{exa}

    The following consequence of \autoref{thm:fractal semantics coincide} is an analog of Kleene's theorem (the correspondence between regular expressions and regular languages~\cite{Kleene56}) for regular subfractals.

    \begin{cor}
        \label{thm:regular subfractals are fixed-points} Fix a complete metric space $M$ and a contraction operator interpretation \(\sigma\colon A \to \Con(M)\).
        A set is a regular subfractal if and only if it is a component of the solution to a finite productive LTS.
    \end{cor}
    
    \begin{proof}
        Let \(S\) be a regular subfractal as defined in~\autoref{def:regular subfractal}.
        So, there exists $e\in\Term$ such that $\sem{e}_{\gamma,\sigma} = S$.
        By~\autoref{def:solution}, for each $f\in\Term$, $\sem{f}_{\gamma,\sigma} = \bigcup_{f \tr{a} g} \sigma_a(\sem{g}_{\gamma,\sigma})$.
        Consider the finite productive LTS $\langle e\rangle$ generated by $e$.  
        Let us write $\gamma_0$ for the restriction of $\gamma$ to $\langle e \rangle$. 
        % Then $\langle e\rangl$ is a sub-LTS of $(\Term,\gamma)$.
        Define $\phi\colon \langle e \rangle \to \K(M)$ by $\phi(g) = \sem{g}_{\gamma,\sigma}$.
        Then $\phi$ is a solution to $\langle e \rangle$.  
        So, by~\autoref{thm:fractal semantics coincide}, $\phi = \sem{-}_{\gamma_0,\sigma}$.  
        And so, $S = \sem{e}_{\gamma,\sigma} = \phi(e) =  \sem{e}_{\gamma_0,\sigma}$.
        Thus, $S$ is a component of  the solution to the finite productive LTS $\langle e \rangle$.

        For the converse, let $(X,\alpha)$ be a finite productive LTS, and take a component of its solution, say  \(\sem{x}_{\alpha,\sigma}\).
        By~\autoref{lem:local finiteness}, there is a term $e\in \Term$ such that $\trace_{\gamma}(e) = \trace_{\alpha}(x)$.  
        By~\autoref{lem:trace stream equivalence},  $\str_{\gamma}(e) = \str_{\alpha}(x)$.  
        By~\autoref{defn:subfractal from process}, $\sem{e}_{\gamma,\sigma} = \sem{x}_{\alpha,\sigma}$.
        Thus, \(\sem{x}_{\alpha,\sigma}\) is a regular subfractal.
    \end{proof}   

    \begin{exa}
        \label{eg:stream system}
        The following example of our setup is essential for the rest of the paper.
        Let \(c \colon A \to (0,1)\) and let \((A^\omega, d_c)\) be a diameter-\(c_\diam\) Cantor set equipped with the \(c\)-metric.
        For each \(a \in A\), recall the \emph{\(a\)-prefixing operator} \(a \colon A^{\omega} \to A^{\omega}\) from (\ref{prefixing}).
        In the \(c\)-metric, \(a(-)\) is a contraction operator with contraction coefficient \(c_a\), so the map \(\sigma \colon A \to \Con(A^\omega, d_c)\) defined \(\sigma_a = a(-)\) is a contraction operator interpretation.
        
        A fact which we shall need is that the map $\sigma_{\omega}$ in this example is the identity on $A^{\omega}$.
        To this end, observe that
        \[
            \sigma_\omega((a_1, a_2, \dots))
            = \lim_{n \to \infty} \sigma_{a_1}\circ \cdots \circ \sigma_{a_n} (p)
            = \lim_{n \to \infty} a_1a_2\cdots a_n(p)
            = (a_1, a_2, \dots)
        \] 
        for any \(p \in A^\omega\). 
        Since \((A^\omega, d_c)\) is a compact metric space, it is in particular complete.
        Thus, \autoref{thm:fractal semantics coincide} applies to this example. 
        Since $\sigma_{\omega}$ is the identity on \(A^\omega\), \(\sem{x}_{\alpha,\sigma}\) is simply \(\str_\alpha(x)\) for a given productive LTS \((X, \alpha)\) and \(x \in X\).
        It follows from \autoref{thm:fractal semantics coincide} that \(\sem{x}_{\alpha,\sigma} = \str_\alpha(x)\), hence \(\str_\alpha \colon X \to A^\omega\) is the unique solution to \((X, \alpha)\) in \(A^\omega\).

        % \begin{rem}
        %     Note that the equality derived above, \(\sem{x}_{\alpha,\sigma} = \str_\alpha(x)\) with \(\sigma\) the prefixing operator, could also be derived from~\autoref{thm:}
        % \end{rem}

     \end{exa}

    \paragraph*{Real-valued Cantor sets}
    Traditional real-valued Cantor sets (like the ternary Cantor set) are also able to faithfully represent streams.
    Here, by \emph{traditional real-valued Cantor set}, we mean any self-similar set corresponding to a contraction operator interpretation \(\sigma \colon A \to \Con(\mathbb R)\) of the form \(\sigma_a(p) = cp + r_a\) for some \(c \in (0,1)\) and distinct \(r_a \in \mathbb R\) for each \(a \in A\).
    We will revisit this point in \cref{sec:fractal and trace}.

    A map \(f \colon \mathbb R \to \mathbb R\) of the form \(f(p) = cp + r\) for some \(c,r \in \mathbb R\) is called \emph{affine}.
    With $|c| < 1$, this operator is a contraction with coefficient \(c\).   
    An \emph{affine operator interpretation} is a  contraction operator interpretation \(\sigma \colon A \to \Con(\mathbb R)\)  where each $\sigma_a$ is affine.

    \begin{prop}
        \label{prop:cantor sets in R}
        Let \(A\) be a finite set of actions.
        There exists an affine operator interpretation \(\sigma \colon A \to \Con(\mathbb R)\) such that \(\sigma_\omega \colon A^\omega \to \mathbb R\) is injective.
    \end{prop}

    \begin{proof}
        Let \(A = \{a_1, \cdots a_n\}\) and write \(c = 1/(2n-1)\). 
        Now consider the affine operator interpretation \(\sigma \colon A \to \Con(\mathbb R)\) given by
        \[
            \sigma_{a_i}(p) = cp + 2c(i-1)
        \]
        for each \(i \le n\).
        %Then \(\sigma\) is a contraction operator consisting of affine maps.
        Note that \(\mathbf S_{\sigma} \subseteq [0,1]\).
        We are going to show that \(\sigma_\omega \colon A^\omega \to \mathbb R\) is injective.

        To this end, suppose that \(d_c((b_i)_{i \in \N}, (b_i')_{i \in \N}) > 0\).
        Then there is an \(i \in \N\) such that \(b_i \neq b_i'\). 
        Let \(m = \min\{i \mid b_i \neq b_i'\}-1\).   We first handle the case 
       \(m = 0\), i.e., \(b_1 \neq b_1'\).
        Let $i$ and $j$ be such that  \(b_1 = a_i\) and \(b_1' = a_j\).   Note that $i \neq j$, and indeed $|i-j|\geq 1$.
        Then for any \(p,q \in \mathbb R\), 
        \[
            |\sigma_{b_1}(p) - \sigma_{b_1'}(q)|
            = |cp + 2c(i - 1) - cq - 2c(j - 1)|
            = c|(p - q) + 2(i - j)|
        \]
        so that for any \(p,q \in [0,1]\), \(|(p - q) + 2(i - j)| > 0\).
        In particular, if \(b_1 \neq b_1'\), then
        \[
            |\sigma_\omega((b_i)_{i \in \N}) - \sigma_\omega((b_i')_{i \in \N})|
            = |\sigma_{b_1}(\sigma_\omega((b_{i+1})_{i \in \N})) - \sigma_{b_1'}(\sigma_\omega((b_{i+1}')_{i \in \N}))|
            > 0
        \]
        When \(m > 0\), we reduce to the first case:
        \begin{align*}
            &|\sigma_\omega((b_i)_{i \in \N}) - \sigma_\omega((b_i')_{i \in \N})| \\
            &= |\sigma_{b_1} \cdots \sigma_{b_m} \sigma_\omega((b_{i+m})_{i \in \N}) - \sigma_{b_1} \cdots \sigma_{b_m} \sigma_\omega((b_{i+m}')_{i \in \N})| \\
            &= c^m|\sigma_\omega((b_{i+m})_{i \in \N}) - \sigma_\omega((b_{i+m}')_{i \in \N})| \\
            &> c^m0 = 0
        \end{align*}
        since \(b_{i+m} \neq b_{i + m}'\).
        It follows that \(\sigma_\omega\) is injective.%
        \footnote{Another way to say all this is that the restrictions of $\sigma_{a_i}$ and $\sigma_{a_j}$ to $[0,1]$ have disjoint image sets.}
    \end{proof}

    \section{Completeness}
    \label{sec:fractal and trace}

    We have seen that finite productive LTSs (LTSs that only emit infinite streams) can be specified by process terms.
    We also introduced a family of fractal sets called regular subfractals, those subsets of self-similar sets obtained from the streams emitted by finite productive LTSs. 
    An LTS itself is representative of a certain system of equations, and set-wise the system of equations is solved by the regular subfractals corresponding to it. 
    Going from process terms to LTSs to regular subfractals, we see that a process term is representative of a sort of \emph{uninterpreted fractal recipe}, which tells us how to obtain a regular subfractal from an interpretation of action symbols as contractions on a complete metric space.

    \begin{defi}
        \label{def:fractal semantics}
        Let \((X, \alpha)\) be a productive LTS.
        Given \(x,y \in X\), we write \(x \approx y\) and say that \(x\) and \(y\) are \emph{fractal equivalent} (or that they are \emph{equivalent fractal recipes}) if \(\sem{x}_{\alpha,\sigma} = \sem{y}_{\alpha,\sigma}\) for every complete metric space \(M\) and every \(\sigma : A \to \Con(M)\).
    \end{defi}

    \begin{thm}
    	\label{thm:fractal equivalence is traced}
    	Let \((X, \alpha)\) be a finitely branching productive LTS and \(x,y \in X\). 
        Then \(x \approx y\) if and only if \(\str_{\alpha}(x) = \str_{\alpha}(y)\).
    \end{thm}

	\begin{proof}
		Clearly, if \(\str_\alpha(x) = \str_\alpha(y)\), then \(x \approx y\) by~\autoref{defn:subfractal from process}.
		Conversely, consider \autoref{eg:stream system}, i.e., for a map \(c\colon A \to (0,1)\) and diameter \(c_\diam > 0\), a diameter-\(c_\diam\) Cantor set equipped with the \(c\)-metric and the contraction operator interpretation given by prefixing.
        In \autoref{eg:stream system}, we argued that the unique solution to \((X, \alpha)\) with this contraction operator interpretation is the map \(\str_\alpha \colon X \to A^\omega\).
        Thus, if \(x \approx y\), then \(\str_\alpha(x) = \str_\alpha(y)\). 
	\end{proof}

    This gives way to a soundness/completeness theorem for our version of Rabinovich's logic with respect to its 
    fractal semantics.
    Our proof relies on the axiomatization of trace equivalence that we saw in \autoref{thm:Rabinovich}.

    \begin{thm}[Soundness/Completeness]
        \label{thm:completeness}
        Let \(e, f \in \Term\). 
        Then 
        \[e \approx f ~\text{ if and only if }~ e \equiv f\]
    \end{thm}

    \begin{proof}
        Let \(e,f \in \Term\).
        Then since \((\Term, \gamma)\) is finitely branching and productive,
        \begin{align*}
        	e \approx f
        	\stackrel{\text{(\autoref{thm:fractal equivalence is traced})}}\iff \str(e) = \str(f) 
        	\stackrel{\text{(\autoref{lem:trace stream equivalence})}}\iff \trace(e) = \trace(f) 
        	\stackrel{\text{(\autoref{thm:Rabinovich})}}\iff e \equiv f
            \tag*{\qedhere}
        \end{align*}
    \end{proof}

    One interesting consequence of Theorems~\ref{thm:completeness} and~\ref{prop:cantor sets in R} is that one may restrict their attention to affine contraction operator interpretations  on \(\mathbb R\) and still obtain completeness.

    \begin{cor}
        \label{corr:affine completeness}
        Let \(e,f \in \Term\). 
        Then \(e \equiv f\) if and only if for every affine contraction operator interpretation \(\sigma \colon A \to \Con(\mathbb R)\),
         \(\sem{e}_{\gamma,\sigma} = \sem{f}_{\gamma,\sigma}\). 
    \end{cor}

    \begin{proof}
        The forward direction is soundness.
        For the reverse direction, by \autoref{prop:cantor sets in R}, there is a contraction operator interpretation \(\sigma \colon A \to \Con(\mathbb R)\) such that \(\sigma_\omega \colon A^\omega \to \mathbb R\) is injective.  
        It follows that the setwise extension of $\sigma_\omega$ (by image sets) is injective.
        This implies that if \(\sem{e}_{\gamma,\sigma} = \sem{f}_{\gamma,\sigma}\), then \(\str_\gamma(e) = \str_\gamma(f)\), and therefore \(e \approx f\). 
        From \autoref{thm:completeness}, we obtain \(e \equiv f\).
    \end{proof}

    % SECTION
  \section{Subfractal Measures}
 % \section{A Calculus of Subfractal Measures}
  \label{sec:calculus of measures}
    
    In addition to showing the existence of self-similar sets and their correspondence with contraction operator interpretations (in Hutchinson's terminology, iterated function systems), Hutchinson also shows that every probability distribution on the contractions corresponds to a unique measure, called the \emph{invariant measure}, that satisfies a certain recursive equation and whose support is the self-similar set.
    In this section, we replay the story up to this point, but with Hutchinson's invariant measure construction instead of the invariant (self-similar) set construction.
    We specify fractal measures using a probabilistic version of LTSs called \emph{labelled Markov chains}, as well as a probabilistic version of Milner's specification language introduced by Stark and Smolka~\cite{Stark2000Probabilistic}.
    Similar to how fractal equivalence coincides with trace equivalence, \emph{fractal measure equivalence} is equivalent to a probabilistic version of trace equivalence due to Kerstan and K\"onig~\cite{Kerstan2013Trace}.
    
    \paragraph*{Invariant measures}
    Recall that a \emph{Borel probability measure} on a metric space \(M\) is a \([0,1]\)-valued function \(\rho\) defined on the Borel subsets of $M$ (the smallest \(\sigma\)-algebra containing the open balls of \(M\)) that is countably additive and assigns \(\rho(\emptyset) = 0\) and \(\rho(M) = 1\).
    A Borel probability measure \(\rho\) is \emph{supported by a point \(p\in M\)} if for any open set \(U\) containing \(p\), \(\rho(U) > 0\).
    The \emph{support} \(\supp(\rho)\) of a Borel measure \(\rho\) is the set of points supporting \(\rho\). 
    A Borel probability measure is \emph{compactly supported} if the set of points supporting it is a compact set. 
    Note that the support of a Borel measure is always a topologically closed set, so a Borel measure is compactly supported if and only if its support is bounded.
    For a more detailed introduction to Borel measures, see for example~\cite{folland}.
    
    Hutchinson shows in~\cite{Hutchinson1981Fractals} that, given \(\sigma\colon A \to \Con(M)\), each probability distribution \(\theta\colon A \to [0,1]\) on \(A\) gives rise to a unique compactly supported Borel probability measure \(\hat \theta_\sigma\) on \(M\), called the \emph{invariant measure}, that is supported by the self-similar set
     \(\mathbf S_\sigma\) from~\eqref{def-S-sigma}
    and such that for any Borel set \(B \subseteq M\),
    \begin{equation}
    	\label{eq:invariant measure 1}
        \hat \theta_\sigma(B) = \sum_{a \in A} \theta(a)~\sigma_a^\#\hat\theta_\sigma(B)
    \end{equation}
    Here and elsewhere, the \emph{pushforward} \(f^\#\rho\) of a Borel measure \(\rho\) with respect to a continuous (or more generally, Borel measurable) map \(f\) is 
    the probability measure defined by \(f^\#\rho(B) = \rho(f^{-1}(B))\) for any Borel subset \(B\) of \(M\)~\cite{folland}.
    The pushforwards \(\sigma_a^\#\hat\theta_\sigma\) in~\eqref{eq:invariant measure 1} are well-defined because contractions are continuous maps.
 	\autoref{eq:invariant measure 1} tells us that, similar to how self-similar sets are the union of their images over a set of contractions, the invariant measure \(\hat\theta_\sigma\) is a convex combination of its pushforwards with respect to a set of contractions.
    We can use the first part of the Banach Fixed Point Theorem, \autoref{thm:Banach fixed-point} (1), to show that~\eqref{eq:invariant measure 1} is satisfied by at most one compactly supported Borel measure on \(M\).

    Fix a complete metric space \(M\).
    
    \begin{defi}        
        \label{KRmetric}
        We write \(\Prob(M)\) for the set of compactly supported Borel probability measures on \(M\) equipped with the \emph{Kantorovich-Rubinstein metric}, defined by
        \[
            d(\rho_1, \rho_2)
            = \sup \Big\{\int_M f~\diff\rho_1 - \int_M f ~\diff\rho_2~\Big|~f \colon M \to \mathbb R \text{ is nonexpanding}\Big\}
        \]
        for any \(\rho_1,\rho_2 \in \Prob(M)\).
        Above, \(\int_M f~\diff\rho_i\) is given by the Lebesgue integral.%
        \footnote{Note that we use the symbol \(\diff\) in the integral to represent the differential, so as not to clash with \(d\) as in \emph{distance}.}
    \end{defi}
    
    It is important to note that finiteness of \(d(\rho_1, \rho_2)\) requires the assumption that \(\rho_1\) and \(\rho_2\) are compactly supported (equivalently, have bounded support).

    \begin{defi}\label{def:Markov}
        Given a fixed probability distribution \(\theta\colon A \to [0,1]\), the \emph{Markov operator} \(F\colon\Prob(M)\to \Prob(M)\) is defined by
        \begin{equation}
            \label{eq:Markov op}
            F(\rho)  = \sum_{a\in A} \theta(a) ~ \sigma_a^{\#} \rho
        \end{equation}
        for each \(\rho \in \Prob(M)\).
    \end{defi}
        
    A measure $\hat{\theta}_{\sigma}$ satisfies~\eqref{eq:invariant measure 1} if and only if it is a fixed point of $F$.
    Thus, we would like to use fixed-point techniques to show the existence and uniqueness of invariant measures.
    Indeed, Hutchinson proves in~\cite[Theorem 4.4(1)]{Hutchinson1981Fractals} that \(F\) is a contraction in the Kantorovich-Rubinstein metric.
    Therefore, if \(F\) has a fixed-point, then \autoref{thm:Banach fixed-point}(1) tells us it is unique, and so the invariant measure also would be unique.
    
    However, it is not true for an arbitrary complete metric space \(M\) that \(\Prob(M)\) is complete, so we cannot apply \autoref{thm:Banach fixed-point}(2) to prove the existence of the invariant measure.\footnote{In his original paper~\cite{Hutchinson1981Fractals}, Hutchinson fallaciously relies on the Banach fixed-point theorem to construct the invariant measure. 
    This minor hiccup can be remedied by either restricting the space of measures in a suitable way~\cite{kravchenko}, by assuming \(M\) is compact, or by constructing the invariant measure explicitly.
    We opt for the latter approach.}
    In the next two paragraphs, we describe an explicit construction of the invariant measure.
    A more general construction and verification of its stated properties can be found in \autoref{thm:LMC unique solution}.
    
    The invariant measure can be concretely described as follows.
    Given \(\sigma\colon A \to \Con(M)\), let \(c(a) = c_a > 0\) be a contraction coefficient for \(\sigma_a\) for each \(a \in A\), and consider the diameter-\(1\) Cantor set equipped with the \(c\)-metric \((A^\omega, d_c)\).
    Now, given a probability distribution \(\theta\colon A \to [0,1]\), we define \(\hat\theta\), called the \emph{stream measure specified by \(\theta\)}, to be the unique Borel probability distribution on \((A^\omega, d_\sigma)\) such that
    \begin{equation}
    	\label{eq:invariant measure 2}
        \hat\theta(B_{w}) = \prod_{i = 1}^n \theta(a_i)
    \end{equation}
    for any \(w = a_1\cdots a_n \in A^*\) (here, we take the empty product to be \(1\)).
    As will be explained shortly, results from~\cite{Kerstan2013Trace} tell us that~\eqref{eq:invariant measure 2} does indeed extend to a unique Borel probability measure on \((A^\omega, d_\sigma)\). 
    This again relies on the fact that any two Borel measures that agree on all basic open sets must agree on all Borel sets.
    The next proposition tells us that the invariant measure is precisely \(\hat\theta_\sigma = \sigma_\omega^\#\hat\theta\), where \(\sigma_\omega\) is the map given by limits which we saw in~\autoref{def:contra operator interpretation}.
    We will prove a more general version of this result in \autoref{thm:LMC unique solution}.

    \begin{prop}\label{prop:Markov}
        Let \(M\) be a complete metric space, let \(\sigma\colon A \to \Con(M)\), and let \(F\) be the Markov operator defined in~\eqref{eq:Markov op}.
        The unique fixed-point of 
        \(F\) is the pushforward \(\sigma_\omega^\#\hat\theta\), where 
        \(\sigma_\omega: A^{\omega}\to M \) is from \autoref{def:contra operator interpretation}, and
        \(\hat\theta\in \Prob(A^{\omega})\) is defined in~\eqref{eq:invariant measure 2}.
    \end{prop}

    In the sequel, we are going to view the specification \(\theta\) of the invariant measure \(\hat \theta_\sigma\)
    (the pushforward of the stream measure) as a probability distribution over the transitions in a one-state LTS.
    This construction generalizes to multiple-state transition systems by moving from probability distributions on \(A\) to \emph{labelled Markov chains}, which are essentially LTSs where each state comes equipped with a probability distribution on its outgoing transitions. 
    Again, the labels will be interpreted as contractions on a complete metric space.

    \paragraph*{Algebras, Coalgebras, and Corecursive Algebras}    
    At this point we make a short digression into the general notions of \emph{universal algebra} and \emph{universal coalgebra}.
    A standard reference for universal algebra is~\cite{SankappanavarBurris1981}, although we will be formulating its notions using category-theoretic language, in the style of~\cite{AdamekTrnkova1989}.
    Universal coalgebra is a category-theoretic framework for studying the many different types of transition systems appearing in computer science in a uniform way.
    A standard reference for universal coalgebra is~\cite{Rutten00}. 
    
    The purpose of this digression is to ease the formulation of \emph{labelled Markov chains} and \emph{coalgebra-to-algebra homomorphisms} that will play a significant role in the rest of the paper.
    Labelled Markov chains are a probabilistic analogue of labelled transition systems.
    While labelled transition systems can mostly be studied without the categorical language of universal coalgebra, the language of coalgebra greatly improves ease-of-use for probabilistic systems~\cite{Sokolova11}.
    Following our digression, we will return to the topic of subfractal measures.
    % To prove this, we are going to make use of \emph{corecursive algebras} and \emph{corecursive maps}.

    % Universal coalgebra is a category-theoretic framework for studying the many different types of transition systems appearing in computer science in a uniform way.
    % A standard reference for universal coalgebra is~\cite{Rutten00}. 

    \begin{defi}
       Let $H\colon \mathcal{C}\to \mathcal{C}$ be an endofunctor on any category.
        An \emph{\(H\)-algebra} is a pair \((C, \gamma)\) consisting of an object $C$ of $\mathcal{C}$ and a morphism \(\gamma \colon HC \to C\). 
        An \emph{\(H\)-algebra homomorphism} \(h\colon (C, \gamma_C) \to (D, \gamma_D)\) is a morphism \(h \colon C \to D\) such that \(\gamma_D \circ H(h) = h \circ \gamma_C\).
         
        An \emph{\(H\)-coalgebra} is a pair \((X, \beta)\) consisting of an object $X$ of $\mathcal{C}$ and a  morphism \(\beta \colon X \to HX\). 
        An \emph{\(H\)-coalgebra homomorphism} \(h\colon (X, \beta_X) \to (Y, \beta_Y)\) is a morphism \(h \colon X \to Y\) such that \(\beta_Y \circ h = H(h) \circ \beta_X\).
    \end{defi}
     
    In this paper, we are mainly interested in two endofunctors on the category $\Set$ of sets and functions.
    First, $FX = \pstar(A\times X)$ where $\pstar$ is the non-empty finite power set functor and $A$ is our set of actions (a fixed finite set in all of our work).
    Given $f\colon X\to Y$, $F(f)\colon \pstar(A\times X)\to \pstar(A\times Y)$ is given by $F(f)(U) = \set{(a,f(x)) \mid (a,x) \in U}$.
    The second functor of interest is $GX = \D(A\times X)$.  
    Here, $\D (X)$ is the set of finitely supported probability distributions on $X$, i.e., functions $\theta\colon X\to [0,1]$ such that $\theta(x) = 0$ for all but finitely many $x\in X$, and $\sum_{x \in X} \theta(x) = 1$.  
    This time, $G(f)(\theta)(y)= \sum \set{\theta(x) \mid f(x) = y}$ for all $y\in Y$.

    \begin{exa} \label{example-LTS-coalgebras}
        Due to~\autoref{lem:trace stream equivalence}(2), productive and finitely branching LTSs are
        exactly the coalgebras of $\pstar$, the non-empty finite power set functor.
    \end{exa}
    
    Later in the paper, we will see algebras and coalgebras of \(\pstar(A\times (-))\) and \(\D(A\times(-))\),
     and in due course we will name them as shown below.
    \[\begin{tabular}{ccc} %FIXED changed to booktabs and removed vertical lines
        Set Functor & Name of Algebra & Name of Coalgebra \\
        \toprule
        $FX = \pstar(A\times X)$ & labelled transition algebra & productive, finitely branching \\
        & & labelled transition system \\
        \midrule
        $FX = \D(A\times X)$ & labelled Markov algebra & labelled Markov chain\\
    \end{tabular}\]
    Aside from the specific examples of labelled transition algebras and labelled Markov algebras that are of direct concern to us in this paper, there are two motivating examples that may help the reader. 
    One such example is the labelled transition algebra of infinite (unordered) trees decorated by labels from \(A\). 
    There, the algebra structure turns a set \(\{(a_1, t_1), \dots, (a_n, t_n)\}\) into the tree whose immediate subtrees are \(t_1, \dots, t_n\) and outgoing root-to-child relations are labelled with \(a_1, \dots, a_n\) respectively.
    A potentially helpful example of a labelled Markov algebra is similar to the infinite trees from before, but whose parent-to-child relationships carry (nonzero) probabilities in addition to labels.

    Coalgebras for \(\pstar(A\times (-))\) and \(\D(A\times(-))\) that are of particular interest to us are the ones that are finite around each state.

    \begin{defi}
        Let \(H \colon \Set \to \Set\) be an endofunctor and \((X, \beta)\) an \(H\)-coalgebra. 
        A \emph{subcoalgebra} of \((X, \beta)\) is an \(H\)-coalgebra \((U, \beta_U)\) such that \(U \subseteq X\) and the inclusion map \(U \hookrightarrow X\) is a coalgebra homomorphism.       
        We say that \((X, \beta)\) is \emph{finite} if \(X\) is finite, and \emph{locally finite} if every state \(x \in X\) is included in a finite subcoalgebra of \((X, \beta)\).
    \end{defi}

    We also have the notion of a \emph{finitely corecursive algebra} (Definition \ref{definitionoffinitelycorecursive}), which is a weaker version of the corecursive algebras first introduced by Capretta et al~\cite{cuv06}.  
    Finitely corecursive algebras are also weaker than 
    \emph{iterative algebras}, and these are weaker than 
    the better-known notion of \emph{completely iterative algebras}~\cite{Bloom1993Iteration}.
    %\footnote{All corecursive algebras in this paper area also completely iterative, but we do not need this fact.}
    Connections between completely iterative algebras and fractals date as far back as~\cite{Milius2006Recursion}.

    \begin{defi}\label{definitionoffinitelycorecursive}
        Let $H\colon \Set \to \Set$ be an endofunctor.
        An algebra $\gamma\colon HC \to C$ is \emph{finitely corecursive} if for every locally finite \(H\)-coalgebra $(X, \beta)$, there is a unique \emph{coalgebra-to-algebra homomorphism} $\beta^\dag\colon (X, \beta) \to (C, \gamma)$, i.e., a map \(\beta^\dag \colon X \to C\) such that 
        % 
        % \footnote{I would use $\alpha: HA \to A$ for an algebra. But we have $A$ as our action set, so we can't do this.} 
        %   
        $\beta^\dag = \gamma\o  H(\beta^\dag) \o \beta$:
        \[\begin{tikzcd}
            X \ar{d}[swap]{\beta} \ar{r}{\beta^\dag} 
            & C \\ 
            HX  \ar{r}{H(\beta^\dag)} &  HC \ar{u}[swap]{\gamma}
        \end{tikzcd}\]
        The map $\beta^\dag$ is also called \emph{the solution of $\beta$
        in the algebra $(C,\gamma)$}.
    \end{defi}

    \begin{lem}
        \label{coalg-to-alg-morphisms}
        Let $\gamma\colon HC \to C$ be a finitely corecursive algebra for the endofunctor $H$.
        Let \((X, \beta_X)\) and \((Y, \beta_Y)\) be locally finite coalgebras, and let $\phi: X\to Y$ be a coalgebra morphism.
        Then $\beta_X^\dag = \beta_Y^\dag \o \phi$.
    \end{lem}
        
    \begin{proof} 
        The proof is indicated in the diagram below.
        \[\begin{tikzcd}
            X\ar{r}{\phi}   
                \ar{d}[swap]{\beta_X}
                & Y \ar{d}[swap]{\beta_Y} \ar{r}{\beta_Y^\dag} 
                & C 
                \\ 
                HX\ar{r}{H(\phi)}       
                & HY \ar{r}{H(\beta_Y^\dag)} 
                & HC \ar{u}[swap]{\gamma}
        \end{tikzcd}\] 
        The commutativity of the square on the left expresses that $\phi$ is a coalgebra morphism, and that the square on the right commutes expresses that $\beta_Y^{\dag}$ is a  solution to $(Y, \beta_Y)$ in $C$.  
        Since both squares commute, so does the overall outside.
        This shows that $\beta_Y^{\dag}\o \phi$ is a solution to $(X, \beta)$ in $C$.  
        By uniqueness of solutions, $\beta_Y^{\dag}\o \phi = \beta_X^{\dag}$.
    \end{proof}    
    
    The following lemma allows one to check finite corecursiveness by only considering finite coalgebras. 

    \begin{lem}
        \label{lem:finite corecursion is finite}
        Let \(H \colon \Set \to \Set\) be an endofunctor and let \(\gamma \colon HC \to C\) be an algebra. 
        Then \((C, \gamma)\) is a finitely corecursive algebra if and only if for any finite \(H\)-coalgebra \((X, \beta)\), there is a unique coalgebra-to-algebra homomorphism \((X, \beta) \to (C, \gamma)\).
    \end{lem}

    \begin{proof}
        The forward direction follows from the fact that finite coalgebras are locally finite. 

        For the reverse direction, let \((X, \beta)\) be a locally finite coalgebra. 
        Every state of \((X, \beta)\) is contained in a finite subcoalgebra and by assumption, finite subcoalgebras admit unique coalgebra-to-algebra homomorphisms into \((C, \gamma)\). 
        We start with the claim that there can be at most one coalgebra-to-algebra homomorphism \((X, \beta) \to (C, \gamma)\).

        Suppose that $\phi,\psi \colon (X, \beta) \to (C, \gamma)$ are coalgebra-to-algebra homomorphisms and let \(x \in X\).
        We aim to show that \(\phi(x) = \psi(x)\). 
        Since \((X, \beta)\) is locally finite, there is a finite subcoalgebra \((U, \beta_U)\) of \((X, \beta)\) such that \(x \in U\). 
        Let \(\iota_U \colon U \hookrightarrow X\) be the inclusion map of \(U\) into \(X\), and recall that (by definition) \(\iota_U \colon (U, \beta_U) \to (X, \beta)\) is a coalgebra homomorphism.
        The composition of a coalgebra homomorphism with a coalgebra-to-algebra homomorphism is a coalgebra-to-algebra homomorphism by \autoref{coalg-to-alg-morphisms}, so in particular this holds for \(\phi \circ \iota_U\) and \(\psi \circ \iota_U\).
        \begin{equation*}
            \begin{tikzcd}
                U 
                    \ar[r, "\iota_U"] 
                    \ar[d, "\beta_U"']
                & X 
                    \ar[r, shift left, "\phi"] 
                    \ar[r, shift right, "\psi"'] 
                    \ar[d, "\beta"]
                & C 
                \\
                HU 
                    \ar[r, "H(\iota_U)"]
                & HX
                    \ar[r, shift left, "H(\phi)"] 
                    \ar[r, shift right, "H(\psi)"'] 
                & HC
                    \ar[u, "\gamma"]
            \end{tikzcd}
        \end{equation*}
        We have assumed that \((U, \beta_U)\) is finite, so there is at most one coalgebra-to-algebra homomorphism \((U, \beta_U) \to (C, \gamma)\).
        It follows that \(\phi \circ \iota_U = \psi \circ \iota_U\), and in particular, 
        \[
            \phi(x) = \phi \circ \iota_U(x) = \psi \circ \iota_U(x) = \psi(x)
        \]
        This shows that there is at most one coalgebra-to-algebra homomorphism \((X, \beta) \to (C, \gamma)\).

        It now suffices to construct a coalgebra-to-algebra homomorphism \((X, \beta) \to (C, \gamma)\).
        For each finite subcoalgebra \((U, \beta_U)\) of \((X, \beta)\), let \(\phi_U \colon (U, \beta_U) \to (C, \gamma)\) be the unique coalgebra-to-algebra homomorphism.
        We define the 
        relation 
        \[
        \phi = \{(x, y) \mid x \in X \text{ and there is a finite subcoalgebra \((U, \beta_U)\) with \(y = \beta_U(x)\)}\}
        \]
        We need to show that \(\phi\) is a function and that it is a coalgebra-to-algebra homomorphism. 
        % We need to show that this is well-defined and that it is a coalgebra-to-algebra homomorphism.
        
        To see that it is a function, suppose \(x \in U \cap V\) for two finite subcoalgebras \((U, \beta_U)\) and \((V, \beta_V)\), and let \((x, y), (x, y') \in \phi\) such that \(y = \phi_U(x)\) and \(y' = \phi_V(x)\).
        The intersection of two subcoalgebras is also a subcoalgebra~\cite[Theorem 3.1]{gummschroeder-bounded}, so if we set \(W = U\cap V\), \(W\) is also the state space of a 
        finite subcoalgebra \((W, \beta_W)\).
        Now, because the composition of a coalgebra-to-algebra homomorphism with a coalgebra homomorphism is always a coalgebra-to-algebra homomorphism (again using \autoref{coalg-to-alg-morphisms}), \(\phi_W\), \(\phi_U \circ \iota_W\), and \(\phi_V \circ \iota_W\) are all coalgebra-to-algebra homomorphisms \((W, \beta_W) \to (C, \gamma)\).
        By uniqueness of coalgebra-to-algebra homomorphisms for finite coalgebras, \(\phi_U \circ \iota_W = \phi_W = \phi_V \circ \iota_W\).
        It follows that 
        \[
            y = \phi_U(x) = \phi_U \circ \iota_W(x) = \phi_V \circ \iota_W(x) =  \phi_V(x)=  y'
        \]
        To see that \(\phi\) is a coalgebra-to-algebra homomorphism, let \(x \in U \subseteq X\) and calculate
        \[
            \begin{tikzcd}[ampersand replacement=\&]
                U \ar[r, "\iota_U"] \ar[d, "\beta"] \ar[rr, bend left, "\phi_U"]
                \& X \ar[r, "\phi"] \ar[d, "\beta"]
                \& C \\
                HU \ar[r, "H(\iota_U)"] \ar[rr, bend right, "H(\phi_U)"]
                \& HX \ar[r, "H(\phi)"]
                \& HC \ar[u, "\gamma"]
            \end{tikzcd}
            \qquad
            \begin{aligned}
                \phi(x)&= \phi_U(x) \\
                &= \gamma \circ H(\phi_U) \circ \beta_U(x) \\
                &= \gamma \circ H(\phi) \circ H(\iota_U) \circ \beta_U(x) \\
                &= \gamma \circ H(\phi) \circ \beta \circ \iota_U(x) \\
                &= \gamma \circ H(\phi) \circ \beta(x)
            \end{aligned}
        \]
        \vspace*{-2.8\baselineskip}
        \begin{flushright}
         \qedsymbol
        \end{flushright}
    \renewcommand{\qedsymbol}{}
    \end{proof}
    \renewcommand\qedsymbol{\usebox{\lmcsQEDSymbol}}
%FIXED Adjusted QEDbox
    \vspace{-\baselineskip}
\paragraph*{An extended example: regular subfractals via finite corecursive algebras}

    %For most of this paper we worked with arbitrary complete metric spaces, but for %\emph{bounded} complete metric space.  We must do this in order that every function set $M^X$ has a well-defined and finite metric,
    % given by
    In this next part of the paper, recall that for a complete metric space \(M\) and a finite set \(X\), the set \(M^X\) obtains a complete metric space structure from the product metric,
    \(
        d^X(f,g) = \max_{x\in X} d(f(x),g(x))
    \).
    Recall also that  $\K(M)$  is the space of non-empty compact subsets of \(M\) using the Hausdorff metric in (\ref{eqH}).
% \textcolor{red}{We use $\mathcal{C}$ for an element of $ \pstar(A\times \K(M))\to \K(M)$.}\marginpar{\textcolor{teal}{TN: I'm still kind of confused about this, it looks like $\mathcal{C}$ is an element of the set $\mathcal{P}_*(A\times {\bf K}(M))$?  So $\mathcal{C}$ is just a set of objects in $A\times {\bf k}(M)$, which is a set? I guess I don't understand why we don't just use $C$ for this instead of $\mathcal{C}$, which previously we used for categories. }LM: it's fine with me to change $\mathcal{C}$ to $C$ here.}
    
    \begin{thm}         
        \label{compact-sets-as-corecursive-algebra}
        Let $M$ be a complete metric space, and let \(\sigma \colon A \to \Con(M)\). 
        For the $\Set$ endofunctor $FX = \pstar(A\times X)$, the algebra $k\colon \pstar(A\times \K(M))\to \K(M)$ given by 
        \begin{equation}\label{eq:k}
            k(C) = \bigcup \set{\sigma_{a}(K) \mid (a,K)\in C}.
        \end{equation} 
        is finitely corecursive.%
        \footnote{
            \label{fn:corecursive}
            It is worth noting that if \(M\) is bounded, then \(M^X\) is a complete metric space as well. 
            This leads to a different version of \autoref{compact-sets-as-corecursive-algebra}, which instead states that if \(M\) is bounded, then \(\K(M)\) is \emph{corecursive} as opposed to just finitely corecursive. 
            We do not pursue this issue here because we are concerned with a number of unbounded metric spaces in this paper.
        }
        Moreover, if an \(F\)-coalgebra \((X, \beta)\) is locally finite, its solution in the algebra $\K(M)$ is $\sem{-}_{\beta,\sigma}\colon X \to \K(M)$.     
        \end{thm}
    \noindent %FIXED indent
    We saw in~\autoref{example-LTS-coalgebras} that an \(F\)-coalgebra structure is the same data as a productive and finitely-branching LTS. 
    For this reason, we generally identify the two notions and refer to an \(F\)-coalgebra as an LTS and vice versa.

    \begin{proof}              
        To see the first statement, by \autoref{lem:finite corecursion is finite}, it suffices to show that finite \(F\)-coalgebras admit unique coalgebra-to-algebra homomorphisms into \(\K(M)\).
        Let \((X, \beta)\) be a finite LTS.
        Recall from \autoref{thm:fractal semantics coincide} that finite productive LTSs have unique solutions in the sense of \autoref{def:solution}, i.e., there is a unique map \(\phi \colon X \to \K(M)\) such that 
        \[
            \phi(x) = \bigcup_{x \tr{a} y} \sigma_a(\phi(y))
        \]
        for any \(x \in X\) (this is the equation from \autoref{def:solution}).
        Chasing the coalgebra-to-algebra homomorphism square around its underside, one finds that an arbitrary map \(\phi \colon X \to \K(M)\) is a coalgebra-to-algebra homomorphism if and only if for all \(x \in X\), the equation on the right below holds:
        \[
            \begin{tikzcd}[column sep = 45]
                X \ar{d}[swap]{\beta} \ar{r}{\phi} 
                &  \K(M) 
                \\ 
                \pstar(A\times X) \ar{r}{\pstar(1_A \times \phi)} 
                & \pstar(A\times \K(M)) \ar{u}[swap]{k}
            \end{tikzcd}       
            \qquad
            \begin{aligned}             
                \phi(x) 
                &= k \circ \pstar(1_A \times \phi) \circ \beta(x) \\
                &= k (\{(a, \phi(y)) \mid x \tr{a} y\}) \\
                &= \bigcup_{x \tr{a} y} \sigma_a(\phi(y))
            \end{aligned}
        \]
        So coalgebra-to-algebra maps are the same thing as  solutions in the sense of \autoref{def:solution}.
        Hence, by \autoref{thm:fractal semantics coincide}, every finite \(F\)-coalgebra does indeed admit a unique coalgebra-to-algebra homomorphism into  \((\K(M), k)\). 
        Once again, we use \autoref{lem:finite corecursion is finite} to conclude that \((\K(M), k)\) is finitely corecursive.

        Moreover, by our characterization of coalgebra-to-algebra maps as solutions to the corresponding productive LTS \((X, \beta)\) above, 
        by \autoref{thm:fractal semantics coincide}(\ref{item-needy}), \(\sem{-}_{\alpha, \sigma} \colon X \to \K(M)\) is the unique coalgebra-to-algebra homomorphism. \qedhere
    \end{proof}

    \autoref{compact-sets-as-corecursive-algebra} tells us that there is a direct connection between the regular subfractals of \cref{sec:fractals from LTS} and finitely corecursive algebras.
    In the next section, we turn to the probabilistic analog of labelled transition systems and their measure-theoretic semantics, where we will see another example of a finitely corecursive algebra.

    \paragraph*{Labelled Markov Chains}
    Labelled Markov chains generalize our specification of invariant measures to multi-state transition systems with branching given by finitely supported probability distributions. 

    \begin{defi}
        \label{def:lmc}
        A \emph{labelled Markov chain} (LMC) is a coalgebra for the functor $\D(A\times -)$.  
        That is, it is a      
        pair \((X, \beta)\) consisting of a set \(X\) of \emph{states} and a \emph{transition function} \(\beta\colon X \to \D(A \times X)\).
             We write \(x \tr{r\mid a}_\beta y\) if \(\beta(x)(a, y) = r\), often dropping the symbol \(\beta\) if it is clear form context.     
        A \emph{homomorphism} of LMCs \(h\colon (X, \beta_X) \to (Y, \beta_Y)\) is thus a map \(h\colon X \to Y\) such that 
        \(
            \D(1_A \times h)\circ \beta_X = \beta_Y \circ h
        \)
        where \(1_A\) is the identity map on \(A\).      
    \end{defi}

    As we have already seen, given a complete metric space $M$ and a contraction operator interpretation \(\sigma\colon A \to \Con(M)\), every state \(x\) of a productive LTS \((X, \alpha)\) with labels in \(A\) corresponds to a regular subfractal \(\sem{x}_{\alpha,\sigma}\) of  the set \(\mathbf S_\sigma\) from~\eqref{def-S-sigma}.
    This regular subfractal is the continuous image of the set \(\str_\alpha(x)\) under the map \(\sigma_\omega \colon (A^\omega, d_\sigma) \to M\).
    The set \(\sem{x}_{\alpha,\sigma}\) is characterized by its satisfaction of the equations representing the LTS \((X, \alpha)\).

    Every LMC \((X, \beta)\) has an \emph{underlying} LTS \((X, \bar\beta)\), where \(\bar\beta(x) = \{(a, y) \mid \beta(x)(a, y) > 0\}\).  Note that since $\beta(x)$ is a probability distribution, our underlying LTS is automatically productive since at least one transition must have positive probability.  
    For each \(x\in X\), we are going to define a probability measure \(\hat\beta_\sigma(x)\) on \(\mathbf S_\sigma\) whose support is 
    \(\sem{x}_{\bar\beta,\sigma}\), and that satisfies a recursive system of equations represented by the LMC \((X, \beta)\).
    Roughly, \(\hat\beta_\sigma(x)\) is the pushforward of a certain Borel probability measure \(\hat\beta(x)\) on \(A^\omega\).

    \begin{defi}\label{trmdef}
        Suppose we are given an  LMC \((X, \beta)\)  and a state  \(x\in X\).
        We follow Kerstan and K\"onig~\cite{Kerstan2013Trace} and define the \emph{trace measure} $\hat\beta(x)\in\Prob(A^{\omega})$ by giving its value on a basic open set \(B_{w}\). 
        Given \(w = a_1\cdots a_n \in A^*\), we define
        \begin{equation}
            \label{eq:trace measure}
            \hat\beta(x)(B_w) = \sum\{r_1\cdots r_n \mid x\tr{r_1\mid a_1} x_1 \tr{\phantom{\ \mid \ }}\cdots \tr{r_n\mid a_n} x_n
            \mbox{ for some $x_1$, $x_2$, $\ldots$, $x_n$}\}
        \end{equation}
        In particular, \(\hat\beta(x)(B_\epsilon) = \hat\beta(x)(A^\omega) = 1\).
    \end{defi}

    We use the following convenient proposition, stated as Theorem 3.5 in Goy and Rot~\cite{GoyRot18}, which tells us that~\eqref{eq:trace measure} defines a unique Borel probability measure on \(A^\omega\).

    \begin{prop}\label{prop:unique measure extension}
        Let \(j\colon A^* \to [0,1]\) satisfy \(j(w) = \sum_{a \in A} j(wa)\) for any \(w \in A^*\) and \(j(\epsilon) = 1\), where \(\epsilon\) is the empty word.
        Then there is a unique compactly supported Borel probability measure \(\rho \in \Prob(A^\omega)\) such that for any \(w \in A^*\), \(\rho(B_w) = j(w)\).
    \end{prop}
    A simple calculation using~\eqref{eq:trace measure} shows that \(\hat\beta(x)(B_w) = \sum_{a \in A} \hat\beta(x)(B_{wa})\) for any LMC \((X, \beta)\). 
    So, if we take \(j(w) = \hat\beta(x)(B_w)\), we immediately have \(j(\epsilon) = 1\) and \(j(w) = \sum_{a \in A} j(wa)\) like in \autoref{prop:unique measure extension}.
    Applying the proposition,
    there is a unique Borel probability measure \(\hat\beta(x)\) on \(A^\omega\) such that~\eqref{eq:trace measure} holds for any basic open set \(B_w\).

    \begin{defi}
        \label{def:regular subfractal measure}
        Let \((X, \beta)\) be an LMC, and \(\sigma \colon A \to \Con(M)\) be a contraction operator interpretation in a complete metric space.
        For each \(x \in X\),  the \emph{regular subfractal measure} corresponding to \(x\) is \(\hat\beta_\sigma(x) = \sigma_\omega^\#\hat\beta(x)\), i.e., \(\hat\beta_\sigma \colon X\to \Prob(M)\) is defined by \(\hat\beta_\sigma = \sigma_\omega^\#\o \hat\beta\).
    \end{defi}

    % Intuitively, the regular subfractal measure of a state in a LMC under a contraction operator interpretation computes the probability that, if run stochastically according to the probabilities labelling the edges, the sequence of points of \(M\) observed in the run eventually lands within a given Borel subset of \(M\).
    
    \paragraph*{Systems of Probabilistic Equations}

    The next theorem is the main result in this section.  Like \autoref{compact-sets-as-corecursive-algebra}, it gives a finitely corecursive algebra
    which may be used to derive some of the basic results on fractals.   
    We would like to prove the two the same way, by using an argument which calls on the Banach fixed point theorem.  
    However, as we noted above, for a complete space $M$, $\Prob(M)$ is not complete in general, so we need a different way.

    \begin{thm}         
        \label{thm:LMC unique solution}
        Let $M$ be a complete metric space, and let \(\sigma \colon A \to \Con(M)\). 
        The algebra $p: \D(A\times \Prob(M))\to \Prob(M)$ given by 
        \begin{equation}
            \label{eq:p}
            p (r_1 (a_1, \theta_1) + \cdots + r_n (a_n,  \theta_n))(B)
            = \sum_{i=1}^n  r_i \sigma_{a_i}^{\#}\theta_i(B)
        \end{equation} 
        is a finitely corecursive algebra for the $\Set$ endofunctor $FX = \D(A\times X)$.
        More specifically, given a locally finite labelled Markov chain $(X,\beta)$, its solution $\beta^{\dag}$ is 
        $\hat \beta_\sigma \colon X\to \Prob(M)$ (see \autoref{def:regular subfractal measure}).
        Also, the support of \(\hat\beta_\sigma(x)\) is \(\sem{x}_{\bar\beta, \sigma}\).
    \end{thm}
    
    \begin{proof}
        Again, we appeal to \autoref{lem:finite corecursion is finite}.
        Fix a finite coalgebra $(X,\beta)$.
        We aim to show that there is a unique coalgebra-to-algebra homomorphism \((X, \beta) \to (\Prob(M), p)\).
        Let us first characterize coalgebra-to-algebra maps among all maps \(\phi\colon X \to \Prob(M)\) in the following way: $\phi$ is a coalgebra-to-algebra map (i.e., the diagram on the left below commutes) if and only if for any \(x \in X\) and any Borel set \(B\subseteq M\), the equations on the right below hold:
        \[
            \begin{tikzcd}[column sep = 45, row sep = 30]
                X \ar{d}[swap]{\beta} \ar{r}{\phi} 
                & \Prob(M) 
                \\ 
                \D(A\times X) \ar{r}{\D(1_A \times \phi)} 
                & \D(A\times \Prob(M)) \ar{u}[swap]{p}
            \end{tikzcd}       
            \qquad     
            \begin{aligned}
                \phi(x)(B) 
                &= p \circ \D(1_A \times \phi) \circ \beta(x)(B) \\
                &= p\Big(\sum_{x \tr{r \mid a} y} r(a, \phi(y))\Big)(B) \\
                &= \sum_{x \tr{r\mid a} y} r\sigma_a^\#(\phi(y))(B)
            \end{aligned}        
        \]
        We are going to show that \(\hat\beta_\sigma\colon X \to \Prob(M)\) is the unique map with this property.  
        This can be done in two steps. 
        \begin{enumerate}[(i)]
            \item \label{item-f-contraction}
            First, we show that the map \(F_\beta\colon\ \Prob(M)^X \to  \Prob(M)^X \) defined by 
            \[
                F_{\beta}(h)(x)(B) 
                = p\o \D(1_A \times h)\o \beta (x)(B)
                = \sum_{x \tr{r\mid a} y} r\sigma_a^\#(h(y))(B)
            \]
            is a contraction mapping on the metric space \((\Prob(M)^X, d^X)\), where the metric is the product metric, \(d^X(f, g) = \max\{d(f(y), g(y)) \mid y \in X\}\), and
            \item \label{item-f-fixed-point}
            that \(\hat\beta_\sigma\) is a fixed-point of this mapping.
        \end{enumerate}
        For~\ref{item-f-contraction}, note that indeed, \(F_\beta(h)(x)\) is a compactly supported Borel probability measure, because it is a finite convex combination of pushforwards of compactly supported Borel probability measures.
        For the rest our proof of~\ref{item-f-contraction}, we reuse the proof technique of~\cite[Theorem 4.4(1)]{Hutchinson1981Fractals}. 
        For each \(a \in A\), let \(c_a\) be a nonzero contraction coefficient for \(\sigma_a\). 
        Let \(c = \max\{c_a \mid a \in A\}\), so that \(0 < c < 1\) and for any \(a \in A\), \(c_a/c \le 1\).  
        For any nonexpanding \(f \colon M \to \mathbb R\) and any \(y \in X\), 
        \begin{gather*} 
            \label{first-display-5-8}
            \begin{aligned}
                \int_M f~\diff(\sigma_a^\#h(y))
                &= c \int_M \frac1{c} f ~\diff(\sigma_a^\# h(y)) \\
                &= c \int_{\sigma_a^{-1}(M)} \frac1{c} f \circ \sigma_a ~\diff h(y) \\
                &= c \int_{M} \frac1{c} f \circ \sigma_a ~\diff h(y)
            \end{aligned}
        \end{gather*}        
        The second equality uses the change-of-variables formula, which in its general form states that \(\int_U f(x)~\diff (g^\#\mu) = \int_{g^{-1}(U)} f \circ g~\diff \mu\)~\cite[Theorem 3.6.1]{Bogachev2007Measure}.
        The last equality holds since $\sigma_a^{-1}(M) = M$.
        Since \(0 < c_a/c \le 1\), \(\frac1{c}f \circ \sigma_a\) is a nonexpanding map into \(\mathbb R\).
        Taking \(d\) to be the Kantorovich-Rubenstein metric, this implies that for any \(\rho_1,\rho_2 \in \Prob(M)\),
        \begin{equation}
            \label{eq:Hutchinson ineq}
            \int_M  \frac1cf\circ \sigma_a~\diff\rho_1 - \int_M \frac1cf\circ \sigma_a~\diff\rho_2
            \le d(\rho_1, \rho_2)
        \end{equation}
        Putting these observations together, we obtain the following for any \(g, h \colon X \to \Prob(M)\). 
        If \(f \colon M \to \mathbb R\) is a nonexpanding map, then
        \begin{align*}
            &\int_M f~\diff F_\beta(g)(x) - \int_M f~\diff F_\beta(h)(x) \\
            &= \int_M \sum_{x\tr{r\mid a} y}r f~\diff (\sigma_a^\#g(y)) - \int_M \sum_{x\tr{r\mid a} y}r f~\diff (\sigma_a^\#h(y)) 
            \tag{definition of \(F\)}\\
            &= \sum_{x\tr{r\mid a} y} r \Big(\int_M  f~\diff (\sigma_a^\#g(y)) - \int_M f~\diff (\sigma_a^\#h(y))\Big) 
            \tag{linearity of \(\int\)}\\
            &= \sum_{x\tr{r\mid a} y} r c\Big(\int_M  \frac1cf\circ \sigma_a~\diff g(y) - \int_M \frac1cf\circ \sigma_a~\diff h(y)\Big) \tag{linearity of \(\int\)}\\
            &\le \sum_{x\tr{r\mid a} y} r c~d(g(y), h(y))  \tag{\ref{eq:Hutchinson ineq}} \\
            &\le c~\max\{d(g(y), h(y)) \mid y \in X \text{ and } x \tr{r \mid a} y\} \tag{\(\sum r = 1\)} \\
            %\tag*{\(\left(\sum_{x \tr{r\mid a} y} r = 1\right)\)}
            %\\
            &\le c~d^X(g, h) \tag{def.~of \(d^X\)}
        \end{align*}
        Since \(d^X(F_\beta(g), F_\beta(h)) = \max\{d(F_\beta(g)(x), F_\beta(h)(x)) \mid x \in X\}\) and \(d(F_\beta(g)(x), F_\beta(h)(x))\) is the supremum over nonexpanding \(f \colon M \to \mathbb R\) in the first expression in the calculation above, this shows that \(F_\beta\) is a contraction mapping in the product metric on \(\Prob(M)^X\). 

        To see item~\ref{item-f-fixed-point}, we begin by showing that for any \(x \in X\) and Borel set \(B\subseteq A^\omega\),
        \begin{equation}
        \label{eq:stream measure solution}
            \hat\beta(x)(B) = \sum_{x \tr{r \mid a} y} r\sigma_a^\#(\beta(y))(B)
        \end{equation}
        By the Identity and Extension Theorems for \(\sigma\)-finite premeasures~\cite{Kerstan2013Trace}, it suffices to show~\eqref{eq:stream measure solution} for each basic open set \(B = B_w\), with \(w \in A^*\). 
        We proceed by induction on the length of \(w\).
        We begin with the base case \(w = \epsilon\).
        Let us use \(a(-)\) again for the prefixing map \(a(a_1, a_2, \dots) = (a,a_1,a_2,\dots)\) on \(A^\omega\).
        Since \(a^{-1}(A^\omega) = A^\omega\) and \(\sum\{r \mid x \tr{r \mid a} y\} = 1\),
        \[
            \hat\beta(x)(B_\epsilon) 
            = \hat\beta(x)(A^\omega) 
            = 1
            = \sum_{x \tr{r \mid a} y} r\hat\beta(y)(a^{-1}(A^\omega))
            = \sum_{x \tr{r \mid a} y} ra^\#\hat\beta(y)(B_\epsilon)
        \]
        Assume our result for \(w = a_1\cdots a_n\), and 
        let \(a \in A\),  and consider  \(aw = a a_1\cdots a_n\). 
        Then
        \begin{align*}
            \hat\beta(x)(B_{aw})
            &= \sum\{r r_1\cdots r_n \mid x \tr{r \mid a} y \tr{r_1 \mid a_1} y_1 \tr{\phantom{r \mid a}} \cdots \tr{\phantom{r \mid a}} x_n\} \\
            &= \sum_{x \tr{r \mid a} y} r ~ \sum\{r_1\cdots r_n \mid y \tr{r_1 \mid a_1} y_1 \tr{\phantom{r \mid a}} \cdots \tr{\phantom{r \mid a}} x_n\} \\
            &= \sum_{x \tr{r \mid a} y} r \hat\beta(y)(B_w) \tag{induction hypothesis} \\
            &= \sum_{x \tr{r \mid a} y} r \hat\beta(y)(a^{-1}(B_{aw})) \tag{\(a^{-1}(B_{aw}) = B_w\)}\\
            &= \sum_{x \tr{r \mid a} y} r a^\#(\hat\beta(y))(B_{aw}) 
        \end{align*}
        This shows~\eqref{eq:stream measure solution}.
        Finally, we will use~\eqref{eq:stream measure solution} to prove~\ref{item-f-fixed-point}.
        First consider a measure \(\rho \in \Prob(A^\omega)\) and a Borel set \(B \subseteq A^\omega\). 
        Since \(\sigma_a \circ \sigma_\omega = \sigma_\omega \circ a(-)\) (this is~\eqref{lem:a dot commutes}), we have 
        \begin{align*}
            \sigma_a^\#\sigma_\omega^\#\rho(B)
            &= \rho(\sigma_\omega^{-1} \circ \sigma_a^{-1}(B)) \\
            &= \rho((\sigma_a \circ \sigma_\omega)^{-1}(B)) \\
            &= \rho((\sigma_\omega \circ a(-))^{-1}(B)) \\
            &= \rho(a(-)^{-1} \circ \sigma_\omega^{-1}(B)) \\
            &= a^\#\rho(\sigma_\omega^{-1}(B)) 
        \end{align*}
        For any \(x \in X\) and Borel set \(B \subseteq M\), if we take \(\rho = \hat\beta(y)\) below, then from the above calculation we obtain
        \begin{align*}
            F_\beta(\hat\beta_\sigma)(x)(B) 
            &= \sum_{x \tr{r \mid a} y} r \sigma_a^\#\sigma_\omega^\#\hat\beta(y)(B) 
            \tag{def.~of \(F_\beta, \hat\beta_\sigma\)} \\
            &= \sum_{x \tr{r \mid a} y} r a^\#\hat\beta(y)(\sigma_\omega^{-1}(B))  \\
            &= \hat\beta(x)(\sigma_\omega^{-1}(B)) 
            \tag{\ref{eq:stream measure solution}}\\
            &= \sigma_\omega^\#\hat\beta(x)(B) \\
            &= \hat\beta_\sigma(x)(B) \tag{def.~of \(\hat\beta_\sigma\)}
        \end{align*}
        At this point, we return to~\ref{item-f-contraction} and~\\ref{item-f-fixed-point} from near the start of this proof.  
        We have verified that \(\hat\beta_\sigma\) is a fixed point of $F_{\beta}$, and we also know that $F_{\beta}$ is contractive and consequently has at most one fixed point.  
        So, \(\hat\beta_\sigma\) is the unique fixed point. 
        It is therefore the unique coalgebra-to-algebra homomorphism.

        For the last statement, let us first verify that the support of \(\hat\beta(x)\) is precisely \(\str_{\bar\beta}(x)\). 
        To see that \(\str_{\bar\beta}(x) \subseteq \supp(\hat\beta(x))\), let \((a_i)_{i \in \N} \in \str_{\bar\beta}(x)\).
        Then there is a path \(x \tr{r_1\mid a_1} x_1 \tr{r_2 \mid a_2} \cdots\) in \((X, \beta)\).
        It suffices to see that for any \(w \in A^*\) such that \((a_i)_{i \in \N} \in B_w\), \(\hat\beta(x)(B_w) > 0\).
        But, if \((a_i)_{i \in \N} \in B_w\), then \(a_1\cdots a_n = w\), so \(\hat\beta(x)(B_w) = \prod_{i=1}^n r_i > 0\).
        Hence, \((a_i)_{i \in \N} \in \supp(\hat\beta(x))\).
        This shows \(\str_{\bar\beta}(x) \subseteq \supp(\hat\beta(x))\).
        
        For the reverse inclusion, given \((a_i)_{i \in \N} \in \supp(\hat\beta(x))\), we know that every \(B_w\) containing \((a_i)_{i \in \N}\) has positive trace measure. 
        Remember that \((a_i)_{i \in \N} \in B_w\) if and only if \(a_1 \cdots a_n = w\).
        But \(\hat\beta(x)(B_w) > 0\) if and only if there is a path \(x \tr{r_1\mid a_1} x_1 \tr{r_2 \mid a_2} \cdots\) in \((X, \beta)\) with \(\hat\beta(x)(B_w) = \prod_{i=1}^n r_i\).
        Since \(w\) was arbitrary, we have just seen that every finite initial segment of \((a_i)_{i \in \N}\) corresponds to a path in \((X, \beta)\).
        By \autoref{lem:trace stream equivalence}, \((a_i)_{i \in \N} \in \str_{\bar\beta}(x)\).
        Thus, \(\str_{\bar\beta}(x) = \supp(\hat\beta(x))\).

        Now let us use the previous paragraph to prove the last statement of the theorem.
        We use the following well-known fact from measure theory, that for any Borel measure \(\rho\) and continuous map \(f\) between metric spaces, \(\supp(f^\#\rho) = \overline{f(\supp(\rho))}\), where \(\overline{(-)}\) denotes topological closure.
        In our case, we have \(f = \sigma_\omega\) and \(\rho = \hat\beta\), which gives
        \begin{align*}
            \supp(\hat\beta_\sigma(x)) 
            &= \supp(\sigma_\omega^\#\hat\beta(x)) \\
            &= \overline{\sigma_\omega(\supp(\hat\beta)(x))} \tag{fact above}\\
            &= \overline{\sigma_\omega(\str_{\bar\beta}(x))} \tag{\(\str_{\bar\beta}(x) = \supp(\hat\beta(x))\)} \\
            &= \overline{\sem{x}_{\bar\beta,\sigma}} \tag{def.~of \(\sem{-}_{\bar\beta,\sigma}\)} \\
            &= \sem{x}_{\bar\beta,\sigma} \tag{\(\sem{x}_{\bar\beta,\sigma}\) is closed by \autoref{thm:fractal semantics coincide}}
        \end{align*}
    \end{proof}

    \paragraph*{An aside: chaos games for regular subfractals and fractal measures}
    
    Every probability distribution \(\theta\) on our set \(A\) of actions determines a probability measure $\hat{\theta}$ on $A^{\omega}$.
    Now suppose we are also given a contraction operator interpretation \(\sigma \colon A \to \Con(M)\) and a fixed $p\in M$.
    Each infinite sequence $(a_1, a_2, \ldots)$ gives us an infinite sequence of points $p_n\in M$ defined recursively by setting $p_0 = p$, and $p_{n+1} = \sigma_{a_n}(p_n)$.  
    We are interested in the case that the starting point $p$ belongs to the self-similar set $\mathbf S_\sigma$ determined by $\sigma$ (see~\autoref{def-S-sigma}).   
    In this case, each point $p_n$ also belongs to $\mathbf S_\sigma$.
    So we obtain a map $\hat{p}\colon A^\omega \to \mathbf S_\sigma^{\omega}$.
    This map $\hat p$ turns out to be measurable.  

    Before we mention some specific events which are of interest, let us recast the situation under discussion in more dynamic terms.  
    Again given the initial point $p$, we imagine generating the points $p_n$ by a random process.
    The point $p_0 = p$ is given, and from $p_{n}$ we determine $p_{n+1}$ by randomly choosing $a$ from $A$ with probability $\theta(a)$ and then setting $p_{n+1} = \sigma_a (p_n)$.  
    All of our choices are independent, and the past has no effect on the present.
    This is called the \emph{random iteration algorithm} (or \emph{chaos game}), and it is a standard topic in the fractals literature (see e.g.~\cite[Chapter 9]{barnsley}).

    In essence, the chaos game selects a stream \((a_1,a_2,\dots) \in A^\omega\) by repeatedly sampling \(\theta\) and considers its image \(\hat p(a_1, a_2, \dots)\).
    As we have seen, the probability that the stream which it selects lies in a given Borel set \(B \subseteq A^\omega\) is \(\hat\theta(B)\).
    For the purposes of the next few paragraphs, we will call a Borel subset of \(A^\omega\) an \emph{event} and refer to the \emph{probability of an event} \(B\) when we mean \(\hat\theta(B)\).   

    Among the notable results about the chaos game are the following. 
    Firstly, if \(\theta(a) > 0\) for all \(a \in A\), then for any point \(p \in \mathbf S_\sigma\), the event in which its corresponding run of the chaos game \(\{p_0, p_1, p_2, \dots\}\) is dense in \(\mathbf S_\sigma\) has probability \(1\)~\cite{Falconer1986Geometry}.
    Additionally, for a given Borel set \(U \subseteq M\) and starting point \(p \in M\) (not necessarily in \(\mathbf S_\sigma\)), the event in which the following equation holds also has probability \(1\) with respect to its corresponding run of the chaos game~\cite[pg.~7]{DiaconisS1984} (also~\cite{Elton87}).
    \begin{equation}
        \label{eq:residence measure again}
        \hat\theta_\sigma(U) = \lim_{n \to \infty} \frac1n |U \cap \{p_1, \cdots, p_n\}|
    \end{equation}
    In words, the value of $\hat\theta_\sigma(U)$ is the limit of the proportion of the first $n$ points in the infinite sequence \((p_i)_{i \in \N}\) that happen to belong to $U$.
    % We want to emphasize that this is just a recasting of the map $\hat{p}\colon A^\omega \to \mathbf S_\sigma^{\omega}$ which we have already mentioned.

    \begin{figure}[!ht]
        \centering
        \includegraphics[scale=0.2]{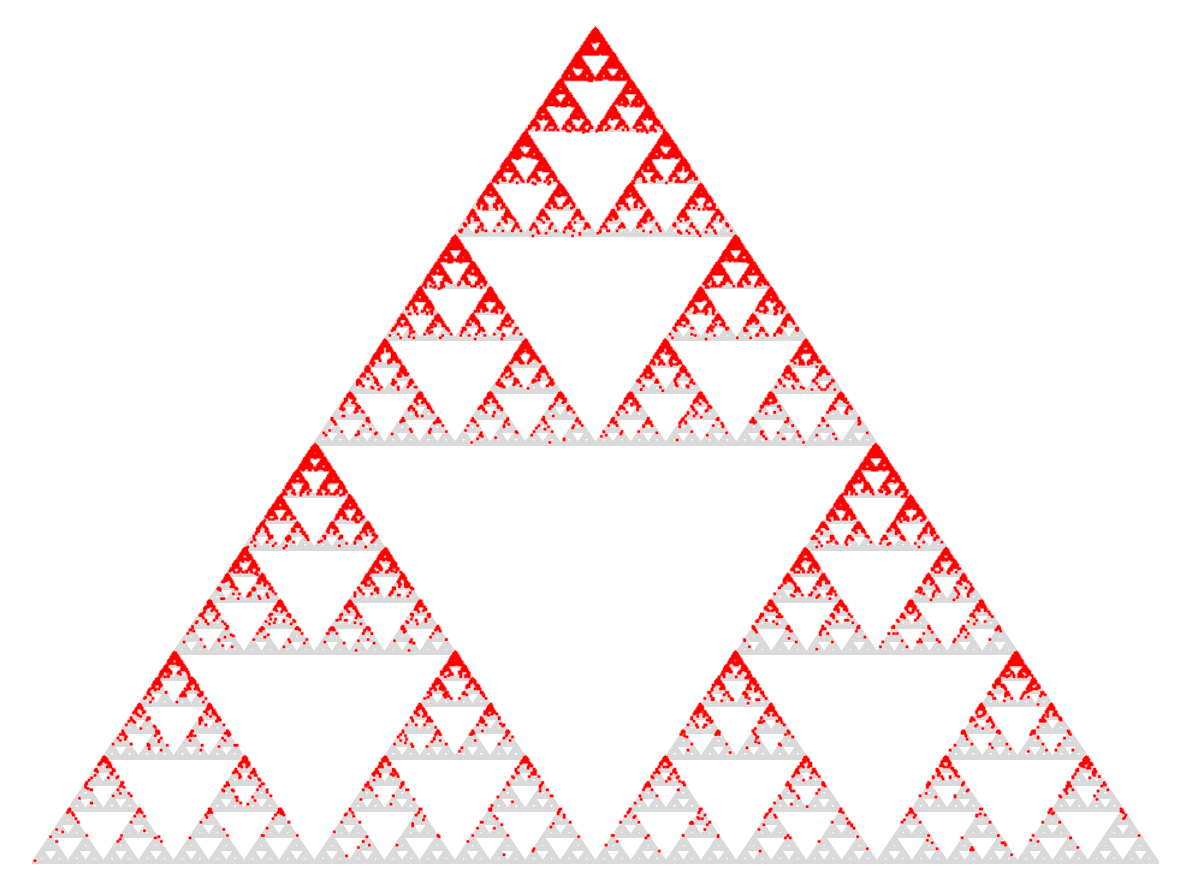}
        \caption{
            \label{fig:weighted Sierpinski gasket}
            One run of the chaos game on the Sierpinski gasket from \autoref{fig:Sierpinski gasket} with probability distribution given by \(\theta(a) = 0.8\), \(\theta(b) = 0.1\), and \(\theta(c) = 0.1\).
            Above, the points generated by the chaos game are drawn in \textcolor{red}{red}.
            In this run, of the \(200,\!000\) points generated, \(160,\!081\) points appeared in the top third of the gasket, i.e., \(\approx 0.8\) of the total.
        }
    \end{figure}

    \begin{exa}
        Let \(A\) and \(\sigma\) be as in \autoref{fig:Sierpinski gasket}. 
        Consider the distribution \(\theta \colon A \to [0,1]\) given by \(\theta(a) = 0.8\), \(\theta(b) = 0.1\), and \(\theta(c) = 0.1\), and let \(\hat\theta_\sigma\) be the invariant measure specified by \(\theta\).
        Running the chaos game with starting point \(p = (0,0)\) we obtained the graph in \autoref{fig:weighted Sierpinski gasket}.
        Since \(80\%\) of the distribution is on \(a\), approximately \(80\%\) of the points generated by the chaos game should be in the top third of the Sierpi\'nski gasket.
    \end{exa}

    The right-hand side of~\eqref{eq:residence measure again} was obtained by generating a sequence of points using a probability distribution on \(A\) and an initial point \(p\) of the underlying complete metric space. 
    For regular submeasures, which are generated by a productive labelled Markov chain instead of a probability distribution on \(A\), experiments indicate that a similar game can be played. 
    
    Let \(\sigma \colon A \to \Con(M)\) be a contraction operator interpretation and let \((X, \beta)\) be a productive labelled Markov chain. 
    Fix \(x \in X\) and let \(K = \sem{x}_{\beta,\sigma}\).
    We define the \emph{multistate chaos game} for \(x\), \(\sigma\), and \(\beta\) with initial point \(p \in \sem{x}_{\beta,\sigma}\) to be the following stochastic procedure: generate a path \(x = x_1 \tr{r_1\mid a_1} x_2 \tr{r_2\mid a_2} \cdots\) by sampling the distribution \(\beta(x_i)\) at each step.
    Now set \(q_1 = p\) and \(q_{i+1} = \sigma_{a_i}(q_i)\) for each \(i \in \N\).
    This defines a sequence of points \((q_i)_{i \in \N}\) in \(M\), usually not all of which are in \(\sem{x}_{\beta,\sigma}\). 
    Notice that \(q_i \in \sem{x_i}_\beta\)  for each \(i \in \N\).
    The outcome of the multistate chaos game is the subsequence \((p_{i_k})_{k \in \N}\) of \((q_i)_{i\in \N}\), i.e., with \(i_1 < i_2 < i_3 < \cdots\), consisting of all the points \(p_{i_k}\) such that at index \(i_k\) the state \(x_{i_k}\) is \(x_{i_k} = x\).
    
    Experimental results so far indicate that the event in which the set \(\{p_{i_k}\}_{k \in \mathbb N}\) is dense in the regular subfractal \(\sem{x}_{\beta,\sigma}\) has probability \(1\).
    Additionally, we suspect that an equation analogous to that of~\eqref{eq:residence measure again} also holds for multistate chaos games: for a Borel set \(U \subseteq M\),
    \begin{equation}
        \label{eq:residence measure multistate}
        \hat\beta_\sigma(x)(U) = \lim_{n \to \infty} \frac1n |U \cap \{p_{i_1}, p_{i_2}, \cdots\}|
    \end{equation}
    where \(\beta\) and \((p_{i_k})_{k \in \N}\) were defined above.
    To the authors' knowledge,~\eqref{eq:residence measure multistate} has not yet been verified or disproven in the literature.
    We will pursue this issue in further work.

\begin{figure}[ht]
        \begin{tabular}{c l}
            \(\begin{gathered}
                \begin{tikzpicture}
                    \node (label) at (-1, 1) {\((X, \beta)\)};
                    \node[state] (0) at(0,0) {\(x\)};
                    \node[state] (1) at (3,0) {\(y\)};
                    \draw[loop] (0) edge[loop left] node[left] {\(0.2\mid a\)} (0);
                    \draw[loop] (1) edge[loop right] node[right] {\(0.8 \mid b\)} (1);
                    \draw (0) edge[bend left] node[above] {\(
                        \begin{gathered}0.5 \mid b\\0.2 \mid c\end{gathered}
                    \)} (1);
                    \draw (1) edge[bend left] node[below] {\(
                        \begin{gathered}0.1 \mid b\\0.1 \mid c\end{gathered}
                    \)} (0);
                \end{tikzpicture}
            \end{gathered}\)
            & 
            \raisebox{-2cm}{\includegraphics[scale=0.2]{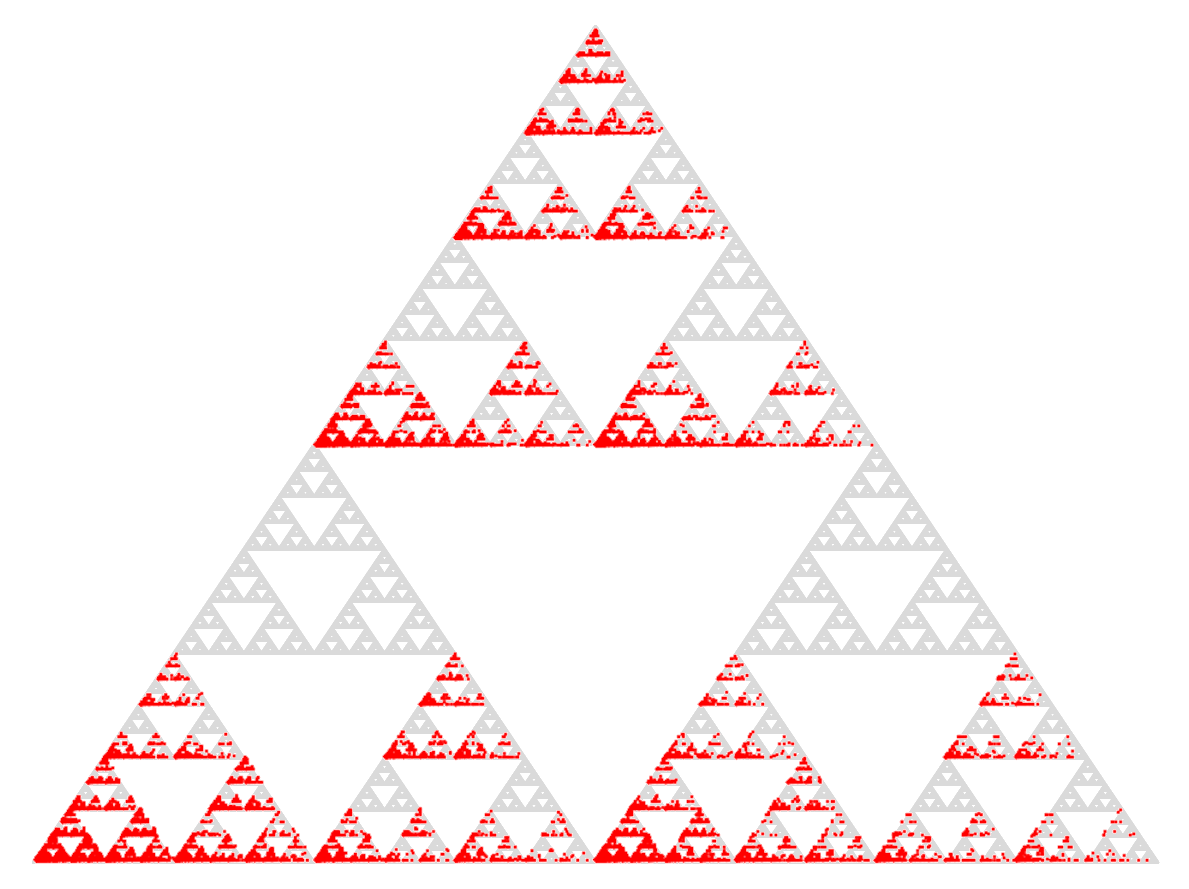}}
        \end{tabular}
        \caption{
            \label{fig:asymmetric chaos sierpinski}
            The sequence of points generated by \(200,\!000\) iterations of the multistate chaos game for \((X, \beta)\), starting from the state \(x\) and the point \(p\).
            Above, the points generated by the chaos game are drawn in \textcolor{red}{red}.
            In this run, \(100,\!120\) out of the \(200,\!000\) points generated appear in the bottom-left third of the regular subfractal, i.e., \(\approx 0.5\) of the total.
        }
    \end{figure}

    \begin{exa}
        \label{eg:asymm chaos sierpinski}
        Consider the labelled Markov chain \((X, \beta)\) displayed in \autoref{fig:asymmetric chaos sierpinski}.
        One run of the multistate chaos game starting at state \(x\) and the point \(p = (0,0)\) generated the sequence of points in the plot on the right in the same figure. 
        The plot suggests that the bottom-left third, i.e., \(\sigma_\omega(B_b)\), carries more weight than the rest of the plot: a raw count of the points in the bottom-left third of the regular subfractal reveals that \(100,\!120\) out of the \(200,\!000\) appear in that third. 
        Thus, the proportion of the points generated that land in \(\sigma_\omega(B_b)\) is very close to the actual value, \(\hat\beta_\sigma(\sigma_\omega(B_b)) = \hat\beta(B_b) = 0.5\). 
        Further experiments have given similar results.
    \end{exa}

    \section{Probabilistic Process Terms}
    \label{sec:syntax for prob}
    Next, we introduce a syntax for specifying LMCs.
    Our specification language is essentially the \emph{productive} fragment of Stark and Smolka's process calculus~\cite{Stark2000Probabilistic}, meaning that the \(\mu\)-expressions do not involve deadlock and all variables are guarded (see Definition~\ref{def:term}).

    \begin{defi}
        \label{def:prob terms}
        The set of \emph{probabilistic \(\mu\)-expressions} is given by the grammar 
        \[
            v \mid ae \mid e_1 \oplus_r e_2 \mid \mu v~e 
        \]
        Here  \(r \in (0,1)\), and otherwise we make the same stipulations as in~\autoref{def:term}.
        The set of \emph{probabilistic process terms} \(\PTerm\) consists of the closed and guarded probabilistic \(\mu\)-expressions.
    \end{defi}
    
    Instead of \emph{languages} of streams, the analog of trace semantics appropriate for probabilistic process terms is a measure-theoretic semantics consisting of trace measures introduced in the previous section (Definition \ref{trmdef}).%\autoref{eq:trace measure}).
    
    We turn next to the semantics.  
    For use there, and in other definitions, we need a preliminary definition.
 
    \begin{defi}  \label{def:un}
        The \emph{unravelling number of $e$}, denoted $\#_{un}(e)$, is defined by the following recursion on guarded (not necessarily closed) terms: for any \(a \in A\), \(e,e_1,e_2\), 
        \[
            \#_{un}(ae) = 0
            \qquad 
            \#_{un}(e_1 \oplus_r e_2) = \max_{i\in\{1,2\}}\{\#_{un}(e_i)\}
            \qquad 
            \#_{un}(\mu v~e) = 1 + \#_{un}(e)
        \]
        Suppose that $e$ were displayed as a syntax tree.
        Then $\#_{un}(e)$ would be the maximum number of $\mu x$ nodes,  with the maximum  taken along all paths from the root $e$ to some internal node labelled with an action prefix \(a\).
    \end{defi}

    For example, \(
        \#_{un}(\mu v~(av \oplus_{\frac12} \mu w~(bw)) )
        = 1 + \max\{\#_{un}(av), \#_{un}(\mu w~(bw)) \}
        = 2
        \).
    
    \begin{prop}
        \label{prop-un}
        For all guarded terms $e$,
        \begin{enumerate}
            \item whenever $e[f/v]$ is defined, $\#_{un}(e) = \#_{un}(e[f/v])$.
            \item Assume that $\mu v~e$ is guarded and closed.  
            Then $\#_{un}(\mu v~e) = 1 + \#_{un}(e[\mu v~e/v])$.
        \end{enumerate}
    \end{prop}
    
    \begin{proof} 
        The first part is proven by induction on $e$.  
        The base case for variables is trivial, since variables are not guarded.
        The only induction step worth mentioning is that of a guarded term $\mu w~e$, with $w$ a variable different from $v$, and such that $(\mu w~e)[f/v]$ is defined. 
        In this case, $e$ also is guarded, and the induction hypothesis applies to it.
        Also $w$ has no free occurrences in $f$, and therefore $e[f/v]$ is well defined.
        We have
        \begin{align*}
            \#_{un}(\mu w~e) 
            &= 1 + \#_{un}(e) \\
            &= 1 +  \#_{un}(e[f/v]) \tag{induction hypothesis}\\
            &= \#_{un}(\mu w~(e[f/v]))\\
            &= \#_{un}((\mu w~e)[f/v])
        \end{align*}
        In the second part, assume that $\mu v~e$ is guarded and closed.  
        Then $e$ is guarded (but not necessarily closed).
        We use the first part, taking $f$ to be $\mu v~e$.
        Now $e[\mu v~e/v]$ is well defined, since $v$ is not free in $\mu v~e$.
        And $\#_{un}(\mu v~e) =1 + \#_{un}(e) = 1 +  \#_{un}(e[\mu v~e/v])$.
    \end{proof}
    
        The important point in~\autoref{prop-un} is the last one, since it enables us to define functions $t$ on $\PTerm$ by recursion on $\prec$ in such a way that $t(\mu v~e)$ only depends on \(t(e[\mu v~e/v])\).

    \newcommand{\unravht}{\mathsf{rht}}

    \begin{defi}
        \label{unraveling-number}
        For each term $e$, we write $\#_{\oplus}(e)$ for the number of $\oplus_r$-operators in $e$.
        We define a function $\unravht:\PTerm\to \omega \times \omega$ by%
        \footnote{The notation \(\unravht\) stands for ``r-height''.}
        \[
            \unravht(e) = (\#_{un}(e), \#_{\oplus}(e))
        \]
        Let $\prec$ be the strict lexicographic order on  $\omega\times\omega$, and recall that this is a strict well-order. 
    \end{defi}

Here is the point of these definitions.  For $i = 1, 2$, $\unravht( e_i) \prec \unravht(e_1 \oplus_r e_2)$. 
And      $\unravht(e[\mu v~e/v]) \prec \unravht(\mu v~e)$.

%    \sout{
  %  Note that in the second and third clauses of the definition, of each term on the right is strictly below the 
  %  value of the term on the left in the well-order $\prec$.} 

    \begin{figure}[!t]
        \begin{align*}
            \delta(ae)(b,f) &= \begin{cases}
                1 & f = e \text{ and }b = a \\
                0 & \text{otherwise}
            \end{cases}
            \\ 
            \delta(e_1 \oplus_r e_2)(b, f)
            &= r\delta(e_1)(b, f) + (1-r)\delta(e_2)(b, f)
            \\
            \delta(\mu v~e)(b, f) &= \delta(e[\mu v~e/v])(b, f)
        \end{align*}
        \caption{\label{fig:syntactic LMC} The LMC structure \((\PTerm, \delta)\). 
        Above, \(a,b \in A\), $r\in (0,1)$, and \(e,e_i,f \in \PTerm\).}
    \end{figure}

    \begin{defi} 
        \label{def:trm}
        We define the LMC \((\PTerm, \delta)\) in \autoref{fig:syntactic LMC} and call it the \emph{syntactic LMC}.
        The \emph{trace measure semantics} \(\meas(e)\) of a probabilistic process term \(e\) is defined to be \(\meas(e) = \hat\delta(e)\) (see \autoref{trmdef}). 
        
        Two probabilistic process terms \(e\) and \(f\) are \emph{trace measure equivalent} if they have  the same trace measure semantics.
        Given \(\sigma\colon A \to \Con(M)\), the \emph{subfractal measure semantics of \(e \in \PTerm\) corresponding to \(\sigma\)} is \(\hat\delta_\sigma(e)\) (see \autoref{def:regular subfractal measure}).   
    \end{defi}
    
    The next result gives a recursive method for computing the trace measure semantics of a process term.
    This recursive method will be useful in \cref{sec:axioms for prob}.

    \begin{lem}
        \label{lem:measure is recog}
        For any \(w \in A^*\), \(a \in A\), \(e,e_i \in \PTerm\), and \(r \in (0,1)\), \(\meas(e)(A^\omega) = 1\) and
        \begin{align*}
            \meas(ae)(B_w) 
            &= \begin{cases}
                \meas(e)(B_{u}) & \text{\(w = au\)} \\
                0 & \text{otherwise}
            \end{cases}
            \\
            \meas(e_1 \oplus_r e_2)(B_w)
            &= r\meas(e_1)(B_w) + (1-r)\meas(e_2)(B_w)
            \\
            \meas(\mu v~e)(B_w) &= \meas(e[\mu v~e/v])(B_w)
        \end{align*}
    \end{lem}

	\begin{proof}
		For each probabilistic process term \(e\), define \(j(e) \colon A^* \to [0,1]\) inductively by setting \(j(e)(\epsilon) = 1\) for all \(e \in \PTerm\), and for \(a \in A\), \(r \in [0,1]\), \(e,e_1,e_2 \in \PTerm\),
		\begin{gather*}
            j(ae)(w) 
            = \begin{cases}
                j(e)(u) & \text{\(w = au\)} \\
                0 & \text{otherwise}
            \end{cases}
            \qquad
			\begin{aligned}
	            j(e_1 \oplus_r e_2)(w)
	            &= r j(e_1)(w) + (1-r)j(e_2)(w)
	            \\
	            j(\mu v~e)(w) &= j(e[\mu v~e/v])(w)
	        \end{aligned}
        \end{gather*}
        We are going to show that \(j(e)(w) = \meas(e)(B_w)\) for every \(e \in \PTerm\) by induction on $w$, and then by a subinduction on $\unravht(e)$ (see \autoref{unraveling-number}).
		Since $\meas(e)$ already satisfies $\meas(e)(B_w) = \sum_{a\in A}\meas(e)(B_{wa})$ and $\meas(e)(B_\epsilon) = \meas(e)(A^\omega) = 1$, we get $j(e)(w) = \sum_{a\in A}j(e)(wa)$ and $j(e)(\epsilon) =1$, allowing us to apply the uniqueness part of \autoref{prop:unique measure extension}.
	
		Here is the base case of induction on $w$:
		For any \(e \in \PTerm\), $j(e)(\epsilon) = 1$ and $\meas(e)(B_\epsilon) = \meas(e)(A^\omega) =1$.
		
		In the induction step, we consider the word \(w = bu\).  
        We assume that \(j(e)(u) = \meas(e)(B_u)\) for every \(e \in \PTerm\) and we show the same thing for $bu$.   
        We argue by induction on $\unravht(e)$.
		\begin{itemize}
			\item Our first case is for $\unravht(e) = (0,0)$, i.e., for a process term \(ae\).  
				If \(b = a\), we recall that  \(w = bu\) and then calculate:
				\begin{align*}
					j(ae)(bu) 	
                    &= j(e)(u)  \\
                    &= \meas(e)(u) \tag{top level induction, on \(w\)} \\
                    &= 1\meas(e)(u) \\
                    &= \delta(ae)(a, e)\meas(e)(u) \\
                    &= \meas(ae)(B_{au}) \tag{\(\meas = \hat \delta\),~\eqref{eq:stream measure solution}}
				\end{align*}
			    If $b\neq a$, then $j(ae)(bu)=0$ and $\meas(ae)(B_{bu}) = 0$, since no string starting with $a$ is in $B_{bu}$. 			
			    Note that we have shown the claim for all terms of the form $ae$.
        \end{itemize}
        \noindent %FIXED indent
        Now let $f\in \PTerm$ where $(0,0) \prec \unravht(f)$ and assume that for all $e$ with $\unravht(e)\prec \unravht(f)$, that $j(e)(w) = \meas (e)(B_w)$.  
        \begin{itemize}
            \item We next consider a term $e_1 \oplus_r e_2$.
            Note that $\unravht(e_i)$ is strictly less than $\unravht(e_1\oplus_r e_2)$ for $i=1,2$, so we have assumed that \(j(e_i)(w) = \meas(e_i)(B_w)\) for \(i \in \{1,2\}\).
            Then
            \begin{align*}
                &j(e_1 \oplus_r e_2)(w) \\
                &= rj(e_1)(w) + (1-r)j(e_2)(w) \\
                &= r \meas(e_1)(B_w) + (1-r) \meas(e_2)(B_w) 
                \tag{ind.~hyp.,~\(\unravht(e_1), \unravht(e_2) \prec \unravht(e_1\oplus_r e_2)\)}\\
                &= r \sum_{g\in \PTerm} \delta(e_1)(b, g) \meas(g)(B_u) \\
                &\hspace{4em} + (1-r) \sum_{g\in \PTerm} \delta(e_2)(b, g) \meas(g)(B_u) 
                \tag*{(\(w = bu\)),~\eqref{eq:stream measure solution}} \\
                &= \sum_{g \in \PTerm} (r \delta(e_1)(b, g) + (1-r) \delta(e_2)(b, g))\meas(g)(B_u) \\
                &= \sum_{g \in \PTerm} \delta(e_1 \oplus_r e_2)(b, g) \meas(g)(B_u) 
                \tag{def.~\(\delta\)}\\					
                &= \meas(e_1 \oplus_r e_2)(B_w) 
            \end{align*}
			
            \item Finally we consider a term \(\mu v~e\).
			Again, since $\unravht(e[\mu v~e/v]) \prec \unravht(\mu v~e)$, we assume that we have $j(e[\mu v~e/v])(w) = \meas(e[\mu v~e /v])(B_w)$.  
            So we have
            \begin{align*}
                j(\mu v~e)(w) &= j(e[\mu v~e/v])(w) \\
                &= \meas(e[\mu v~e/v])(B_w) \tag{ind.~hyp.~\( \unravht(e[\mu v~e/v])  \prec \unravht(\mu v~e)\)} \\
                &= \sum_{g \in \PTerm} \delta(e[\mu v~e/v])(b, g) ~\meas(g)(B_u) \tag{\(w = bu\),~\eqref{eq:stream measure solution}}\\
                &= \sum_{g \in \PTerm} \delta(\mu v~e)(b, g)  \meas(g)(B_u) \tag{def.~of \(\delta\)}\\
                &= \meas(\mu v~e)(B_{w}) \tag*{\eqref{eq:stream measure solution}}
            \end{align*}
			%\item The multiple summand case is exactly as in the base case of the induction on words. \qedhere
		\end{itemize}
		\noindent %FIXED indent
        This completes the inductive step on $w$, hence the overall proof.	
    \end{proof}

    Two process terms have the same trace measure semantics iff they have the same semantics in any contraction operator interpretation.
    This is an easy consequence of our definitions.   
    We have a similar result for probabilistic process terms.

    \begin{thm}
        \label{thm:trace meas is fractal meas}
        Let \(e,f \in \PTerm\).
        Then \(\meas(e) = \meas(f)\) if and only if for any contraction operator interpretation \(\sigma\colon A \to \Con(M)\), \(\hat\delta_\sigma(e) = \hat\delta_\sigma(f)\).
    \end{thm}

    \begin{proof}
        Let \(e,f \in \PTerm\).
        Suppose \(\meas(e) = \meas(f)\), and let \(\sigma \colon A \to \Con(M)\) be any contraction operator interpretation. 
        Then\[
            \hat\delta_\sigma(e) 
            = \sigma_\omega^\#\hat\delta(e)
            = \sigma_\omega^\#\hat\delta(f)
            = \hat\delta_\sigma(f)
        \]
        where the first and third equalities are by definition.
        As for the second equality, we assumed that  \(\meas(e) = \meas(f)\),
        and so by \autoref{def:trm}, \(\hat\delta(e) = \hat\delta(f)\).  
        Hence, \(\sigma_\omega^\#\hat\delta(e) = \sigma_\omega^\#\hat\delta(f)\).

        Conversely, suppose that \(\hat\delta_\sigma(e) = \hat\delta_\sigma(f)\) for every contraction operator interpretation \(\sigma\).
        Then this specifically holds for the Cantor set \((A^\omega, d_{\frac12})\) with the \(1/2\)-metric.
        We also have the prefixing operator interpretation \(\sigma\colon A \to \Con(A^\omega)\) given by \(\sigma_a = a(-)\).
        Indeed, each \(a(-)\) has a contraction coefficient of \(1/2\). 
        It suffices to show that \(\meas = \hat\delta_\sigma\), since this would mean that \(\meas(e) = \hat\delta_\sigma(e) = \hat\delta_\sigma(f) = \meas(f)\).
        As we saw in \autoref{eg:stream system}, \(\sigma_\omega\) in this setting is the identity map.  
        So are its inverse and pushforward, therefore, and so
        \begin{align*}
            \hat\delta_\sigma(e)(B_w)
            = \sigma_\omega^\#\hat\delta(e)(B_w) 
            = \hat\delta(e)(B_w)
            = \meas(e)(B_w)
        \end{align*}
        for any \(e \in \PTerm\) and \(w \in A^*\).
        It follows that \(\meas = \hat\delta_\sigma\), as requested.
    \end{proof}

    % Axiomatization
    \section{Axioms for Subfractal Measure Equivalence}
    \label{sec:axioms for prob}

    In this section, we propose an inference system for deriving equations between trace measure equivalent (and therefore, subfractal measure equivalent) probabilistic process terms. 
    Our inference system can be found in \autoref{fig:probabilistic axioms}.
    We show that it is \emph{sound} with respect to trace measure semantics in this section, in the sense that if \(e = f\) can be proven from the axioms, then \(\meas(e) = \meas(f)\). 
    We also establish \emph{completeness} with respect to trace measure semantics, the converse of soundness.
    The completeness theorem relies on a recent result due to C\^irstea, Sokolova, Silva, and the authors~\cite{cirstea_et_al2025}. 
 
    Similar to how Rabinovich's axiomatization of trace semantics for process terms~\cite{Rabinovich1993Traces} adds a left-distributivity axiom to Milner's axioms for bisimilarity~\cite{Milner1984Complete}, the inference system in \autoref{fig:probabilistic axioms} adds a left-distributivity axiom (\textsf{D}) to Stark and Smolka's inference system~\cite{Stark2000Probabilistic} axiomatizing so-called \emph{probabilistic bisimilarity}.%
    \footnote{The same approach to axiomatizing finite trace semantics for probabilistic systems was taken by Silva and Sokolova in~\cite{Silva2011Probabilistic}.}
    Due to the intimate connection between our axioms and bisimilarity of probabilistic systems, we discuss probabilistic bisimilarity next.
    
    \paragraph*{Probabilistic Bisimilarity and Trace Measure Equivalence}
    Fix an LMC \((X, \beta)\).

    \begin{defi}\label{def:prob_bisimilarity}
        A \emph{bisimulation equivalence} on an LMC \((X, \beta)\) is an equivalence relation \(R \subseteq X \times X\) such that for any \((x, y) \in R\), \(a \in A\), and equivalence class \(S \in X/R\),
        \[
            \sum_{z \in S} \beta(x)(a, z) = \sum_{z \in S} \beta(y)(a, z)
        \]
        For states \(x,y \in X\), if \((x, y) \in R\) for some bisimulation equivalence \(R\) on \((X, \beta)\), then we write \(x \bisim y\) and say \(x\) and \(y\) are \emph{bisimilar}. 
        The relation \({\bisim} \subseteq X \times X\) is called \emph{probabilistic bisimilarity}. 
    \end{defi}
    The following lemma gives a well-known characterization of probabilistic bisimilarity that will play a role in subsequent proofs.
    
    \begin{lem}
        \label{lem:prob bisim quotient}
        Let \((X, \beta)\) be an LMC.
        Given \(x,y \in X\), \(x\) and \(y\) are bisimilar if and only if there is an LMC \( (Y, \beta_Y)\) and a homomorphism \(h \colon (X, \beta) \to (Y, \beta_Y)\) such that \(h(x) = h(y)\).%
        \footnote{The latter condition is often called \emph{behavioural equivalence} in the  literature. See, for eg.,~\cite{RotBBPRS17,Sokolova11}.}
    \end{lem}
        
    \begin{proof}
        This is standard in the coalgebra literature. 
        See, for example,~\cite[Corollary 2.3]{Sokolova11}.
    \end{proof}
    
    While both bisimilarity and trace measure equivalence can be seen as observational equivalences for LMCs, probabilistic bisimilarity is a particularly granular notion of equivalence.
    For example, it is not true that for any LMC \((X, \beta)\), \(\hat\beta \colon X \to \Prob(A^\omega)\) is a homomorphism of LMCs:
    if \(\hat\beta\) were a homomorphism of LMCs, then states which are trace measure equivalent would always be bisimilar. 
    As we can see from \autoref{fig:not bisim}, however, there are examples of trace measure equivalent states that are not bisimilar.

    Consequently, the set  \(\Prob(A^\omega)\) of Borel probability measures on \(A^\omega\) therefore does not carry a coalgebraic structure that is universal among LMCs. 
    On the other hand, it does carry an \emph{algebra} structure that is universal among the class of LMCs relevant to us in this paper.

    \begin{figure}[ht]
        \begin{gather*}
            \begin{tikzpicture}
                \node[state] (0) at (0,0) {\(x\)};
                \node[state] (00) at (-1, -1) {\(z\)};
                \node[state] (01) at (1, -1) {};
                \node[state] (000) at (-1, -2.3) {};
                \node[state] (011) at (1, -2.3) {};
                \draw (0) edge[bend right] node[above left] {\(\frac12\mid a\)} (00);
                \draw (0) edge[bend left] node[above right] {\(\frac12\mid a\)} (01);
                \draw (00) edge node[left] {\(1\mid b\)} (000);
                \draw (01) edge node[left] {\(1\mid c\)} (011);
                \draw[loop] (000) edge[loop below] node[below] {\(1\mid d\)} (000);
                \draw[loop] (011) edge[loop below] node[below] {\(1\mid d\)} (011);
            \end{tikzpicture}
            \hspace*{5em}
            \begin{tikzpicture}
                \node[state] (0) at (0,0) {\(y\)};
                \node[state] (00) at (0, -1.3) {\(z'\)};
                \node[state] (000) at (-1, -2.3) {};
                \node[state] (001) at (1, -2.3) {};
                \draw (00) edge[bend right] node[above left] {\(\frac12\mid b\)} (000);
                \draw (00) edge[bend left] node[above right] {\(\frac12\mid c\)} (001);
                \draw (0) edge node[left] {\(1\mid a\)} (00);
                \draw[loop] (000) edge[loop below] node[below] {\(1\mid d\)} (000);
                \draw[loop] (011) edge[loop below] node[below] {\(1\mid d\)} (011);
            \end{tikzpicture}
        \end{gather*}
        \caption{
            \label{fig:not bisim} 
            A standard example of trace equivalent states \(x,y\) that are not bisimilar~\cite{LarsenS91}.
            If \(x\) and \(y\) were bisimilar, then there would be a bisimulation equivalence that relates them. 
            Any such bisimulation equivalence must also relate \(z\) and \(z'\), which are not bisimilar.
        }
    \end{figure}

    \begin{cor}\label{cor:prob corecursive}
        Define the labelled Markov algebra \((\Prob(A^\omega), \alpha)\) by
        \[
            \alpha(\theta)(B)
            = \sum_{(a, \rho) \in \supp(\theta)} \theta(a,\rho)~a^\#\rho(B)
        \]
        Then \((\Prob(A^\omega), \alpha)\) is a finitely corecursive algebra.%
        \footnote{
            Recall \autoref{fn:corecursive}. 
            Since \((A^\omega, d_c)\) is bounded (for any choice of \(c\)-metric \(d_c\)), \((\Prob(A^\omega), \alpha)\) is in fact corecursive, as opposed to just finitely corecursive. Again, we will not need this stronger result.
        }
        Furthermore, given a locally finite LMC \((X, \beta)\), the corecursive map induced by \(\beta\) is \(\hat\beta\), given by \autoref{eq:trace measure}.
    \end{cor}

    \begin{proof}
        The first part follows from \autoref{thm:LMC unique solution}, taking $M$ to be the Cantor set.
        % , regarded as a complete metric space with $c_a = \frac{1}{2}$ for all $a$ in Example~\ref{eg:stream system}.
        We also take each $\sigma_a: A^{\omega} \to A^{\omega}$ to be prefixing by $a$; we have also written this map as $a(-)$.    
        For these choices, the map $\sigma_{\omega}\colon A^{\omega} \to A^{\omega}$ is the identity (see~\autoref{eg:stream system}).
        As Theorem~\ref{thm:LMC unique solution} shows, the induced corecursive map is $\hat{\beta}_{\sigma}$.
        Recall that $\hat{\beta}_{\sigma}$ is the pushforward $\sigma_{\omega}^{\#}\hat{\beta}$.
        Since $\sigma_{\omega} = \mathsf{id}$, we see that $\hat{\beta}_{\sigma} = \hat{\beta}$.
    \end{proof}

    \begin{thm}\label{lem:bisim implies meas equiv}
        Let \((X, \beta_X)\) be an LMC, \(x,x' \in X\). 
        If \(x \bisim x'\), then \(\meas(x) = \meas(x' )\).
    \end{thm}
       
    \begin{proof}
        Let $(Y,\beta_Y)$  and $h:(X,\beta_X)\rightarrow (Y,\beta_Y)$ be as in~\autoref{def:prob_bisimilarity}.
        By~\autoref{cor:prob corecursive},
        the maps $\widehat{\beta_X}: X\to \Prob(A^{\omega})$ and $\widehat{\beta_Y}: Y\to \Prob(A^{\omega})$ are the corecursive maps induced by $\beta_X$ and $\beta_Y$.
        So by \autoref{coalg-to-alg-morphisms},
        $\widehat{\beta_X} = \widehat{\beta_Y}\o h$.
        Then 
        \begin{equation*}
            \meas(x)  = \widehat{\beta_X}(x) = \widehat{\beta_Y} \circ h(x) =  \widehat{\beta_Y} \circ h(x' ) = \widehat{\beta_X}(x' ) = \meas(x').
            \tag*{\qedhere}
        \end{equation*}
    \end{proof}  

    \paragraph*{Axiomatization}
    We are ready to present an inference system, which can be found in \autoref{fig:probabilistic axioms}, and show that it is sound with respect to the subfractal measure semantics of probabilistic process terms.

    \begin{defi}
        Given \(e, f \in\PTerm\), write \(e \equiv f\) and say that \(e\) and \(f\) are \emph{provably equivalent} if the equation \(e \equiv f\) can be derived from 
        the inference rules in \autoref{fig:probabilistic axioms}.
    \end{defi}

    \begin{figure}[ht]
        \begin{gather*}
            \begin{array}{c r l}
                (\axiom{I})
                & e \oplus_r e &\hspace{-0.8em}\equiv e \\
                (\axiom{C}) 
                & e_1 \oplus_r e_2 &\hspace{-0.8em}\equiv e_2 \oplus_{1-r} e_1 \\
                (\axiom{A}) 
                & (e_1 \oplus_r e_2) \oplus_s e_3 &\hspace{-0.8em}\equiv e_1 \oplus_{rs} (e_2 \oplus_{\frac{s(1-r)}{1-rs}} e_3) \\
                (\axiom{D})
                & a(e_1 \oplus_r e_2) &\hspace{-0.8em}\equiv ae_1 \oplus_r ae_2 \\
                (\axiom{F})
                & \mu v~e &\hspace{-0.8em}\equiv e[\mu v~e/v] 
            \end{array}
            \qquad
            \begin{array}{c c}
                (\axiom{\alpha}) & \mu w~e \equiv \mu v~e[v/w] \\
                (\axiom{Cong}) & \infer{e_1 \equiv f_1\ \cdots\ e_n \equiv f_n}{g[\vec e/\vec v] \equiv g[\vec f/\vec v]}  \\
                (\axiom{RSP}) & \infer{g \equiv e[g/v]}{g \equiv \mu v~e}
            \end{array}
        \end{gather*}
        \caption{
            \label{fig:probabilistic axioms} 
            Axioms for subfractal measure equivalence (equational logic axioms not depicted).
            Here, \(e, g, e_i, f_i \in \PTerm\) for all \(i\).
            In \((\axiom{Cong})\), \(g\) has precisely the free variables \(v_1, \dots, v_n\), and no variable that appears free in \(f_i\) is bound in \(g\) for any \(i\). 
            In \((\axiom{\alpha})\), \(v\) does not appear free in \(e\). 
        }
    \end{figure}

    \begin{thm}[Soundness]
        \label{thm:probabilistic completeness}
        For any \(e,f \in \PTerm\), if \(e \equiv f\), then for any complete metric space \(M\) and any \(\sigma \colon A \to \Con(M)\), \(\hat\delta_\sigma(e) = \hat\delta_\sigma(f)\).
    \end{thm}

    \begin{proof}
        Stark and Smolka~\cite{Stark2000Probabilistic} show that all the inference rules in \autoref{fig:probabilistic axioms} except for \textax{D} are sound with respect to bisimilarity in \((\PTerm, \delta)\).  
        
        Therefore, by \autoref{lem:bisim implies meas equiv}, all the inference rules in \autoref{fig:probabilistic axioms} except for \textax{D} are sound with respect to trace measure equivalence.
        It therefore suffices to show that \textax{D} is sound with respect to trace measure equivalence. 

        Let \(e_1,e_2 \in \PTerm\), \(a \in A\), and \(r \in (0,1)\). 
        Then for each \(w \in A^*\),
        \begin{align*}
            \meas(a(e_1\oplus_r e_2))(B_w)
            &= \begin{cases}
                \meas(e_1 \oplus_r e_2)(B_u) & w = au \\
                0 &\text{otherwise}
            \end{cases} \\
            &= \begin{cases}
                r \meas(e_1)(B_u) + (1-r)~ \meas(e_2)(B_u) & w = au \\
                0 &\text{otherwise}
            \end{cases} \\
            &= r \meas(ae_1)(B_w) + (1-r)~\meas(ae_2)(B_w) \\
            &= \meas(ae_1 \oplus_r ae_2)(B_w) 
        \end{align*}
        Note that the second equality used \autoref{lem:measure is recog}.  Thus the trace measures associated to \(a(e_1\oplus_r e_2)\) and \(a e_1\oplus_r a e_2\) agree on basic open sets.
        By the Identity theorem~\cite[Corollary 2.5]{Kerstan2013Trace}, \(\meas(a(e_1\oplus_r e_2))\) and \(\meas(a e_1\oplus_r a e_2)\) agree on on all Borel sets.
    \end{proof}

    \begin{thm}[Completeness]\cite[Theorem 3.6]{cirstea_et_al2025}
        \label{con:completeness}     
        The logical system in  \autoref{fig:probabilistic axioms} provides a complete axiomatization of trace measure semantics.
        That is, for any \(e,f \in \PTerm\), if for any complete metric space \(M\) and any \(\sigma \colon A \to \Con(M)\) we have \(\hat\delta_\sigma(e) = \hat\delta_\sigma(f)\), then \(e \equiv f\).
    \end{thm}
    \noindent %FIXED indent
    We use~\autoref{con:completeness} to prove another completeness theorem, the probabilistic analog of~\autoref{corr:affine completeness}.
    We need the lemma below. 

    \begin{lem}
        \label{lem:pushforward preserves injective} 
        Let \(f \colon A^\omega \to \mathbb R\) be injective and continuous.    
        Write \(f^\# \colon \Prob(A^\omega) \to \Prob(\mathbb R)\) for the pushforward map.
        Then \(f^\#\) is injective.
    \end{lem}

    \begin{proof}
        Let \(\rho_1,\rho_2 \in \Prob(A^\omega)\), and suppose that \(f^\#(\rho_1) = f^\#(\rho_2)\).
        For any subset \(B \subseteq A^\omega\), \(f^{-1}\circ f(B) = B\), because \(f\) is injective.
        Furthermore, by the Lusin-Suslin theorem~\cite[Theorem 15.1]{Kechris1995Classical}, \(f(B)\) is Borel if \(B\) is Borel.
        This allows for the following calculation: for any Borel set \(B \subseteq A^\omega\),
        \begin{align*}
            \rho_1(B) 
            &= \rho_1(f^{-1}\circ f(B)) \tag{\(f\) injective}\\
            &= f_\#\rho_1(f(B)) \tag{\(f(B)\) Borel, def.~of \(f^\#\)}\\
            &= f_\#\rho_2(f(B)) \tag{by assumption}\\
            &= \rho_2(f^{-1} \circ f(B)) \tag{def.~of \(f^\#\)}\\
            &= \rho_2(B) \tag{\(f\) injective}
        \end{align*}
        Hence, \(\rho_1 = \rho_2\).
    \end{proof}

    \begin{thm}
        \label{thm:affine probabilistic completeness}
        The logical system in  \autoref{fig:probabilistic axioms} provides a complete axiomatization of trace measure equivalence for affine contraction operator interpretations, i.e., for any \(e,f \in \PTerm\), \(e \equiv f\) if and only if for any affine contraction interpretation \(\sigma \colon A \to \Con(\mathbb R)\), \(\hat\delta_\sigma(e) = \hat\delta_\sigma(f)\).
    \end{thm}

    \begin{proof}
        Recall from \autoref{rem:sigma-continuous} that \(\sigma_\omega\) is continuous for any contraction operator interpretation \(\sigma\).
        In \autoref{prop:cantor sets in R}, we constructed a contraction operator interpretation \(\sigma \colon A \to \Con(\mathbb R)\) such that \(\sigma_\omega \colon A^\omega \to \mathbb R\) is injective. 
        This provides us with an injective continuous map \(\sigma_\omega \colon \Prob(A^\omega) \to \Prob(\mathbb R)\).

        Given \(e, f \in \PTerm\) such that \(e \equiv f\), \(\hat\delta_\sigma(e) = \hat\delta_\sigma(f)\) follows from soundness.
        For the reverse direction, suppose \(\hat\delta_\sigma(e) = \hat\delta_\sigma(f)\). 
        Then \(\sigma_\omega^\#\hat\delta(e) = \sigma_\omega^\#\hat\delta(f)\).
        From \autoref{lem:pushforward preserves injective} we see that
        \(\sigma_\omega^\#\) is injective because \(\sigma_\omega\) is continuous and injective.
        It follows that \(\hat\delta(e) = \hat\delta(f)\).
        Finally, by \autoref{con:completeness}, from \(\hat\delta(e) = \hat\delta(f)\) we know that \(e \equiv f\). 
    \end{proof}
    
    So we see that  the subfractal measure axioms and rules in \autoref{fig:probabilistic axioms} are complete for affine real-valued contraction operator interpretations.   

    \section{Related Work}
    \label{sec:related work}

    This paper is part of a larger effort of examining topics in continuous mathematics from the standpoint of coalgebra and theoretical computer science.
    The topic itself is quite old, and originates perhaps with  Pavlovic and Escard\'o's paper ``Calculus in Coinductive Form''~\cite{PavlovicE98}.
    Another early contribution is Pavlovic and Pratt~\cite{PavlovicP02}.  
    These papers proposed viewing some structures in continuous 
    mathematics---the real numbers, for example, and power series expansions---in terms of final coalgebras and streams.   
    The next stage in this line of work was a set of papers specifically about fractal sets and final coalgebras.
    For example, Leinster~\cite{Leinster} offered a very general theory of self-similarity that used categorical modules in connection with the kind of gluing that is prominent in constructions of self-similar sets.
    In a different direction, papers like~\cite{BhattacharyaMRR14} showed that for some very simple fractals (such as the Sierpi\'nski gasket treated here), the final coalgebras were Cauchy completions of the initial algebras. \medskip

    % Generalized Iterated Function systems
    \paragraph*{Other generalizations of iterated function systems}
    Many generalizations of Hutchinson's self-similar sets have appeared in the literature.
    We mention two closely related generalizations below.
    \medskip

    \noindent\textbf{Graph IFSs.}
    The generalization that most closely resembles our own is that of an attractor for a \emph{directed-graph iterated function system} (or \emph{graph IFS})~\cite{Mauldin1988Hausdorff}.
    An LTS paired with a contraction operator interpretation on a complete metric space is equivalent data to that of a graph IFS, and equivalent statements to \autoref{thm:fractal semantics coincide} can be found for example in~\cite{Mauldin1988Hausdorff,Edgar1990Fractal,Mihail2010GIFS}.
    As opposed to the regular subfractal corresponding to one state, as we have studied above, the geometric object studied in the graph IFSs literature (the \emph{attractor} for the graph IFS) is typically the union of the regular subfractals corresponding to all the states (in our terminology), and geometric properties such as Hausdorff dimension and connectivity are emphasized~\cite{Mauldin1988Hausdorff,Falconer1986Geometry,Edgar1990Fractal,Edgar1992Multifractal,Boore2011GIFS}.
    We have taken this work in a slightly different direction by presenting a coalgebraic perspective on graph IFSs, seeing each state of a labelled transition system as a ``recipe'' for constructing fractal sets on its own. 
    We have also allowed the interpretations of the labels to vary to obtain a semantics for process terms.

    In a certain sense, regular subfractals are a more general class of objects than attractors of graph IFSs. 
    Call a language of streams \(L \subseteq A^\omega\) a \emph{graph IFS language} if there is a finite LTS \((X, \alpha)\) such that \(L = \bigcup_{x \in X} \str_\alpha(x)\).
    Given a contraction operator interpretation \(\sigma \colon A \to \Con(M)\) on a complete metric space \(M\), the fractal generated by a graph IFS consisting of \((X,\alpha)\) and \(\sigma\) is precisely the image of \(L\) under \(\sigma_\omega\). 
    Every graph IFS language is the stream language emitted by a state in a finite LTS, but not conversely.

    \begin{lem}
        \label{lem:graph IFS gen}
        Let \(L\) be a graph IFS language. 
        There is a finite LTS \((X,\alpha)\) and a state \(x \in X\) such that \(L = \str_\alpha(x)\).
    \end{lem}

    \begin{proof}
        Let \((Y,\alpha_Y)\) be a finite LTS such that \(L = \bigcup_{y \in Y} \str_{\alpha_Y}(y)\). 
        Define \(X = Y + \{x_0\}\), and \[
          \alpha(z) = \begin{cases}
            \alpha_Y(z) &z \in Y \\
            \bigcup_{y \in Y} \alpha_Y(y) &z = x_0
          \end{cases}  
        \]
        In other words, \(x_0 \tr{a} y\) if and only if there is a \(y' \in Y\) such that \(y' \tr{a} y\).
        Then 
        \[
            \str_\alpha(x_0) 
            = \bigcup_{x_0 \tr{a} y} a ~\str_{\alpha_Y}(y)
            = \bigcup_{y' \tr{a} y} a ~\str_{\alpha_Y}(y)
            = \bigcup_{y \in Y} \str_{\alpha_Y}(y)
            = L
            \qedhere
        \]
    \end{proof}

    To show the converse of the above lemma is false, consider the language \(\{(a, b, b, b, \dots)\}\) with one stream in it. 
    This is not a graph IFS language for a simple reason: a graph IFS language includes the streams emitted by every state of an LTS, and therefore is closed under deleting initial segments.
    For example, if a state \(x\) emits \((a, b, b, b, \dots)\), then it must have an outoing transition \(x \tr{a} y\) such that \(y\) emits \((b, b, b,\dots)\).
    It follows that regular stream languages, and by extension regular subfractals, properly generalize graph IFS languages. \bigskip
    
    \noindent\textbf{Generalized IFSs.} Another generalization of self-similar sets is Mihail and Miculescu's notion of attractor for a \emph{generalized iterated function system}~\cite{Mihail2010GIFS}.
    A generalized IFS is essentially one of Hutchinson's IFSs but with multi-arity contractions \(\sigma_a \colon M^n \to M\).
    Attractors for generalized IFSs have been shown to be a proper generalization of self-similar sets~\cite{strobin_attractors_2015}, as well as limits of attractors for infinite IFSs~\cite{oliveira2023hutchinsonbarnsley}. 
    A common generalization of graph IFSs and generalized IFSs might be achieved by considering coalgebras of the form \(X \to \mathcal P(\coprod_{n \in \mathbb N} A_n\times X^n)\) and interpreting each \(a \in A_n\) as an \(n\)-ary contraction.
    We suspect that a similar story to the one we have outlined in this paper is possible for this common generalization.

    % Rabinovich
    \paragraph*{Process algebra}
    The process terms we use to specify labelled transition systems and labelled Markov chains are fragments of known specification languages.
    Milner used process terms to specify LTSs in~\cite{Milner1984Complete}, and we have repurposed his small-step semantics here.
    Stark and Smolka use probabilistic process terms to specify labelled Markov chains (in our terminology) in~\cite{Stark2000Probabilistic}, and we have used them for the same purpose.
    Both of these papers also include complete axiomatizations of bisimilarity, and we have also repurposed their axioms.
    
    However, fractal semantics is strictly coarser than bisimilarity, and in particular, fractal equivalence of process terms is trace equivalence.
    Rabinovich added a single axiom to Milner's axiomatization to obtain a sound and complete axiomatization of trace equivalence of \(\mu\)-expressions~\cite{Rabinovich1993Traces}. We used this axiom in our proof of \autoref{thm:completeness}.
    In contrast, the axiomatization of trace equivalence for probabilistic processes is only well-understood for \emph{finite} traces, see Silva and Sokolova's~\cite{Silva2011Probabilistic}, which our probabilistic process terms do not exhibit.  
    We use the trace semantics of Kerstan and K\"onig~\cite{Kerstan2013Trace} because it takes into account infinite traces.
    Infinite trace semantics of probabilistic systems has yet to see a complete axiomatization in the literature, although a complete axiomatization of the total variation distance for these systems has been obtained in the form of a so-called \emph{quantitative equational theory}~\cite{BACCI201827}.
    
    \paragraph*{Other types of syntax}
    In this paper, we used the specification language of $\mu$-terms as our basic syntax.   
    As it happens, there are two other flavors of syntax that we could have employed.  
    These are \emph{iteration theories}~\cite{Bloom1993Iteration}, and terms in the Formal Language of Recursion $FLR$, especially its $FLR_0$ fragment.  
    The three flavors of syntax for fixed point terms are compared in a number of papers: In~\cite{Hurkens}, it was shown that there is an equivalence of categories between $FLR_0$ structures and iteration theories, and Bloom and \'Esik make a similar connection between iteration theories and the $\mu$-calculus in~\cite{BloomEsik94}.
    Again, these results describe general matters of equivalence, but it is not completely clear that for a specific space or class of spaces that they are equally powerful or equally convenient specification languages.   
    We feel this matter deserves some investigation.
   
    \paragraph*{Equivalence under hypotheses}   
    A specification language fairly close to iteration theories was used by Milius and Moss to reason about fractal constructions in~\cite{MiliusMoss09} under the guise of \emph{interpreted solutions} to recursive program schemes~\cite{Milius2006Recursion}. 
    Moreover,~\cite{MiliusMoss09} contains important examples of reasoning about the equality of fractal sets under assumptions about the contractions.   
    Based on the general negative results on reasoning from hypotheses in the logic of recursion~\cite{Hurkens}, we would not expect a completeness theorem for fractal equivalence under hypotheses.  
    However, we do expect to find sound logical systems which account for interesting phenomena in the area.
    
    \section{A Question about Regular Subfractals}
    \label{sec:questions and future work}

    Before we end this paper, we would like to pose a question regarding the relationship between regular subfractals and self-similar sets.
    
    Certain regular subfractals that have been generated by LTSs with multiple states happen to coincide with self-similar sets using a different alphabet of action symbols and under a different contraction operator interpretation.
    For example, the twisted Sierpi\'nski gasket in \autoref{fig:Twisted Sierpinski gasket} is the self-similar set generated by the iterated function system consisting of the compositions \(\sigma_a, \sigma_b\sigma_b, \sigma_b\sigma_c, \sigma_c\sigma_b\), and \(\sigma_c\sigma_c\).

    \begin{qu}
    \label{q1} 
        Is every regular subfractal a self-similar set?
        In other words, are there regular subfractals which can only be generated by a multi-state LTS?
    \end{qu}
    
    \begin{exa}
        To illustrate the subtlety of this question, consider the following LTS.
        \[ 
            \begin{tikzpicture}
                \node[state] (0) at (0,0) {\(x\)};
                \node[state] (1) at (2,0) {};
                \draw (0) edge[] node[above] {\(b\)} (1);
                \draw[loop] (0) edge[loop left] node[left] {\(a\)} (0);
                \draw[loop] (1) edge[loop right] node[right] {\(b\)} (1);
            \end{tikzpicture} 
            \qquad \qquad 
            \begin{tikzpicture}[scale=3]
                \draw (0.25, 0) edge[->, thin] (2.25, 0);
                \draw (-.25, 0) edge[<-, thin] (0, 0);
                \draw (0, 0) edge[-, dashed, thin] (0.25, 0);
                \node[red] (0) at (2, 0) {\(\bullet\)};
                \node[red] (1) at (1, 0) {\(\bullet\)};
                \node[red] (2) at (0.5, 0) {\(\bullet\)};
                \node[red] (3) at (0.25, 0) {\(\bullet\)};
                % \node[red] (4) at (0.125, 0) {\(\bullet\)};
                \node at (2, 0.15) {\(1\)};
                \node at (1, 0.15) {\(\frac12\)};
                \node at (0.5, 0.15) {\(\frac1{2^2}\)};
                \node at (0.25, 0.15) {\(\frac1{2^3}\)};
                \node at (0, 0.15) {\(0\)};
                %\node at (0.125, 0.18) {\(\frac1{2^4}\)};
                %\node at (0.125, 0.1) {\(\cdots\)};
                % \node (6) at (0., 0) {\(\bullet\)};
                % \node (7) at (2, 0) {\(\bullet\)};
                % \node (8) at (2, 0) {\(\bullet\)};
                \node[red] (inf) at (0, 0) {\(\bullet\)};
            \end{tikzpicture}
        \] 
        The state $x$ emits $(a,a,\ldots)$ (an infinite stream of $a$'s) and $(a,\ldots,a,b,b,\ldots)$, a stream with some finite number (possibly $0$) of $a$'s followed by an infinite stream of $b$'s. 
        Now let $M = \reals$ with Euclidean distance and consider the contraction operator interpretation $\sigma_a(r) = \frac{1}{2}r$ and $\sigma_b(r) = \frac{1}{2}r + \frac{1}{2}$.  
        Let \(K =\{0\} \cup \{\frac{1}{2^n}|n\geq 0\}\) (depicted above).
        Then \(K\) is the component of the solution at \(x\). 
        This example is interesting because unlike the Twisted Sierpi\'nski gasket in~\autoref{fig:Twisted Sierpinski gasket}, there is no obvious finite set of compositions $\sigma_a$ and $\sigma_b$ such that $K$ is the self-similar set generated by that iterated function system.  
    
        There is an LTS  \((X,\alpha)\) with \( X \) a singleton set 
        \(\set{x}\), and a contraction operator interpretation whose solution is \(K\).   We take the set of action labels underlying \(X\) to be 
        \(B = \{f,g,h\}\) and use the contraction operator interpretation $\sigma_f(r) = 0$, $\sigma_g(r) = 1$ and $\sigma_h(r) = \frac{1}{2}r$.  
            It is easy to verify that \(K = \bigcup_{i\in \{f,g,h\}} \sigma_i(K)\).

        But we claim that \( K \)   cannot be obtained using a single-state LTS and \emph{the same contractions} 
        \(\sigma_a(r) = \frac{1}{2}r\) and $\sigma_b(r) = \frac{1}{2}r + \frac{1}{2}$, or  using \emph{any (finite) compositions}
        of \( \sigma_a\) and \(\sigma_b\).
        Indeed, suppose there were such a finite collection $\sigma_1,\ldots,\sigma_n$ consisting of (finite) compositions of $\sigma_a$ and $\sigma_b$ such that $K = \bigcup_{i=1}^n\sigma_i(K)$.  
        Since $1 \in K$, we must be using the stream \( (b,b,b,\ldots)\) (since if there is an \(a\) at position \(n\), the limit corresponding to the stream would be $\leq 1-\frac{1}{2^n}<1$), so some $\sigma_i$ must consist of a composition of $\sigma_b$ some number $m\geq 1$ of times with itself.
        Similarly, the only way to obtain $0$ is with $(a,a,a,\ldots)$, so there must be some $\sigma_j$ which is a composition of $\sigma_a$ some number of times $p\geq 1$ with itself.       
        But then $\lim_{n\to\infty}\sigma_i\circ \sigma_j\circ\sigma^n_i (r) = 1-(\frac{2^{p}-1}{2^{m+p}}) > \frac{1}{2}$, since $m,p\geq 1$.  
        That point must be in the subset of $\reals$ generated by this LTS. 
        However, it is not in $K$, since $\frac{1}{2}<1-(\frac{2^{p}-1}{2^{m+p}})<1$.  
        More generally, we cannot obtain \(K\) using a single-state LTS even if we allowed finite sums of compositions of \(\sigma_a\) and \(\sigma_b\).

        Once again, it is possible to find a single state LTS whose corresponding subset of $\reals$ is $K$, but to do this we needed to \emph{change the alphabet and also the contractions}.  
        Perhaps un-coincidentally, the constant operators are exactly the limits of the two contractions from the original interpretation.  
        Our question is whether this can always be done. 
    \end{exa}

    Towards understanding \autoref{q1}, it may be useful to focus on a specific complete metric space \(M\), such as \(M = K = \{\frac1{2^n} \mid n \ge 0\} \cup \{0\}\) above, and give a characterization of all contractions on \(M\).
    This would allow us to characterize its corresponding regular subfractals and self-similar sets.

    Boore and Falconer have answered a restricted version of \autoref{q1} in~\cite{boore_falconer_2013}.
    There, they present a graph IFS whose attractor is not the self-similar set generated by any iterated function system consisting of \emph{similitudes}.%
    \footnote{A similitude is a contraction \(f\) such that for some \(c \in [0,1)\), \(d(f(p), g(q)) = cd(p, q)\) for all \(p,q \in M\).}
    As we saw in \autoref{lem:graph IFS gen}, every graph IFS generates a regular subfractal, so the example of Boore and Falconer is indeed a regular subfractal that is not the self-similar set generated by an IFS consisting of similitudes.  In other words, this is a regular subfractal which cannot be generated with a single state LTS whose contraction operators are similitudes.  
    But not every contraction is a similitude, so the example given in~\cite{boore_falconer_2013} does not fully answer \autoref{q1}. 
    It is unknown if their example could be generated by a single state LTS with contractions that are not similitudes.

 \section{Conclusion}
    This paper connects fractals to trace semantics, a topic originating in process algebra.  
    This connection is our main contribution, because it opens up a line of communication between two very different areas of study.
    The study of fractals is a well-developed area, and like most of mathematics it is pursued without a special-purpose specification language.   
    When we viewed process terms as recipes for fractals, we provided a specification language that was not present in the fractals literature.  
    Of course, one also needs a contraction operator interpretation to actually define a fractal, but the separation of syntax (the process terms) and semantics (the fractals obtained using contraction operator interpretations of the syntax) is something that comes from the tradition of logic and theoretical computer science.  
    Similarly, the use of a logical system and the emphasis on soundness and completeness is a new contribution here.

    All of the above opens questions about fractals and their specifications.
    Our most concrete question was posed in \cref{sec:questions and future work}.
    A more logic oriented question is inspired by \autoref{corr:affine completeness}, which states that the axioms in \autoref{fig:axioms} are complete for a specific contraction operator interpretation corresponding to the real-valued Cantor set.
    That is, no additional equations are imposed by the Cantor set interpretation. 
    We would also like to know if we can obtain completeness theorems for other self-similar sets by allowing for extra equations in the axiomatization.
    For example, several additional equations between process terms are sound for the contraction operator interpretation corresponding to the Sierpinski gasket in \autoref{fig:Sierpinski gasket}, including
    \begin{gather*}
        a~\mu v~(bv) = b~\mu v~(av)
        \qquad
        b~\mu v~(cv) = c~\mu v~(bv)
        \qquad 
        c~\mu v~(av) = a~\mu v~(cv)
    \end{gather*} 
    We would like to know if adding these equations gives a complete axiomatization of the Sierpinski gasket interpretation.
    Lastly, and most speculatively, since LTSs (and other automata) appear so frequently in decision procedures from process algebra and verification, we would like to know if our semantics perspective on fractals can provide new complexity results in fractal geometry.
    We hope we have initiated a line of research where questions and answers come from both the analytic side and from theoretical computer science.

    \paragraph*{Acknowledgements}
    Todd Schmid was partially funded by ERC Grant Autoprobe (grant agreement 10100269).
    Lawrence Moss was supported by grant \#586136 from the Simons Foundation.
    We would like to thank Alexandra Silva and Dylan Thurston for helpful discussions.
    The images in Figures~\ref{fig:Sierpinski gasket}, \ref{fig:Twisted Sierpinski gasket},
    \ref{fig:weighted Sierpinski gasket}, and \ref{fig:asymmetric chaos sierpinski} were made using \href{https://www.sagemath.org/}{SageMath} and \href{https://www.gimp.org/}{GIMP}.
    The SageMath code used to generate Figures~\ref{fig:weighted Sierpinski gasket} and~\ref{fig:asymmetric chaos sierpinski} was based on several scripts written by Connor Li during an undergraduate research project at Saint Mary's College of California in the Spring of 2024.

    \bibliographystyle{alphaurl}% the mandatory bibstyle
    \bibliography{refs}

\newcommand{\etalchar}[1]{$^{#1}$}
\begin{thebibliography}{HMM{\etalchar{+}}98}

\bibitem[AT89]{AdamekTrnkova1989}
Jir\'i Ad\'amek and Vera Trnková.
\newblock {\em Automata and Algebras in Categories}.
\newblock Springer Dordrecht, 1989.

\bibitem[Bae05]{Baeten2005History}
Jos C.~M. Baeten.
\newblock A brief history of process algebra.
\newblock {\em Theor. Comput. Sci.}, 335(2-3):131--146, 2005.
\newblock \href {https://doi.org/10.1016/j.tcs.2004.07.036}
  {\path{doi:10.1016/j.tcs.2004.07.036}}.

\bibitem[Ban22]{Banach1922Fixedpoint}
Stefan Banach.
\newblock Sur les op{\'e}rations dans les ensembles abstraits et leur
  application aux {\'e}quations int{\'e}grales.
\newblock {\em Fundamenta Mathematicae}, 3:133--181, 1922.

\bibitem[Bar88]{barnsley}
Michael Barnsley.
\newblock {\em Fractals everywhere}.
\newblock Academic Press, Inc., Boston, MA, 1988.

\bibitem[BBLM18]{BACCI201827}
Giorgio Bacci, Giovanni Bacci, Kim~G. Larsen, and Radu Mardare.
\newblock Complete axiomatization for the total variation distance of markov
  chains.
\newblock {\em Electronic Notes in Theoretical Computer Science}, 336:27--39,
  2018.
\newblock The Thirty-third Conference on the Mathematical Foundations of
  Programming Semantics (MFPS XXXIII).
\newblock \href {https://doi.org/10.1016/j.entcs.2018.03.014}
  {\path{doi:10.1016/j.entcs.2018.03.014}}.

\bibitem[B{\'{E}}93]{Bloom1993Iteration}
Stephen~L. Bloom and Zolt{\'{a}}n {\'{E}}sik.
\newblock {\em Iteration Theories - The Equational Logic of Iterative
  Processes}.
\newblock {EATCS} Monographs on Theoretical Computer Science. Springer, 1993.
\newblock \href {https://doi.org/10.1007/978-3-642-78034-9}
  {\path{doi:10.1007/978-3-642-78034-9}}.

\bibitem[BE94]{BloomEsik94}
Stephen~L. Bloom and Zolt\'{a}n \'{E}sik.
\newblock Solving polynomial fixed point equations.
\newblock In {\em Mathematical Foundations of Computer Science 1994
  ({K}o\v{s}ice, 1994)}, volume 841 of {\em Lecture Notes in Comput. Sci.},
  pages 52--67. Springer, Berlin, 1994.

\bibitem[BF13]{boore_falconer_2013}
G.~C. Boore and K.~J. Falconer.
\newblock Attractors of directed graph {IFS}s that are not standard {IFS}
  attractors and their {H}ausdorff measure.
\newblock {\em Mathematical Proceedings of the Cambridge Philosophical
  Society}, 154(2):325–349, 2013.
\newblock \href {https://doi.org/10.1017/S0305004112000576}
  {\path{doi:10.1017/S0305004112000576}}.

\bibitem[BMRR14]{BhattacharyaMRR14}
Prasit Bhattacharya, Lawrence~S. Moss, Jayampathy Ratnayake, and Robert Rose.
\newblock Fractal sets as final coalgebras obtained by completing an initial
  algebra.
\newblock In Franck van Breugel, Elham Kashefi, Catuscia Palamidessi, and Jan
  Rutten, editors, {\em Horizons of the Mind. {A} Tribute to Prakash Panangaden
  - Essays Dedicated to Prakash Panangaden on the Occasion of His 60th
  Birthday}, volume 8464 of {\em Lecture Notes in Computer Science}, pages
  146--167. Springer, 2014.

\bibitem[Boo11]{Boore2011GIFS}
Graeme~C. Boore.
\newblock {\em Directed Graph Iterated Function Systems}.
\newblock PhD thesis, University of St. Andrews, 2011.

\bibitem[BPS01]{BPS2001Handbook}
Jan~A. Bergstra, Alban Ponse, and Scott~A. Smolka, editors.
\newblock {\em Handbook of Process Algebra}.
\newblock North-Holland / Elsevier, 2001.
\newblock \href {https://doi.org/10.1016/b978-0-444-82830-9.x5017-6}
  {\path{doi:10.1016/b978-0-444-82830-9.x5017-6}}.

\bibitem[BR07]{Bogachev2007Measure}
Vladimir~Igorevich Bogachev and Maria Aparecida~Soares Ruas.
\newblock {\em Measure Theory}, volume~1.
\newblock Springer, 2007.
\newblock \href {https://doi.org/10.1007/978-3-540-34514-5}
  {\path{doi:10.1007/978-3-540-34514-5}}.

\bibitem[BS81]{SankappanavarBurris1981}
S.~Burris and H.~P. Sankappanavar.
\newblock {\em A Course in Universal Algebra}.
\newblock Springer New York, NY, 1981.

\bibitem[CMN{\etalchar{+}}25]{cirstea_et_al2025}
Corina C\^{i}rstea, Lawrence~S. Moss, Victoria Noquez, Todd Schmid, Alexandra
  Silva, and Ana Sokolova.
\newblock {A Complete Inference System for Probabilistic Infinite Trace
  Equivalence}.
\newblock In J\"{o}rg Endrullis and Sylvain Schmitz, editors, {\em 33rd EACSL
  Annual Conference on Computer Science Logic (CSL 2025)}, volume 326 of {\em
  Leibniz International Proceedings in Informatics (LIPIcs)}, pages
  30:1--30:23, Dagstuhl, Germany, 2025. Schloss Dagstuhl -- Leibniz-Zentrum
  f{\"u}r Informatik.
\newblock \href {https://doi.org/10.4230/LIPIcs.CSL.2025.30}
  {\path{doi:10.4230/LIPIcs.CSL.2025.30}}.

\bibitem[CUV06]{cuv06}
Venanzio Capretta, Tarmo Uustalu, and Varmo Vene.
\newblock Recursive coalgebras from comonads.
\newblock {\em Inform.~Comput.}, 204:437--468, 2006.

\bibitem[DS84]{DiaconisS1984}
Percy Diaconis and Mehrdad Shahshahani.
\newblock Products of random matrices and computer image generation.
\newblock Technical report, Stanford University, 1984.

\bibitem[Edg90]{Edgar1990Fractal}
Gerald~A. Edgar.
\newblock {\em Measure, Topology, and Fractal Geometry}.
\newblock Springer New York, NY, 1 edition, 1990.
\newblock \href {https://doi.org/10.1007/978-1-4757-4134-6}
  {\path{doi:10.1007/978-1-4757-4134-6}}.

\bibitem[Elt87]{Elton87}
John~H. Elton.
\newblock An erogodic theorem for iterated maps.
\newblock {\em Ergodic Theory and Dynamical Systems}, 7(4), 1987.
\newblock \href {https://doi.org/10.1017/S0143385700004168}
  {\path{doi:10.1017/S0143385700004168}}.

\bibitem[EM92]{Edgar1992Multifractal}
G.~A. Edgar and R.~Daniel Mauldin.
\newblock Multifractal decompositions of digraph recursive fractals.
\newblock {\em Proceedings of the London Mathematical Society},
  s3-65(3):604--628, 1992.
\newblock \href {https://doi.org/10.1112/plms/s3-65.3.604}
  {\path{doi:10.1112/plms/s3-65.3.604}}.

\bibitem[Fal86]{Falconer1986Geometry}
Kenneth~J. Falconer.
\newblock {\em The Geometry of Fractal Sets}.
\newblock Cambridge Tracts in Mathematics. Cambridge University Press, 1986.
\newblock \href {https://doi.org/10.1017/CBO9780511623738}
  {\path{doi:10.1017/CBO9780511623738}}.

\bibitem[Fol99]{folland}
Gerald~B. Folland.
\newblock {\em Real Analysis: Modern Techniques and Their Applications}.
\newblock Wiley, 2nd edition, 1999.

\bibitem[GR18]{GoyRot18}
Alexandre Goy and Jurriaan Rot.
\newblock ({I}n)finite trace equivalence of probabilistic transition systems.
\newblock In Corina C{\^{\i}}rstea, editor, {\em Coalgebraic Methods in
  Computer Science - 14th {IFIP} {WG} 1.3 International Workshop, {CMCS} 2018,
  Colocated with {ETAPS} 2018, Thessaloniki, Greece, April 14-15, 2018, Revised
  Selected Papers}, volume 11202 of {\em Lecture Notes in Computer Science},
  pages 100--121. Springer, 2018.
\newblock \href {https://doi.org/10.1007/978-3-030-00389-0\_7}
  {\path{doi:10.1007/978-3-030-00389-0\_7}}.

\bibitem[Gro15]{Hausdorff}
Wilh Gro{\ss}.
\newblock Grundz{\"u}ge der mengenlehre.
\newblock {\em Monatshefte f{\"u}r Mathematik und Physik}, 26(1):A34--A35,
  1915.
\newblock \href {https://doi.org/10.1007/BF01999507}
  {\path{doi:10.1007/BF01999507}}.

\bibitem[GS02]{gummschroeder-bounded}
H.~Peter Gumm and Tobias Schr{\"{o}}der.
\newblock Coalgebras of bounded type.
\newblock {\em Math. Struct. Comput. Sci.}, 12(5):565--578, 2002.
\newblock \href {https://doi.org/10.1017/S0960129501003590}
  {\path{doi:10.1017/S0960129501003590}}.

\bibitem[HMM{\etalchar{+}}98]{Hurkens}
A.~J.~C. Hurkens, Monica McArthur, Yiannis~N. Moschovakis, Lawrence~S. Moss,
  and Glen~T. Whitney.
\newblock The logic of recursive equations.
\newblock {\em J. Symbolic Logic}, 63(2):451--478, 1998.

\bibitem[Hoa78]{Hoare1978Traces}
C.~A.~R. Hoare.
\newblock Communicating sequential processes.
\newblock {\em Commun. {ACM}}, 21(8):666--677, 1978.
\newblock \href {https://doi.org/10.1145/359576.359585}
  {\path{doi:10.1145/359576.359585}}.

\bibitem[Hut81]{Hutchinson1981Fractals}
John~E. Hutchinson.
\newblock Fractals and self similarity.
\newblock {\em Indiana University Mathematics Journal}, 30(5):713--747, 1981.

\bibitem[Kec95]{Kechris1995Classical}
A.~Kechris.
\newblock {\em Classical Descriptive Set Theory}.
\newblock Graduate Texts in Mathematics. Springer New York, 1995.

\bibitem[KK13]{Kerstan2013Trace}
Henning Kerstan and Barbara König.
\newblock {Coalgebraic trace semantics for continuous probabilistic transition
  systems}.
\newblock {\em {Logical Methods in Computer Science}}, {Volume 9, Issue 4}, Dec
  2013.
\newblock \href {https://doi.org/10.2168/LMCS-9(4:16)2013}
  {\path{doi:10.2168/LMCS-9(4:16)2013}}.

\bibitem[Kle56]{Kleene56}
S.~C. Kleene.
\newblock Representation of events in nerve nets and finite automata.
\newblock In Claude Shannon and John McCarthy, editors, {\em Automata Studies},
  pages 3--41. Princeton University Press, Princeton, NJ, 1956.

\bibitem[Kle02]{kleene2002mathematical}
Stephen~Cole Kleene.
\newblock {\em Mathematical Logic}.
\newblock Courier Corporation, 2002.

\bibitem[Kra06]{kravchenko}
A.~S. Kravchenko.
\newblock Completeness of the space of separable measures in the
  kantorovich-rubenste\u{i}n metric.
\newblock {\em Siberian Mathematical Journal}, 47(1), 2006.

\bibitem[LAS78]{counterexamples}
J.~Arthur~Seebach Lynn Arthur~Steen.
\newblock {\em Counterexamples in Topology}.
\newblock Springer New York, NY, 1978.
\newblock \href {https://doi.org/10.1007/978-1-4612-6290-9}
  {\path{doi:10.1007/978-1-4612-6290-9}}.

\bibitem[Lei11]{Leinster}
Tom Leinster.
\newblock A general theory of self-similarity.
\newblock {\em Adv. Math.}, 226(4):2935--3017, 2011.

\bibitem[LS91]{LarsenS91}
Kim~Guldstrand Larsen and Arne Skou.
\newblock Bisimulation through probabilistic testing.
\newblock {\em Inf. Comput.}, 94(1):1--28, 1991.
\newblock \href {https://doi.org/10.1016/0890-5401(91)90030-6}
  {\path{doi:10.1016/0890-5401(91)90030-6}}.

\bibitem[Man77]{Mandelbrot1977Fractals}
Benoit~B. Mandelbrot.
\newblock {\em Fractals: Form, Chance, and Dimension}.
\newblock Mathematics Series. W. H. Freeman, 1977.

\bibitem[Mil78]{Milner78}
Robin Milner.
\newblock Synthesis of communicating behaviour.
\newblock In J{\'{o}}zef Winkowski, editor, {\em Mathematical Foundations of
  Computer Science 1978, Proceedings, 7th Symposium, Zakopane, Poland,
  September 4-8, 1978}, volume~64 of {\em Lecture Notes in Computer Science},
  pages 71--83. Springer, 1978.
\newblock \href {https://doi.org/10.1007/3-540-08921-7\_57}
  {\path{doi:10.1007/3-540-08921-7\_57}}.

\bibitem[Mil84]{Milner1984Complete}
Robin Milner.
\newblock A complete inference system for a class of regular behaviours.
\newblock {\em J. Comput. Syst. Sci.}, 28(3):439--466, 1984.
\newblock \href {https://doi.org/10.1016/0022-0000(84)90023-0}
  {\path{doi:10.1016/0022-0000(84)90023-0}}.

\bibitem[MM06]{Milius2006Recursion}
Stefan Milius and Lawrence~S. Moss.
\newblock The category-theoretic solution of recursive program schemes.
\newblock {\em Theor. Comput. Sci.}, 366(1-2):3--59, 2006.
\newblock \href {https://doi.org/10.1016/j.tcs.2006.07.002}
  {\path{doi:10.1016/j.tcs.2006.07.002}}.

\bibitem[MM09]{MiliusMoss09}
Stefan Milius and Lawrence~S. Moss.
\newblock Equational properties of recursive program scheme solutions.
\newblock {\em Cah. Topol. G\'{e}om. Diff\'{e}r. Cat\'{e}g.}, 50(1):23--66,
  2009.

\bibitem[MM10]{Mihail2010GIFS}
Alexandru Mihail and Radu Miculescu.
\newblock Generalized {IFS}s on noncompact spaces.
\newblock {\em Fixed Point Theory and Applications}, 1(584215), 2010.
\newblock \href {https://doi.org/10.1155/2010/584215}
  {\path{doi:10.1155/2010/584215}}.

\bibitem[MW88]{Mauldin1988Hausdorff}
R~Daniel Mauldin and S.~C. Williams.
\newblock Hausdorff dimension in graph directed constructions.
\newblock {\em Transactions of the American Mathematical Society},
  309(2):811--829, 1988.
\newblock \href {https://doi.org/10.1090/S0002-9947-1988-0961615-4}
  {\path{doi:10.1090/S0002-9947-1988-0961615-4}}.

\bibitem[Oli23]{oliveira2023hutchinsonbarnsley}
Elismar~R. Oliveira.
\newblock The {H}utchinson-{B}arnsley theory for generalized iterated function
  systems by means of infinite iterated function systems, 2023.
\newblock \href {https://arxiv.org/abs/2204.00373} {\path{arXiv:2204.00373}}.

\bibitem[PE98]{PavlovicE98}
Dusko Pavlovic and Mart{\'{\i}}n~H{\"{o}}tzel Escard{\'{o}}.
\newblock Calculus in coinductive form.
\newblock In {\em Thirteenth Annual {IEEE} Symposium on Logic in Computer
  Science, Indianapolis, Indiana, USA, June 21-24, 1998}, pages 408--417.
  {IEEE} Computer Society, 1998.

\bibitem[PP02]{PavlovicP02}
Dusko Pavlovic and Vaughan~R. Pratt.
\newblock The continuum as a final coalgebra.
\newblock {\em Theor. Comput. Sci.}, 280(1-2):105--122, 2002.

\bibitem[Rab93]{Rabinovich1993Traces}
Alexander~Moshe Rabinovich.
\newblock A complete axiomatisation for trace congruence of finite state
  behaviors.
\newblock In Stephen~D. Brookes, Michael~G. Main, Austin Melton, Michael~W.
  Mislove, and David~A. Schmidt, editors, {\em Mathematical Foundations of
  Programming Semantics, 9th International Conference, New Orleans, LA, USA,
  April 7-10, 1993, Proceedings}, volume 802 of {\em Lecture Notes in Computer
  Science}, pages 530--543. Springer, 1993.
\newblock \href {https://doi.org/10.1007/3-540-58027-1\_25}
  {\path{doi:10.1007/3-540-58027-1\_25}}.

\bibitem[RBB{\etalchar{+}}17]{RotBBPRS17}
Jurriaan Rot, Filippo Bonchi, Marcello~M. Bonsangue, Damien Pous, Jan Rutten,
  and Alexandra Silva.
\newblock Enhanced coalgebraic bisimulation.
\newblock {\em Math. Struct. Comput. Sci.}, 27(7):1236--1264, 2017.
\newblock \href {https://doi.org/10.1017/S0960129515000523}
  {\path{doi:10.1017/S0960129515000523}}.

\bibitem[Rut00]{Rutten00}
Jan J. M.~M. Rutten.
\newblock Universal coalgebra: a theory of systems.
\newblock {\em Theor. Comput. Sci.}, 249(1):3--80, 2000.
\newblock \href {https://doi.org/10.1016/S0304-3975(00)00056-6}
  {\path{doi:10.1016/S0304-3975(00)00056-6}}.

\bibitem[Sok11]{Sokolova11}
Ana Sokolova.
\newblock Probabilistic systems coalgebraically: {A} survey.
\newblock {\em Theor. Comput. Sci.}, 412(38):5095--5110, 2011.
\newblock \href {https://doi.org/10.1016/J.TCS.2011.05.008}
  {\path{doi:10.1016/J.TCS.2011.05.008}}.

\bibitem[SS00]{Stark2000Probabilistic}
Eugene~W. Stark and Scott~A. Smolka.
\newblock A complete axiom system for finite-state probabilistic processes.
\newblock In Gordon~D. Plotkin, Colin Stirling, and Mads Tofte, editors, {\em
  Proof, Language, and Interaction, Essays in Honour of Robin Milner}, pages
  571--596. The {MIT} Press, 2000.

\bibitem[SS11]{Silva2011Probabilistic}
Alexandra Silva and Ana Sokolova.
\newblock Sound and complete axiomatization of trace semantics for
  probabilistic systems.
\newblock In Michael~W. Mislove and Jo{\"{e}}l Ouaknine, editors, {\em
  Twenty-seventh Conference on the Mathematical Foundations of Programming
  Semantics, {MFPS} 2011, Pittsburgh, PA, USA, May 25-28, 2011}, volume 276 of
  {\em Electronic Notes in Theoretical Computer Science}, pages 291--311.
  Elsevier, 2011.
\newblock \href {https://doi.org/10.1016/j.entcs.2011.09.027}
  {\path{doi:10.1016/j.entcs.2011.09.027}}.

\bibitem[Str15]{strobin_attractors_2015}
Filip Strobin.
\newblock Attractors of generalized {IFSs} that are not attractors of {IFSs}.
\newblock {\em Journal of Mathematical Analysis and Applications},
  422(1):99--108, 2015.
\newblock \href {https://doi.org/10.1016/j.jmaa.2014.08.029}
  {\path{doi:10.1016/j.jmaa.2014.08.029}}.

\end{thebibliography}
    
\end{document}